\documentclass[%
 superscriptaddress,
 amsmath,amssymb,
 aps,
 pra,
 longbibliography,
 twocolumn
]{revtex4-2}
\pdfoutput=1

\usepackage[utf8]{inputenc}
\usepackage{graphicx}% Include figure files
\usepackage{dcolumn}% Align table columns on decimal point
\usepackage{bm}% bold math
\usepackage{physics}
\usepackage{comment}
\usepackage[ruled]{algorithm2e}
\usepackage[svgnames]{xcolor} %for marking text with different color, useful for revisions.
\usepackage[colorlinks=true,citecolor=blue,linkcolor=blue]{hyperref}
\hypersetup{
    allcolors  = {blue},
}
\usepackage{listings}
\usepackage{tikz}

\usepackage{soul}
\usepackage{color}

\definecolor{ColorNoteDenis}{RGB}{100, 0, 200}
\definecolor{ColorModDenis}{RGB}{100, 0, 200}
\definecolor{ColorRemove}{RGB}{255, 0, 0}

\definecolor{ColorOption1}{RGB}{0, 0, 255}
\definecolor{ColorOption2}{RGB}{0, 150, 0}
\definecolor{ColorDuplicate}{RGB}{255,140,0}

\begin{document}

\title{Training continuously-coupled reconfigurable photonic chips with quantum machine learning}
\author{Denis Stanev}
\affiliation{Gran Sasso Science Institute, Viale Francesco Crispi 7, I-67100 L’Aquila, Italy}
\author{Nicol\`o Spagnolo}
\affiliation{Dipartimento di Fisica, Sapienza Universit\`{a} di Roma, Piazzale Aldo Moro 5, I-00185 Roma, Italy}
\author{Fabio Sciarrino}
\email{fabio.sciarrino@uniroma1.it}
\affiliation{Dipartimento di Fisica, Sapienza Universit\`{a} di Roma, Piazzale Aldo Moro 5, I-00185 Roma, Italy}

\begin{abstract}
Integrated photonic technologies have recently shown significant advances, enabling the possibility to implement reconfigurable interferometers with increasing size. One of the main tasks to fully exploit the capabilities of reconfigurable integrated interferometers is the possibility to precisely program their operation to perform a desired target unitary. While recipes are known for circuit layouts based on a cascade of beam-splitter and phase-shifter operations, a methodology applicable for reconfigurable continuously-coupled waveguide arrays is currently missing. Here, we devise a machine learning based approach for this task, using a black box methodology that does not rely on precise a-priori modeling of the circuit internal architectures. We verify the effectiveness and the robustness of this approach via numerical simulations on different continuously-coupled waveguides layouts, either with planar or 3D structures. The proposed method makes use of a limited number of single- and two-photon measurements, making it suitable for optical quantum information processing. The obtained results open the perspective of employing this methodology as an effective tool to program the operation of integrated interferometers designed via different architectures. 
\end{abstract}

\maketitle

\section{Introduction}

Over the last decade, quantum devices have shown a rapid growth and evolution, with some having already claimed the first instances of quantum advantage \cite{QuantumSupGoogle,QuantumSupChina,QuantumSupGBSCHN,QuantumSupSuCHN,XanaduQAdv, Jiuzhang3, QuantumSupIBM, GoogleWillow}. This technological development has been accompanied by the adoption of different architectures and technologies for the fabrication of quantum devices, each with its own set of advantages and drawbacks. Among the different experimental platforms, photonic systems have been shown to be a relevant approach to scale-up to large dimensional quantum systems \cite{XanaduQAdv, Jiuzhang3}. The technological development has been also accompanied to the theoretical definition of different encodings and protocols for quantum information processing, specifically suitable for photonic platforms \cite{PhotonicTech, LoopLinear, BartlettLoop, TimeBinCPHASE, TimeBinCarosini, flamini18, QuantuCompCV, GKP_Orig, LinearOpticalUniversalQuantumGate, PsiquantumFusion, LoopLinearMultiTime, BosonSamplingPaesani}. Some of the more common approaches to process quantum information with photons involve using linear optical components, such as beamsplitters and phase shifters, as a building block for complex interferometric networks. Several recipes are already known to decompose an arbitrary target unitary into a sequence of elementary components \cite{ReckDecomposition, ClementsDecomposition}, with the number of elements scaling typically as $O(m^2)$, being $m$ the number of optical modes. As an effective platform to implement multi-mode interferometer with this architecture, recent experiments have shown the realization of progressively larger devices with integrated photonics \cite{IntegratedReview}.

A different approach to implement multi-mode interferometric networks with integrated photonics involves using an architecture inspired by continuous-time quantum walks. More specifically, a set of continuously-coupled waveguides permits to implement a linear unitary dynamics, where the effective transformation depends on the mutual distance between the modes and on the interaction length \cite{peruzzo2010}. Furthermore, recent technological advances enabled the integration of reconfiguration capabilities via either thermal or elecro-optic phase shifters \cite{3DChipLab, CCElectro1, CCElectro2}. This architecture allows for compact designs, which enable coupling of several modes with a reduced circuit length, thus potentially reducing the impact of losses. Such a property have been further enhanced with recent demonstration of continuously-coupled multi-mode interferometers with a 3D structure, which enables improved connectivity between modes with respect to planar designs \cite{CC3D1, CC3D2, CC3D3, CC3D4, CC3D5, 3DChipLab}. However, there are still challenges in the adoption of this architecture to obtain fully-programmable and controllable devices, given that there is currently no known general algorithm that links the circuit parameters to the implementation of specific target unitaries.

Recently, machine learning has seen a surge in uses in all types of fields \cite{MLPhysics, MLMed, TransformerNN, LLMDeepseek, MLComputerVision}, and has proven to be a useful tool when used in conjunction with quantum devices \cite{DenisCloning, QONN, QMLReview, DenisAnomaly, BartlettUniversal, QuantumLabExtremeLearning, QuantumLabAdaptiveBS, PhotonicNonlinearity}. The ability of machine learning techniques to work even on systems that are not well characterized, using a black box approach, makes them suitable for scenario in which a complete characterization by standard means might be challenging and expensive. Hence, these methods may prove to be applicable to the aforementioned task of programming multi-mode interferometers.

In this work, we devise a machine learning based approach to program reconfigurable integrated interferometers based on continuously-coupled waveguides. The advantages of this approach relies on the lack of needing the full reconstruction of the unitary matrices for a comprehensive set of system parameters, and on employing a limited number of single- and two-photon measurements. We then showcase that the method can be effective for different circuit layouts and programmabilities, including some of the recently demonstrated architectures based on 3D geometries \cite{3DChipLab}.

This work is organized as follows. In Sec. \ref{sec:ContinuousCoupling} we discuss the theoretical framework for continuously-coupled integrated interferometers. Then in Sec. \ref{sec:Methods} we describe the methods at the basis of the devised machine-learning approach acting as local optimization. In Sec. \ref{sec:Results} we discuss the results of numerical simulations of circuits with different layouts used to test the effectiveness of the method. In Sec. \ref{sec:LocalToGlobal}, we then discuss possible approaches to implement the initial step of the optimization, complementing the local optimization approach. Finally in Sec. \ref{sec:Discussion} we provide some concluding remarks.

\section{Implementing linear interferometers via continuously coupled waveguides}
\label{sec:ContinuousCoupling}

\begin{figure}[ht!]
\includegraphics[width=0.49\textwidth]{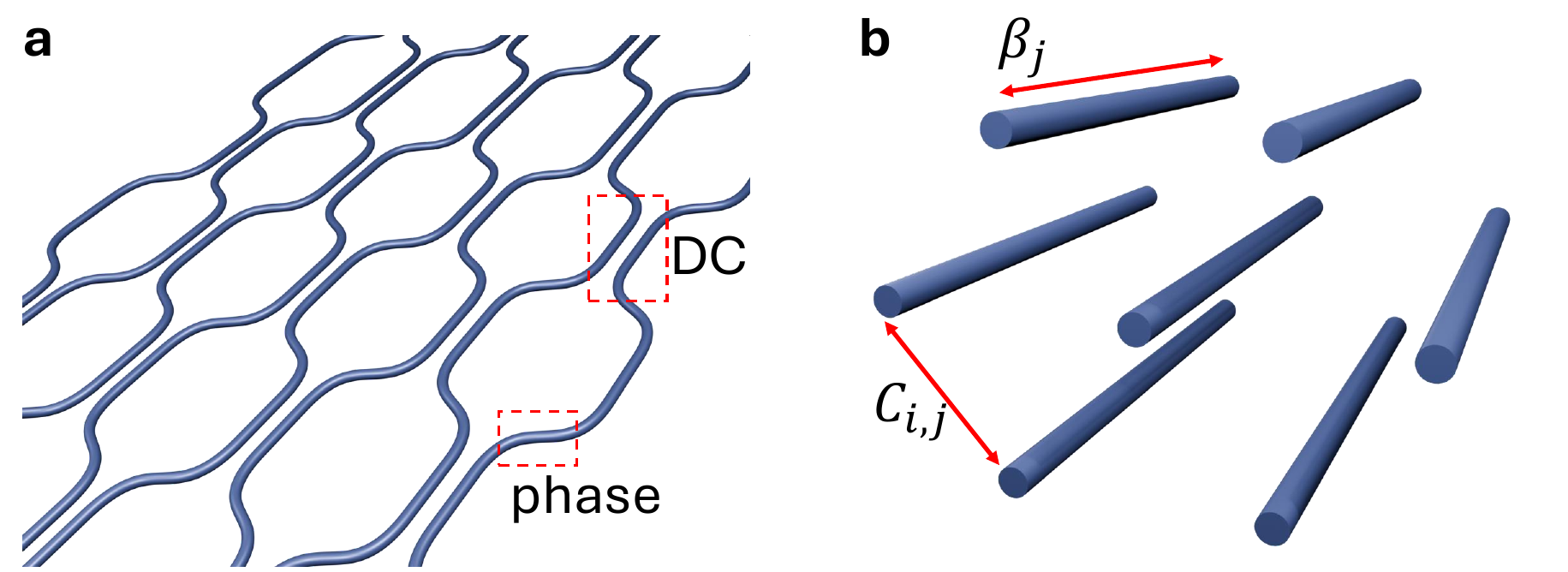}
\centering
\caption{\textbf{Implementation of linear optical unitary transformation with integrated photonics}. (a) Schematics of unitary decomposition via a discrete approach. The unitary matrix is obtained by a set of $2 \times 2$ elementary cells, composed of a directional coupler (DC) with arbitrary transmittivity and a phase shift, arranged in a suitable layout. (b) Interferometer composed by a network of continuously-coupled waveguides. The relevant parameters describing the evolution are represented by the propagation constants $\{ \beta_{i}\}$, and by the coupling coefficients $\{C_{i,j}\}$.}
\label{fig:discrete_continuous}
\end{figure}

As briefly discussed in the introduction, implementation of linear optical unitary transformations can be performed by following two different approaches. As a first approach, a generic $m \times m$ transformation between optical modes can be implemented via a structure composed of discrete $2 \times 2$ cells, where each cell is composed by a beam-splitter of arbitrary transmittivity and a phase shifter in one of the modes (see Fig. \ref{fig:discrete_continuous}a). Different layouts \cite{ReckDecomposition, ClementsDecomposition} have been identified leading to implementation of arbitrary unitary transformations, requiring $O(m^2)$ elementary cells arranged in $O(m)$ layers, each accompanied by a related algorithm to deterministically find the circuit parameters to implement a chosen overall transformation. This approach has the advantage of having a specific recipe for its programming, and is accompanied by a well defined scaling with losses as $O(\eta^{m})$, where $\eta$ is the transmission per element.

As a second approach, integrated devices composed of continuously coupled waveguides do not rely on the composition of multiple discrete optical units to achieve a final target unitary. Conversely, coupling between the modes is obtained by having the set of optical waveguides arranged so that their distance allows for mutual coupling for the complete circuit length (see Fig. \ref{fig:discrete_continuous}b). This can be described by an Hamiltonian $\hat{\mathcal{H}}$ between the mode operators $\{\hat{a}_{i}, \hat{a}^{\dag}_{i}\}$ of the form:
\begin{equation}
\label{eq:ham}
\hat{\mathcal{H}} = \sum_{i=1}^{m} \beta_{i} \hat{a}^{\dag}_{i} \hat{a}_{i} + \sum_{i,j \neq i}^{m} C_{i,j} \hat{a}^{\dag}_{i} \hat{a}_{j},
\end{equation} 
where the coefficients $\beta_i$ are the propagation constants of the modes, while the off-diagonal elements $C_{i,j}$ are the coupling coefficients between different modes. In the case of a chip with $m$ modes with a planar structure and first-neighbour connectivity, the coupling coefficients are non-zero only if $j=\{i-1,i+1\}$. The effective unitary operator $\hat{U}(z)$ induced by a circuit of length $z$ can be obtained from the Hamiltonian \eqref{eq:ham}. It is important to note that the value of the parameters of the Hamiltonian can change over the length of the circuit, based on its structure. In that case, one needs to divide the circuit in $k$ blocks of length $\Delta z$, with $z = k \Delta z$, each described by an Hamiltonian $\hat{\mathcal{H}}_{l}$ and a corresponding unitary operator $\hat{U}_{l}(\Delta z)$. The overall unitary operator is obtained $\hat{U}(z) = \hat{U}_{m}(\Delta z) \cdots \hat{U}_{1}(\Delta z)$. Considering the form of $\hat{\mathcal{H}}$, the transformation induced by this class of systems is linear between the modes, and is thus described by a unitary matrix $U_{i,j}$ that defines the input-output relations for the mode operators $\hat{b}_{i} = \sum_{j} U_{i,j} \hat{a}_{i}$ which are obtained from $\hat{b}_{i} = \hat{U}(z) \hat{a}_{i} \hat{U}^{\dag}(z)$. 

The off-diagonal $C_{i,j}$ elements of the Hamiltonian are given by the structure of the chip, since they depend on the distance between the waveguides, and are thus typically fixed in the fabrication process. Conversely, the diagonal elements $\beta_i$ can be actively modified, through the use of the thermo-optic \cite{3DChipLab} or of the electro-optic \cite{CCElectro1, CCElectro2} effect. Modifying the propagation constants of the modes using thermo-optic effects is usually achieved by placing resistors on the chip surface that, when given a certain input current, change by heating the propagation constants $\beta_i$ of the affected modes.

Albeit some recipes \cite{Tang22} based on stochastic quantum walks \cite{Whit10} to generate Haar-randomness from continuously-coupled waveguide arrays are known, the challenge is the lack of a known procedure to translate the coefficients of a desired unitary matrix $U$, of elements $U_{i,j}$ in a set of parameters to program a given interferometer. In this work, we intend to tackle this issue by using machine learning techniques to train continuously-coupled arrays with thermo-optic tuning to achieve specific unitaries. We test our methods on different waveguide structures, both planar and 3D. Furthermore, we will also consider a scenario where thermo-optic tuning does not allow to control only the propagation constant $\beta_{i}$ of a single given mode due to the presence of thermal cross-talks. This will be accompanied also by an analysis of the impact on the performance caused by performing the training process with a finite number of measurements.

\section{Methods}
\label{sec:Methods}

\begin{figure}[ht!]
\includegraphics[width=0.49\textwidth]{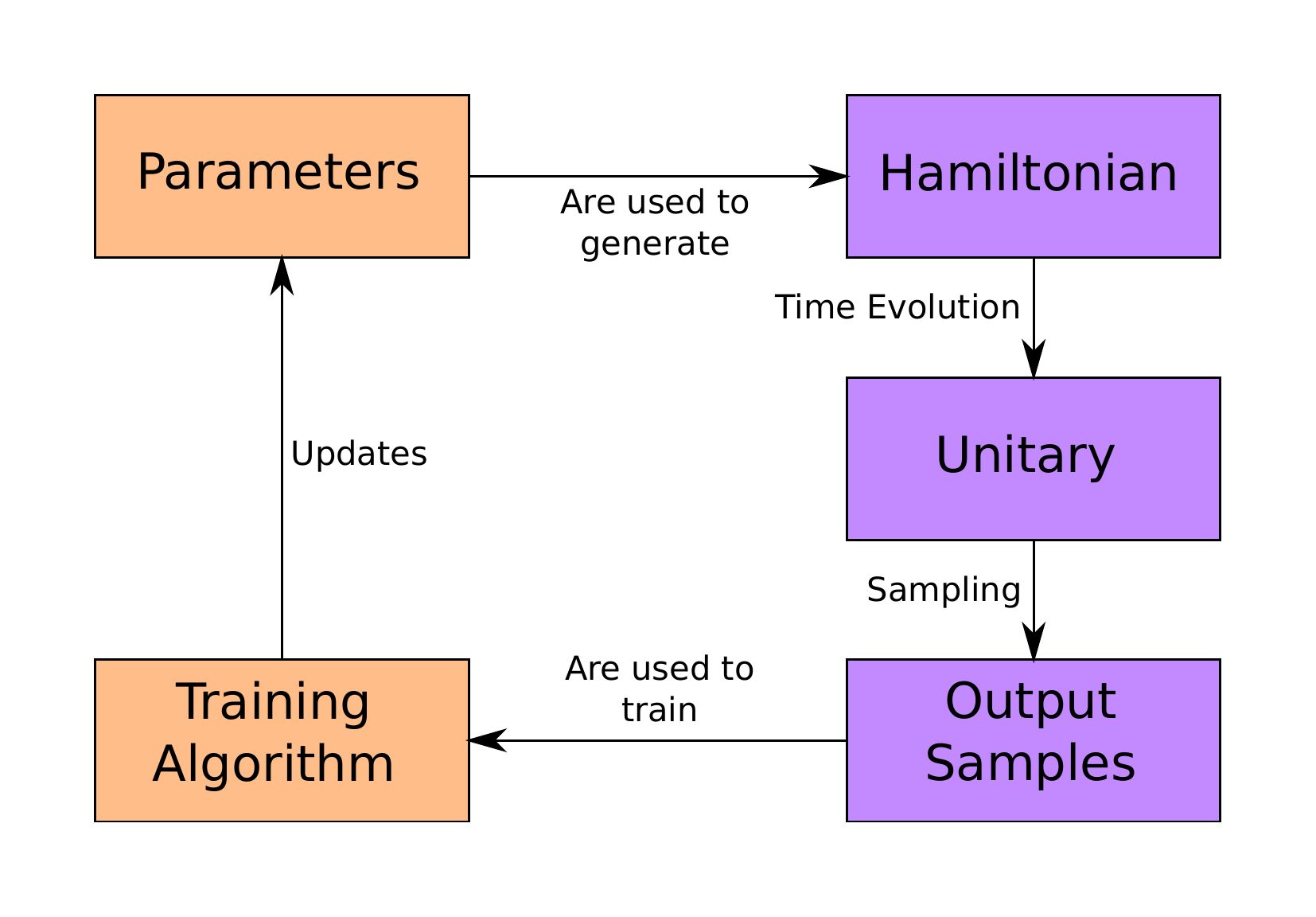}
\centering
\caption{\textbf{Schematic of the training cycle}. Logical sequence of operations to perform the training process in the algorithm. In purple we highlight the parts that are performed on the quantum hardware, while in orange the parts performed on classical hardware.}
\label{fig:Schematic}
\end{figure}

Our approach to program reconfigurable interferometers based on continuously-coupled waveguide arrays is based on a machine learning approach, where the set of parameters to be applied to obtain a given unitary transformation is determined through a training phase. A schematic of the training cycle can be seen in Fig. \ref{fig:Schematic}. More specifically, the cycle embeds a sequence composed of (i) choosing a set of system parameters, (ii) applying them to the reconfigurable interferometer, (iii) measuring a finite sample at its output after feeding the circuit with a chosen set of input states, and (iv) determining through the training algorithm the set to be applied to the circuit in the subsequent step. The training phase stops after a given halting condition is met. 

As shown in Fig. \ref{fig:Schematic}, this cycle involves two main blocks depending on the hardware employed at the corresponding state. A part of the process is performed on the quantum hardware (A), namely setting the parameters and measuring the samples, and a part of the process is performed on a classical hardware (B), namely the training algorithm which decides the set of parameters to be inserted at the next step. In our case, part (A) was performed by testing our approach on numerical simulations of the quantum hardware. The latter is performed by exploiting the Python libraries Qutip \cite{Qutip} and Perceval \cite{Perceval}. Conversely, in part (B) the training code is mostly ad-hoc implemented for this analysis. We describe below all the relative steps carried out in the performed analysis.

\subsection{Simulation of continuously-coupled waveguide arrays}

\begin{figure}[h!]
\centering
\includegraphics[width=0.49\textwidth]{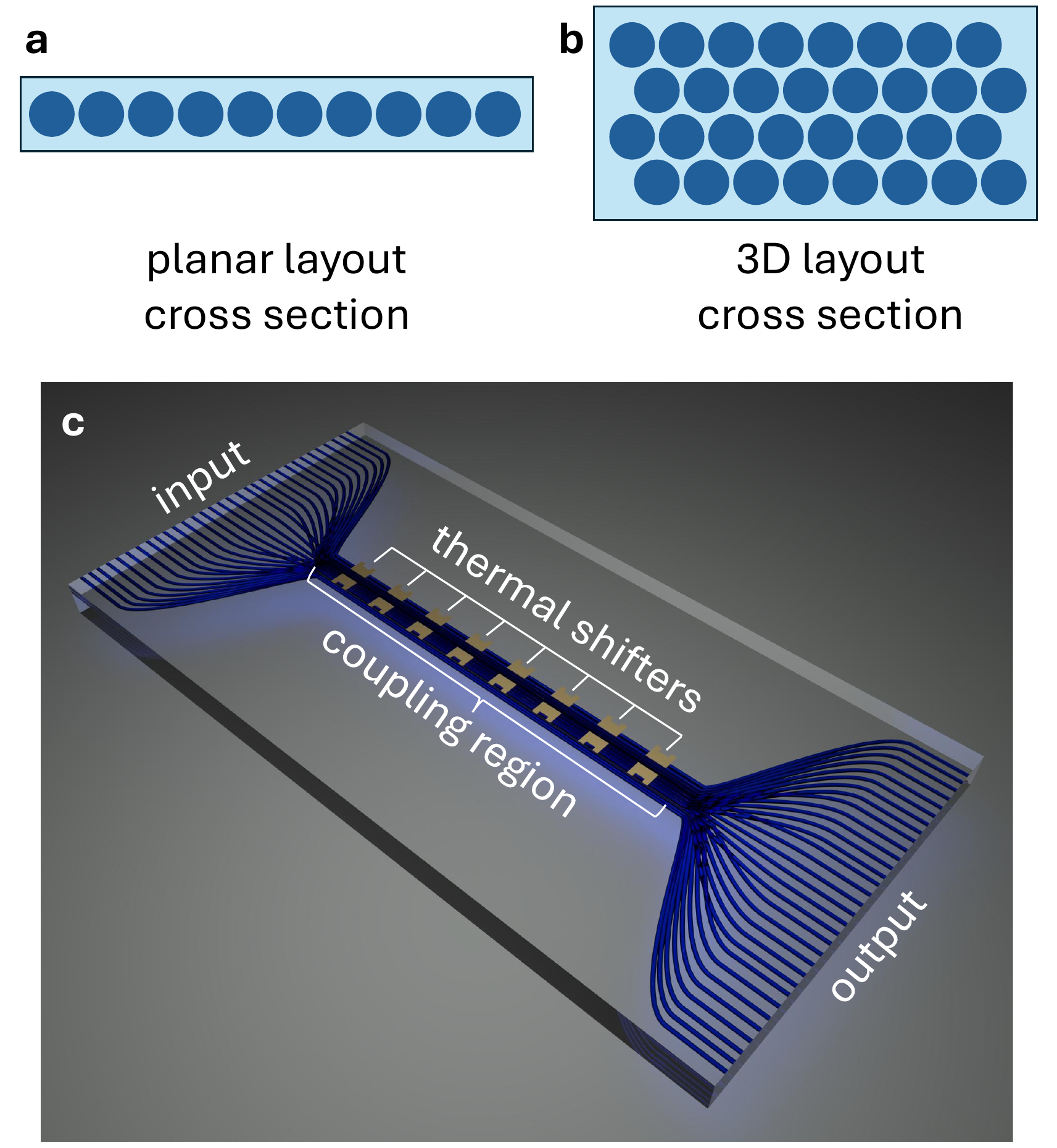}
\caption{\textbf{Structure of the simulated architectures}. \textbf{a,b}, Cross-section of the different interferometer layout, where circle represents a mode. \textbf{a}, Planar design, in this case representing the scenario with 10 modes. \textbf{b}, 3D structure with triangular mesh with 32 modes. \textbf{c}, Visual representation of the architecture for 3D interferometers with multiple segments. The thermal resistors are placed on top of the device surface, and control the propagation constant of the underlying modes.}
\label{fig:MeshCrosssection}
\end{figure}

In order to address the capability of the algorithm to find a chosen target unitary for different device architectures, we simulate various sizes and designs of optical continuously-coupled waveguide arrays. More specifically, we first started by considering integrated interferometers with a planar structure, such as those reported in \cite{peruzzo2010, CCElectro1, CCElectro2}, with varying number of optical modes. Then we also considered integrated interferometers with a 3D structure \cite{CC3D1, CC3D2, CC3D3, CC3D4, CC3D5}, more precisely a triangular structure reproducing the one of Ref. \cite{3DChipLab}. We show in Figs. \ref{fig:MeshCrosssection}a-b the cross-section of the two different geometric architectures simulated in the present study. 

A crucial aspect of the simulation process is how we introduce the active reconfigurability in the device through thermo-optic resistors. These elements inserted modifications in the index of refraction of the underlying waveguide(s) which depend on the amount of dissipated power in the resistor, due to the thermo-optic effect. Hence, thermo-optic resistors can be in principle used to modify the propagation constants $\beta_i$ in Eq. \eqref{eq:ham}. However, one needs to consider that in general, due to the heat propagation process, thermal cross-talks can be present. This lead to inducing modifications of $\beta_{j\neq i}$ even if the resistor is placed on top of waveguide $i$. This is particularly relevant for 3D geometries, where the heat propagation progress can affect not only the waveguide placed right below the resistors, but also those buried deeper below in the structure. To take into account this relevant aspect for actual implementations, we progressively performed a more physically-detailed simulation of the process. More specifically, we first started by considering the case in which we have individual control over all of the modes. Then, in the case of the 3D structure, we also considered the case where tuning the current in a single resistor does not address only the propagation constant of a single mode, but affects multiple modes simultaneously (see also Ref. \cite{3DChipLab}). Finally, we modified the architecture of the simulated interferometers taking into account more complex geometries for the resistors \cite{3DChipLab}, namely the scenario where the devices is performed via multiple connected segments which can be controlled independently (see Fig. \ref{fig:MeshCrosssection}c). More details can be found in App. \ref{app:continuously-coupled}. 

The numerical simulation process of the interferometer evolution is performed with the following approach. First, we generate the Hamiltonian, representing the transformation performed by a cross-section of the chip, with a chosen set of parameters. Then, we calculate the unitary matrix $U$ resulting from a chosen length $z$ for the chip. Then we use the obtained unitary matrix to simulate a set of output events, with a specified number of samples, resulting from injecting the circuits with the set of input states used for the training. For the training process, we employ single- and two-photon input states. More specifically, the complete set of single-photon input states is probed, equal to the number of modes, following the consideration that this set provides information on the moduli of the unitary matrix elements ($\vert U_{i,j} \vert$). For the two-photon input states, we use a limited set of input configurations to also obtain information on the complex phases of the transformation, where the chosen set depends on the structure of the chip and on the training method. 

\subsection{Training}

We now describe the training procedure for our method. 
A schematic description of the used procedure is reported in Algorithm \ref{alg:Training}. The aim is to program the circuit to perform a specific target unitary $U^{t}$, defined initially before the starting point. In principle, for the training to be possible in an exact form, there should exist a set of parameters $\{ \beta^{t}_{i}, C^{t}_{i,j} \}$ which correspond to the unitary $U^{t}$ according to the Hamiltonian description of Eq. \eqref{eq:ham}. In the experimental platform, the circuit can be programmed via a set of control knobs, provided by the currents applied to the resistors placed on the surface of the interferometer. As discussed previously, application of a current $I_{k}$ in the $k$-th resistor modifies the propagation constants $\{ \beta_{i} \}$ via the thermo-optic effect, according to heat propagation mechanisms. In general, this means that changing the current $I_{k}$ on the resistor placed on top of waveguide $i$ does not imply changing only a single $\beta_{i}$ in the presence of thermal cross-talks. Conversely, the coupling parameters $\{ C^{t}_{i,j} \}$ are unchanged by thermo-optic phase shifters, and thus we will consider them fixed in all performed numerical simulations.

\begin{algorithm}
\caption{Training of the continuously coupled optical chip}\label{alg:Training}
\KwIn{ Chip and training settings, initial value of trainable parameters  $\{\beta_i^{(0)}\} = \{\beta_1^{(0)}, \beta_2^{(0)}, ..., \beta_n^{(0)}\} $ }
\KwResult{Trained parameters $\{\beta_i^{(f)}\} $, where $f$ indicates the last epoch of the training.}
$l$ = 0 \;
$f$ = number of epochs \;
\While { $l < f$}{
    $\{\beta_i^{(l+1)}\} = \{\beta_i^{(l)}\}$ \;
    Choose index $i0$ of the parameter to modify at this iteration \;
    Choose subset of available two-photon input pairs to use for this epoch \;
    Compute $l_{prev}$ = $L(\{ \beta_{i}^{(l)} \})$  \;
    Compute $l_{up}$ = $L(\{ \beta_{i}^{(l)} \} + \delta_{i0} )$ \;
    Compute $l_{down}$ = $L(\{ \beta_{i}^{(l)} \} - \delta_{i0})$ \;
  \tcc{Choose in which direction to shift $\beta_{i0}^{(l)}$ to minimize the loss. The exact value of $\delta^{shift}$ depends on the training settings.}
  \eIf{$(l_{up} < l_{prev})$ and $(l_{down} < l_{prev})$}{  
    \eIf{($l_{down} < l_{up})$}{
        $\beta_{i0}^{(l+1)}$ = $\beta_{i0}^{(l)}$ - $\delta^{shift}$ \; 
    }
    {
        $\beta_{i0}^{(l+1)}$ = $\beta_{i0}^{(l)}$ + $\delta^{shift}$ \; 
    }
  }
  {\eIf{$(l_{up} < l_{prev})$}{
      $\beta_{i0}^{(l+1)}$ = $\beta_{i0}^{(l)}$ + $\delta^{shift}$ \; 
    }
    {
     \eIf{$(l_{down} < l_{prev})$}{
        $\beta_{i0}^{(l+1)}$ = $\beta_{i0}^{(l)}$ - $\delta^{shift}$ \; 
        }
        {
        \tcc{Do not shift the parameter if both directions led to higher loss than its current position}
        $\beta_{i0}^{(l+1)}$ = $\beta_{i0}^{(l)}$ \;   
        }
    }
  }
  $l = l+1$ \;
}
\Return{$\{\beta_i^{(f)}\}$} \;
\end{algorithm}

Our training method works by training directly the parameters of the quantum system. In fact, the only classical part used in the approach is the optimizer, which takes the output distributions from the quantum systems, and decides how to update the parameters based on a cost function. These updated parameters are then provided to the quantum system for the successive epoch of training, repeating the process until a predetermined number of epochs has passed. This means that, when working with real quantum hardware, there is no need to simulate the behaviour of the quantum system classically to perform the training process. This allows in principle to apply our training method to devices of arbitrary size.

The proposed training method is then tailored to allow for training directly on-chip, both by making use of a limited number of single- and two-photon measurements, and by ensuring that the techniques used can be applied directly on an actual device. This also means that we avoid using in the training process techniques that would be possible only in a simulated environment, such as backpropagation, to make our training method suitable for implementation on real hardware. Crucially, our training method treats the optical chip as a black box, and thus does not require extensive characterization and a precise a-priori modeling of the circuit internal architectures. This approach also allows the devised method to work for chips with various layouts, without requiring modifications to the training process. We also implemented several optimizations, such as reducing the number of input pairs that are used in each training epoch, to accelerate the training process.

We now give a short overview of the training process, which will be described below in more detail. We start with a desired target unitary, and a set of starting parameters which will be used as the starting point for the training. We then choose a set of single- and two-photon input states, and collect $N_s$ samples for each chosen input state, from which we obtain a set of output distributions. These output distributions are then used to evaluate a chosen loss function $L(\{ \beta_{i} \})$ between the measured samples and the expected target probabilities. After that, we choose one of the trainable parameters, shift its value, and evaluate the loss at the new position. This allows us to use an adapted version of Finite Differences Stochastic Approximation (FDSA) \cite{FDSA1, FDSA2} to determine how to modify the chosen parameter to reduce the loss value. We then repeat this process a chosen number of times, obtaining as output the parameters to be set on the device to perform our desired transformation. We will now proceed to show in more detail how each of these steps works, including the methods used to increase the efficiency of the training process.

The starting point of the implemented approach is thus an initial set of currents $\{ I^{(0)}_{k} \}$. In the numerical simulation of the method performed here to test its functioning, this corresponds to a set of initial propagation constants $\{ \beta^{(0)}_{i} \}$, while as said the coupling parameters are fixed to $\{ C^{t}_{i,j} \}$. The values of $\{ \beta^{(0)}_{i} \}$ ($\{ I^{(0)}_{k} \}$ in an actual experiment) reflect the initial knowledge on the system before the training algorithm is run. Specifically, the starting parameters $\{ \beta^{(0)}_{i} \}$ are fixed by taking the true parameter values $\{ \beta^{t}_{i} \}$, and applying a set of random shifts $\{ \delta \beta_{i} \}$, whose value depend on the size and mesh of the circuit, and on the amount of initial knowledge on the circuit operation.

The training process then starts by choosing a set of single- and two-photon input states, and collect $N_{s}$ samples for each input state. A relevant aspect of the method is the criteria used to choose the set of states. As discussed above, in all tested scenarios the full set of single-photon inputs is employed. For the two-photon states, we considered different rules depending on the circuit geometry. In the case of the planar design, we use all of the input pairs of first neighbours $(i, i+1)$, plus the input pair $(1, m)$ composed by the first and last mode. This corresponds to an overall amount of $M_1 = m$ single-photon and $M_2 = m$ two-photon input states. In the case of the triangular mesh, we first considered the set of all first neighbour states, which is larger than $m$ due to the higher connectivity of the design. For instance, this corresponds to $M_2 = 73$ distinct mode pairs for a $m=32$ mode device. As a second trial, we tested the performance of the training when reducing the size of this set. This can be seen as similar to the approach used by Stochastic Gradient Descent \cite{SGD}, where only a subset of the dataset is used to approximate the gradient at each step. While in each epoch the training uses only a subset of all the input pairs, if this subset is of an appropriate size it is still informative on the complete evolution. This corresponds to an approximately correct identification of the parameter shift. Also, since the elements of the subset change at each epoch, over the length of the whole training all the input pairs are used to train the model, ensuring that the trained model is not biased toward a specific subset of input pairs. This leads to a significant speed up in the training process, while maintaining a similar final fidelity. We will see in the results section how using different numbers of pairs per epoch influences the training process.

The exploited set of output distributions used at each epoch is then composed by (i) estimates $\{\tilde{p}_{i;j}\}$ of the $M_1$ single-photon probabilities from input $i$ to output $j$, obtained as $\tilde{p}_{i;j} = \tilde{N}_{i;j}/N_{s}$ where $\tilde{N}_{i;j}$ is the number of events obtained for the given configuration, and (ii) estimates $\{ \tilde{p}_{i,k;j,l} \}$ of the $M_2$ two-photon probabilities from input pair $(i,k)$ to output pair $(j,l)$, also in this case obtained as $\tilde{p}_{i,k;j,l} = \tilde{N}_{i,k;j,k}/N_{s}$ where $\tilde{N}_{i,k;j,l}$ is the number of events obtained for the given configuration. In an experimental implementation, each output distribution is collected by directly preparing the chosen states and measuring output samples of the chosen size $N_{s}$ for each input. In the numerical simulation performed here to test the algorithm operation, the estimates $\{ \tilde{p}_{i;j} \}$ and $\{ \tilde{p}_{i,k;j,l} \}$ of the single- and two-photon probabilities are obtained by sampling $N_{s}$ events from the simulated output probabilities  $\{ p^{(l)}_{i;j} \}$ and $\{ p^{(l)}_{i,k;j,l} \}$. The latter are evaluated via the permanent formula from the unitary matrix $U^{(l)}$ corresponding to $\{ \beta^{(l)}_{i}, C^{t}_{i,j} \}$, where $\{ \beta^{(l)}_{i} \}$ are the propagation constants at step $l$, as $p^{(l)}_{i;j} = \vert U_{j,i} \vert^{2}$ and $p^{(l)}_{i,k;j,l} = \vert U_{j,i} U_{k,l} + U_{j,l} U_{k,i} \vert^{2}$. Note that explicit knowledge of $\{ \beta^{(l)}_{i} \}$ and $\{ C^{t}_{i,j} \}$ is needed to perform the numerical simulation, while this is not necessary in an actual experiment since only the currents $\{ I_{k} \}$ are modified, and the training dataset is estimated from the measurements.
 
The loss function $L(\{ \beta_{i} \})$ of the algorithm is chosen to be the Mean Absolute Error (MAE) between the measured samples and the expected probabilities, calculated as:
\begin{equation}
L(\{ \beta_{i} \}) = \frac{1}{M_1}\sum_{i,j}^{M_1} \vert \tilde{p}_{i;j} - p^{t}_{i;j} \vert +  \frac{1}{M_2}\sum_{i,k,j,l}^{M_2} \vert \tilde{p}_{i,k;j,l} - p^{t}_{i,k;j,l} \vert,
\end{equation}
where the sums are extended over the measured input-output combinations, and the target probabilities are calculated directly from $U^{t}$. Note that using this loss function, rather than comparing the fidelity $F = \vert \mathrm{Tr}[(U^{t})^{\dag}  U^{(l)}] \vert/m$ between the target unitary $U^t$ and the actual unitary $U^{(l)}$, can be less expensive in terms of measurement overheads since estimating $F$ requires performing the tomography of the implemented unitary at each step to reconstruct also the phase contributions \cite{UnitaryTomography1, UnitaryTomography2, UnitaryTomography3, UnitaryTomography4}. While we do compute the fidelity $F$ during the simulations to prove that the training is performing well, this information is not used to train the circuit, and is thus not needed when performing this training method on actual quantum hardware. An analysis on the loss landscape with respect to individual parameters is discussed in App. \ref{app:CCLossLandscape}.

One significant challenge present in training this model is that backpropagation, or even just backpropagation-like scaling, is generally not possible, at least without using multiple copies of a state \cite{QuantumBackprop}. Thus, we need to use a different method to estimate the gradient required by the training process. In the anayzed scenario, we found out that an adapted version of Finite Differences Stochastic Approximation (FDSA) \cite{FDSA1, FDSA2} suffices for our scope. For each epoch $l$, the starting point is a set of parameters $\{ \beta^{(l)}_{i} \}$ (or, for an actual experiment, the currents \{$I^{(l)}_{k}\}$). The algorithm then chooses a parameter specified by a randomly-sampled index $i_0$, and requires to estimate $L(\{ \beta_{i} \})$ in three different cases, namely for (1) $\{ \beta^{(l)}_{i} \}$, (2) by shifting $\beta^{(l)}_{i_0}$ of a quantity $+ \delta$ and (3) by shifting $\beta^{(l)}_{i_0}$ of a quantity $- \delta$, where $\delta$ is the learning rate that needs to be adapted during the algorithm. Knowledge of these values for the loss permits to estimate the correct direction to shift $\beta^{(l)}_{i_0}$ and define the set of parameters $\{ \beta^{(l+1)}_{i} \}$ at epoch $l+1$. This approach can be also further adapted to involve in some epochs the estimate of $L(\{ \beta_{i} \})$ in only 2 points. Thus, an epoch requires 2 or 3 full evaluation of the training dataset. In App. \ref{app:CCTrainingDetails} we discuss in more details this adapted version of FDSA, and also discuss prospects of applying other approaches such as Simultaneous Perturbation Stochastic Approximation (SPSA) \cite{SAReview}. 

During the numerical simulation we included also some refinements to the discussed training algorithm to match the physical implementation. For instance, we included an upper and lower bound to the propagation constants, which reflect the finite range of operation of thermo-optic resistors. Furthermore, as discussed above, for the triangular mesh we included the effect that shifting a parameter $\beta_{i}$ produces a cross-talk effect that modifies also neighbours $\beta_{j \neq i}$. In our simulation, we included this effect in a way that mimics heat propagation mechanisms (see App. \ref{app:continuously-coupled}). Finally, we also tested the performance of the algorithm when the target unitary is not necessarily achievable by the device, namely that the thermo-optic shifters do not allow to program the set of parameters $\{ \beta^{t}_{i}, C^{t}_{i,j} \}$ for the exact implementation of $U^{t}$. This was done by slightly changing the values of the coupling parameters $\{ C^{t}_{i,j} \}$, and then using throughout the algorithm those modified parameters during the training.

\section{Results}
\label{sec:Results}

In this section we discuss the results of the different simulations performed to test the method described in the previous sections. We simulated the training process for different systems, namely interferometers ranging from 10 to 50 modes with a planar structure, and interferometers with 32 modes using a 3D triangular mesh. In our analysis, we included in the simulations the effect of the sample size, by varying the number of samples $N_{s}$ for each state in the range $N_{s} \in [10^{3}, 2 \times 10^{5}]$. The training has been shown for all these cases to highlight its capability to work with different structures and limitations, and thus show the flexibility of this method. As figures of merit, we report the values of the loss function $L(\{ \beta_{i} \})$, which is effectively estimated during the training process, and of the infidelity $1 - F$, being $F$ the fidelity between the target unitary and the trained one as defined above. The latter parameter is not used in the training process due to the need of performing expensive unitary tomography at each step, while it expresses how close the trained unitary is to the target unitary. We also show that the training algorithm works properly since, although the fidelity $F$ is not computed during the process, minimization of the loss function corresponds to simultaneous minimization of $1 - F$.

\subsection{Planar layout}

\begin{figure*}[ht!]
\includegraphics[width=0.99\textwidth]{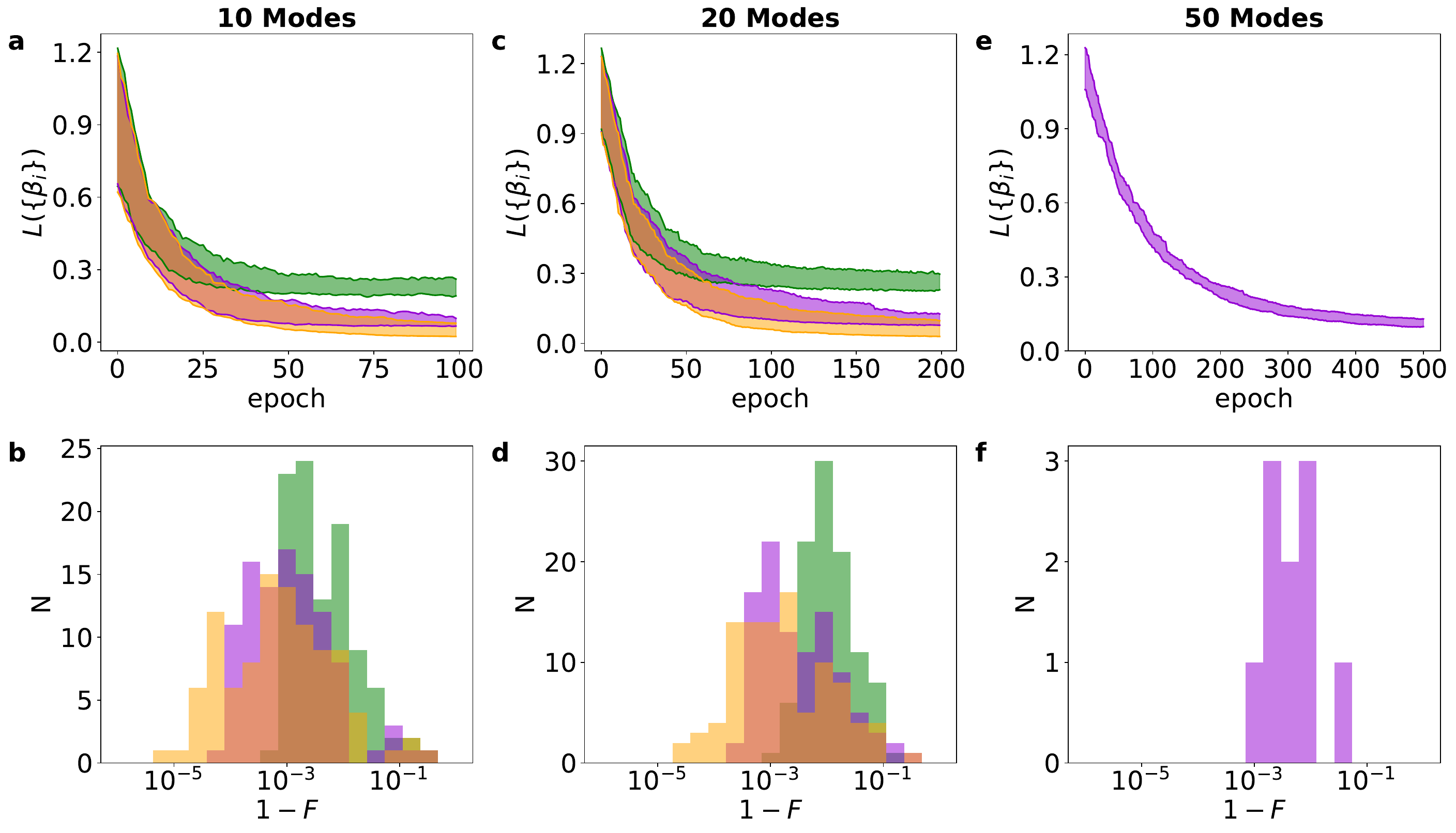}
\centering
\caption{\textbf{Numerical simulations for the planar layout}. \textbf{a}, Loss and \textbf{b}, infidelity for the 10-mode planar layout, with initial fidelity $F_{\mathrm{in}} \sim 0.7$. Green data sets: $N_{s} = 10^3$. Purple data sets: $N_{s} = 10^4$. Orange data sets: $N_{s} = 10^5$. \textbf{c}, Loss and \textbf{d}, infidelity for the 20-mode planar layout, with initial fidelity $F_{\mathrm{in}} \sim 0.7$. Green data sets: $N_{s} = 2 \times 10^3$. Purple data sets: $N_{s} = 2 \times 10^4$. Orange data sets: $N_{s} = 2 \times 10^5$. \textbf{e}, Loss and \textbf{f}, infidelity for the 50-mode planar layout, with initial fidelity $F_{\mathrm{in}} \sim 0.7$. Purple data sets: $N_{s} = 5 \times 10^4$. For the 10 and 20 modes cases, 100 simulations were done for each case, while for the 50 mode case only 10 simulations were performed. For the loss, the shaded zone represents the area between the 10th and 90th percentile of the runs. The histogram x axis is in logarithmic scale.}
\label{fig:PlanarPlots}
\end{figure*}

First, we start by considering interferometers with a planar geometry, assuming that the propagation constant of each mode can be controlled directly via phase shifters placed on top of the surface above the waveguide. We have started by testing our method via numerical simulations for 10-mode interferometers, investigating the role of the sample size statistics per input state. In Fig. \ref{fig:PlanarPlots} a-b we show specifically different trainings for several unitaries, by varying the number of samples $N_{s}$. Each run was performed by having a starting fidelity of $F_{\mathrm{in}} \sim 0.7$, achieved in our simulation with a random shift of $\delta \beta_{i} \in [-0.1,+0.1]$ mm$^{-1}$ for the propagation constants. For an experimental implementation, this corresponds to the need of having a starting knowledge of the circuit operation corresponding to that value of the fidelity. The final average fidelity over 100 runs is reported in Tab. \ref{tab:flatChipFidelity}. From this analysis, we observe that for this system size, an average fidelity $>0.99$ can be achieve already with $N_{s} = 10^{3}$, while considering larger values can lead to further improving the average values of the programming fidelity. Ultimately, the choice of $N_{s}$ has to take into account collecting a larger number of samples $N_{s}$ increases the time required for each epoch of the training, while at a certain point increasing its value $N_{s}$ stops yielding significant gains in the final fidelity of the trained optical chip.

\begin{table}[h!]
\begin{center}
\begin{tabular}{|c | c | c | c |} 
 \hline
 $m$ & $N_{s}$ & $\overline{F}$ & $F_{\mathrm{min}}$ \\
 \hline
 10 & $  10^3$ & 0.9915 &  0.9249 \\
 \hline
 10 & $  10^4$ & 0.9955 &  0.9073 \\
 \hline
 10 & $  10^5$ & 0.9966 &  0.9009 \\
 \hline
 20 & $ 2 \times 10^3$ & 0.9824 &  0.9008 \\
 \hline
 20 & $ 2 \times 10^4$ & 0.9921 &  0.8664 \\
 \hline
 20 & $ 2 \times 10^5$ & 0.9939 &  0.9113 \\
 \hline
 50 & $ 5 \times 10^4$ & 0.9919 &  0.9637 \\
 \hline
\end{tabular}
\caption{\textbf{Summary table of the output fidelity for the simulations with the planar layout}. Average ($\overline{F}$) and minimum ($F_{\mathrm{min}}$) fidelity for the planar layout, for different number of modes and values of $N_{s}$. All results are averaged over $100$ different runs, except for the 50 mode case, for which only $10$ different runs were performed. The starting fidelity for the training was of $F_{\mathrm{in}} \sim 0.7$. For all results, except for the 50 mode one, a couple of the runs were discarded, based on a high loss at the end of training, such as the one shown in Fig. \ref{fig:LossFidelityComparison}.}
\label{tab:flatChipFidelity}
\end{center}
\end{table}

\begin{figure}[ht!]
\includegraphics[width=0.49\textwidth]{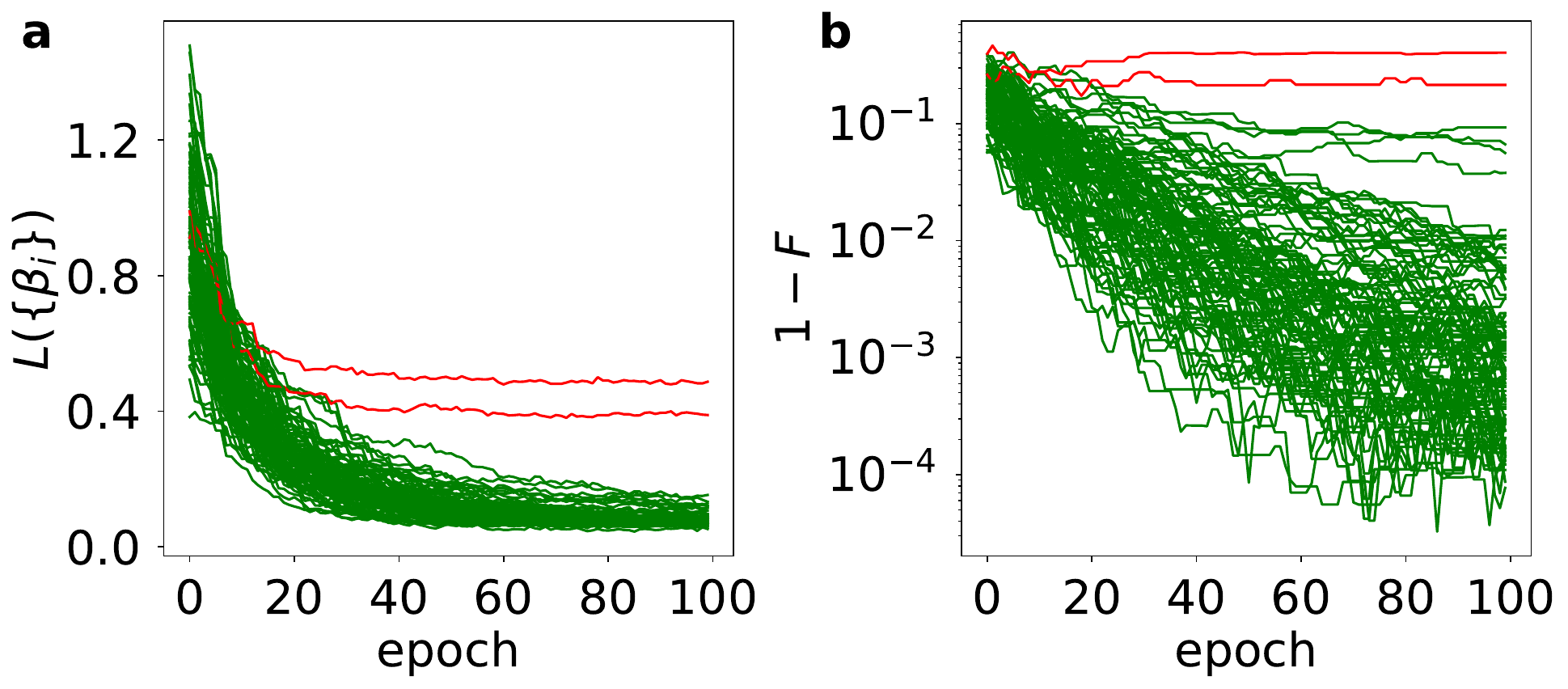}
\centering
\caption{\textbf{Comparison between loss and fidelity for the 10-mode planar layout.} \textbf{a}, Loss and \textbf{b}, infidelity for the 10-mode planar layout, with $N_{s} = 1 \times 10^4$. Each line corresponds to a different run. In the plots, we observe the presence of a few runs characterized by a value of the loss above a certain threshold, marked in red. These runs are also characterized by a low fidelity value. This highlights that runs characterized by low value of the fidelity can be identified from the loss values achieved during the training process, thus enabling the adoption of different techniques to avoid the evolution getting stuck in a local minimum.}
\label{fig:LossFidelityComparison}
\end{figure}

We have then performed the same numerical simulation for the 20-mode scenario, using the same amount of initial shift $\delta \beta_{i}$ in the propagation constants, which corresponds to a starting fidelity in the range $F_{\mathrm{in}} \sim 0.4-0.5$. We observed a similar behaviour to the case with 10 modes in terms of scaling with $N_{s}$, although for this size and starting point we found runs where the training got stuck in a local minimum. However, we can also notice that the lower fidelity reflects in a higher loss. This analysis is shown for the 10 mode case in Fig. \ref{fig:LossFidelityComparison}, and analogous results are obtained in all considered system sizes. Hence, it is possible to recognize early during the process if the algorithm gets stuck in a local minimum. As a result, this permits the use of various techniques to correct the training, either stopping the training early and restarting, or shifting the parameters of a sufficient quantity to move outside the local minimum. We have performed additional analyses by fixing the starting fidelity to $F_{\mathrm{in}} \sim 0.7$ for all circuit sizes. More specifically, we have then performed some numerical simulations in this condition for 20 and 50 modes, whose results are shown in Fig. \ref{fig:PlanarPlots} and in Tab. \ref{tab:flatChipFidelity}.

These analyses show that the method is effective in reaching the target unitary even when scaling up to larger instances, although requiring increasing resources ($N_{s}$, number of epochs) with the system size. Furthermore, we observe that the chosen loss function is appropriate for the training process, since a reduction of the loss corresponds in parallel to an improvement in the fidelity. 

\subsection{3D layout with direct control on the propagation constants}

\begin{figure}[h]
\includegraphics[width=0.49\textwidth]{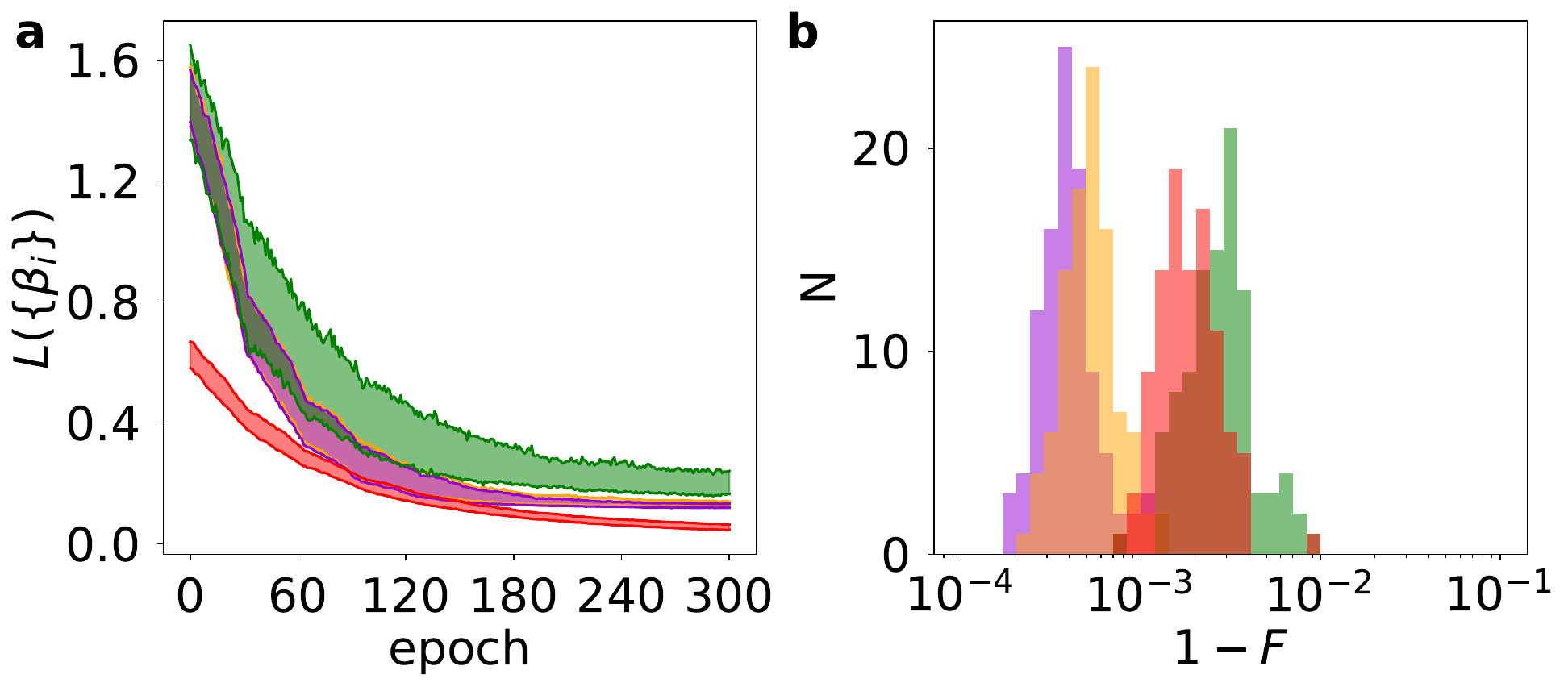}
\centering
\caption{\textbf{Training process for 32-mode, 3D triangular layout interferometers, for different number of two-photon input combinations per epoch.} \textbf{a}, Loss and \textbf{b}, infidelity for the training process. Input pairs for each epoch are chosen randomly from the list of first-neighbour pairs. For each tested number of input combinations per epoch, we report the results for 100 runs. Purple data sets: scenario with all 73 input pairs. Orange data sets: scenario with 10 input pairs. Green data sets: scenario with 1 input pair. Red data sets: scenario by using only single-photon combinations. For the loss, the shaded zone represents the area between the 10th and 90th percentile of the runs. The histogram x axis is in logarithmic scale.}
\label{fig:TriangularFullLimited}
\end{figure}

We then proceeded to test numerically our method on interferometers with 3D triangular layout having $m=32$ modes, and assuming direct control on the single propagation constants without cross-talks. As a first step, we performed some runs with $F_{\mathrm{in}} \sim 0.7$ starting point. As discussed previously, in this scenario we also tested the possibility to reduce the overhead required for the training procedure. More specifically, we tested how the method performed by reducing the number of two-photon input states combinations. Initially, we tested the scenario where, for each epoch, we used all 73 two-photon combinations corresponding to first-neighbour pairs, using $N_{s} = 3 \times 10^4$ samples per input, and performing the training process for 300 epochs. The results are shown in Fig. \ref{fig:TriangularFullLimited}. We observe that the training procedure provides promising results, reaching convergence on average after 200 epochs. We then proceeded to perform the training process by reducing the number of input two-photon pairs for each epoch. This is performed, as previously discussed, by randomly picking at each epoch a subset of the first-neighbour input combinations. The results over 100 runs are shown in Fig. \ref{fig:TriangularFullLimited} and in Tab. \ref{tab:TriangularLimited}. We find that, even with a limited number of 10 input pairs per epoch, the training procedure provides close performances to the case where all 73 combinations are used. However, it is worth noting that a limited number of two-photon combinations is required, since reducing to $1$ or $0$ two-photon pairs per epoch leads to significantly worse performances.

\begin{table}[ht!]
\begin{center}
\begin{tabular}{|c | c | c |} 
 \hline
 \# input pairs & $\overline{F}$ & $F_{\mathrm{min}}$ \\
 \hline
 73(all) & 0.9996 & 0.9988  \\
 \hline
 10 & 0.9994 & 0.9986  \\
 \hline
 1 & 0.9969 & 0.9914  \\
 \hline
 0 & 0.9980 & 0.9911  \\
 \hline
\end{tabular}
\end{center}
\caption{\textbf{Summary table of the output fidelity for the simulations with the triangular layout with direct control}. Average ($\overline{F}$) and minimum ($F_{\mathrm{min}}$) fidelity for the triangular layout with direct control, for different numbers of input pairs used per epoch. All results are averaged over $100$ different instances, with a starting fidelity of $F \sim 0.7$.}
\label{tab:TriangularLimited}
\end{table}

\subsection{3D layout with multiple phase control}

As the next step, we ran simulations by using the model, described above, to consider a scenario similar to the one of Ref. \cite{3DChipLab}. More specifically, we separate the chip in multiple sections, we place resistors on top of the device surface, and introduce the effect of thermal cross-talks due to heat propagation mechanisms. We performed 100 trainings with different starting parameters, starting from a completely random set of propagation constants $\beta_{i}$. 
\begin{figure}[ht!]
\includegraphics[width=0.49\textwidth]{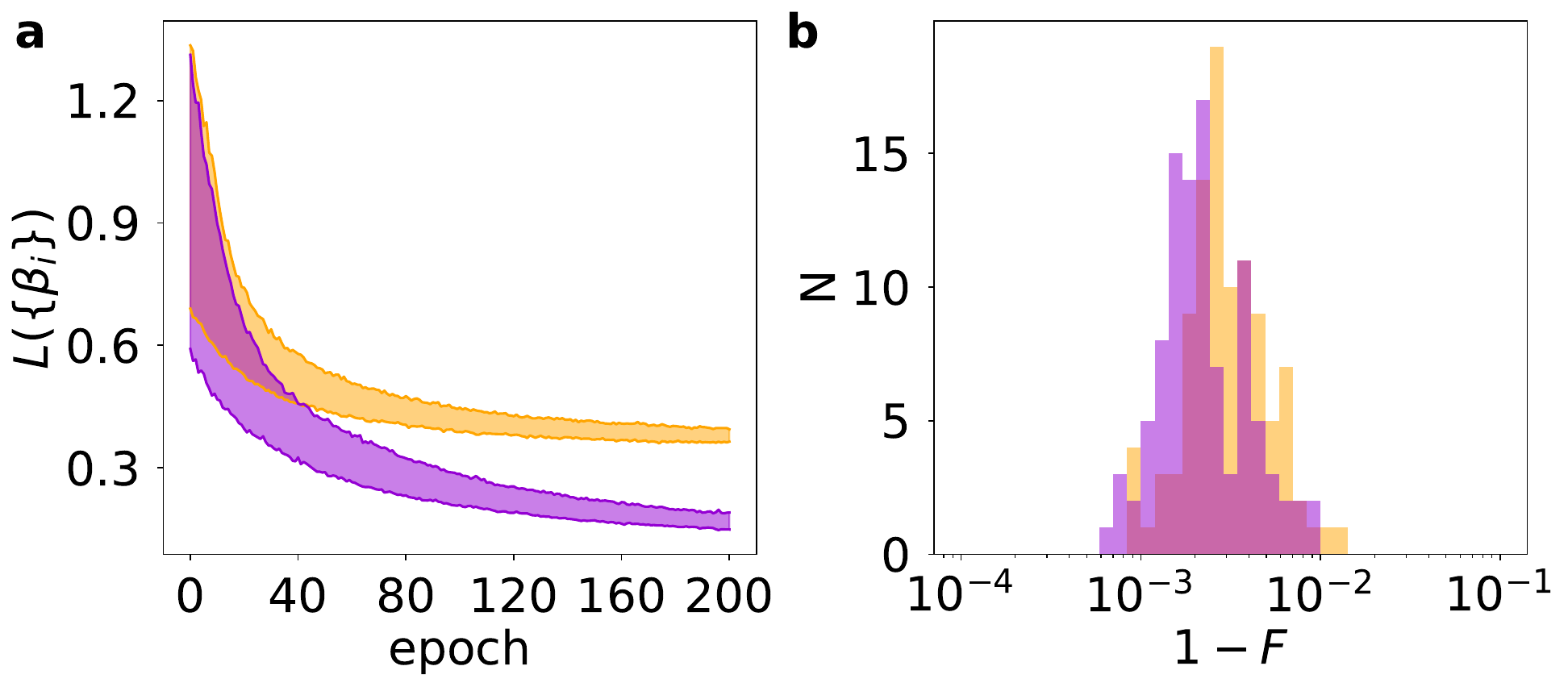}
\centering
\caption{\textbf{Training process for 32-mode, 3D triangular layout interferometers, with multiple phase control.} \textbf{a}, Loss and \textbf{b}, infidelity reached in the training process. The runs were done using only 10 two-photon input configuration per epoch, chosen randomly from the set of first-neighbour mode pairs. Orange data sets: $N_{s} = 3 \times 10^3$ per input. Purple data sets: $N_{s} = 3 \times 10^4$ shots per input. For each tested configuration of number of samples $N_{s}$, we report the results for 100 runs. For the loss, the shaded zone represents the area between the 10th and 90th percentile of the runs. The histogram x axis is in logarithmic scale.}
\label{fig:32ModeChip}
\end{figure} 
Each training was carried out for 200 epochs. The loss and fidelity per epoch of the trainings are shown in Fig. \ref{fig:32ModeChip}, while the obtained values of the fidelities are reported in Tab. \ref{tab:TriangularCrossTalk}. In this case, we observe that, with this kind of model, the training process allows us to obtain convergence regardless of the choice of starting propagation constants $\beta_{i}$, assuming that the target unitary is within the set achievable with the device structure. One of the reasons can be found in the presence of chip structure, where likely multiple parameter configurations can lead to the same target unitary. This potentially helps the training algorithm to find a path that leads to the desired output when starting from different position. 

\begin{table}[ht!]
\begin{center}
\begin{tabular}{|c | c | c |} 
 \hline
 $N_{s}$ & $\overline{F}$ & $F_{\mathrm{min}}$ \\
 \hline
 $3 \times 10^3$ & 0.9966 & 0.9871  \\
 \hline
 $3 \times 10^4$ & 0.9974 & 0.9902  \\
 \hline
\end{tabular}
\end{center}
\caption{\textbf{Summary table of the output fidelity for the simulations with the triangular layout with multiple phase control}. Average ($\overline{F}$) and minimum ($F_{\mathrm{min}}$) fidelity for the triangular layout with multiple phase control, for different values of $N_{s}$. All results are averaged over $100$ different instances, with starting propagation constants $\beta^{(0)}_{i}$ chosen randomly.}
\label{tab:TriangularCrossTalk}
\end{table}

\subsection{3D layouts with shifts in the estimation of the coupling parameters}

\begin{figure*}[ht!]
\includegraphics[width=0.99\textwidth]{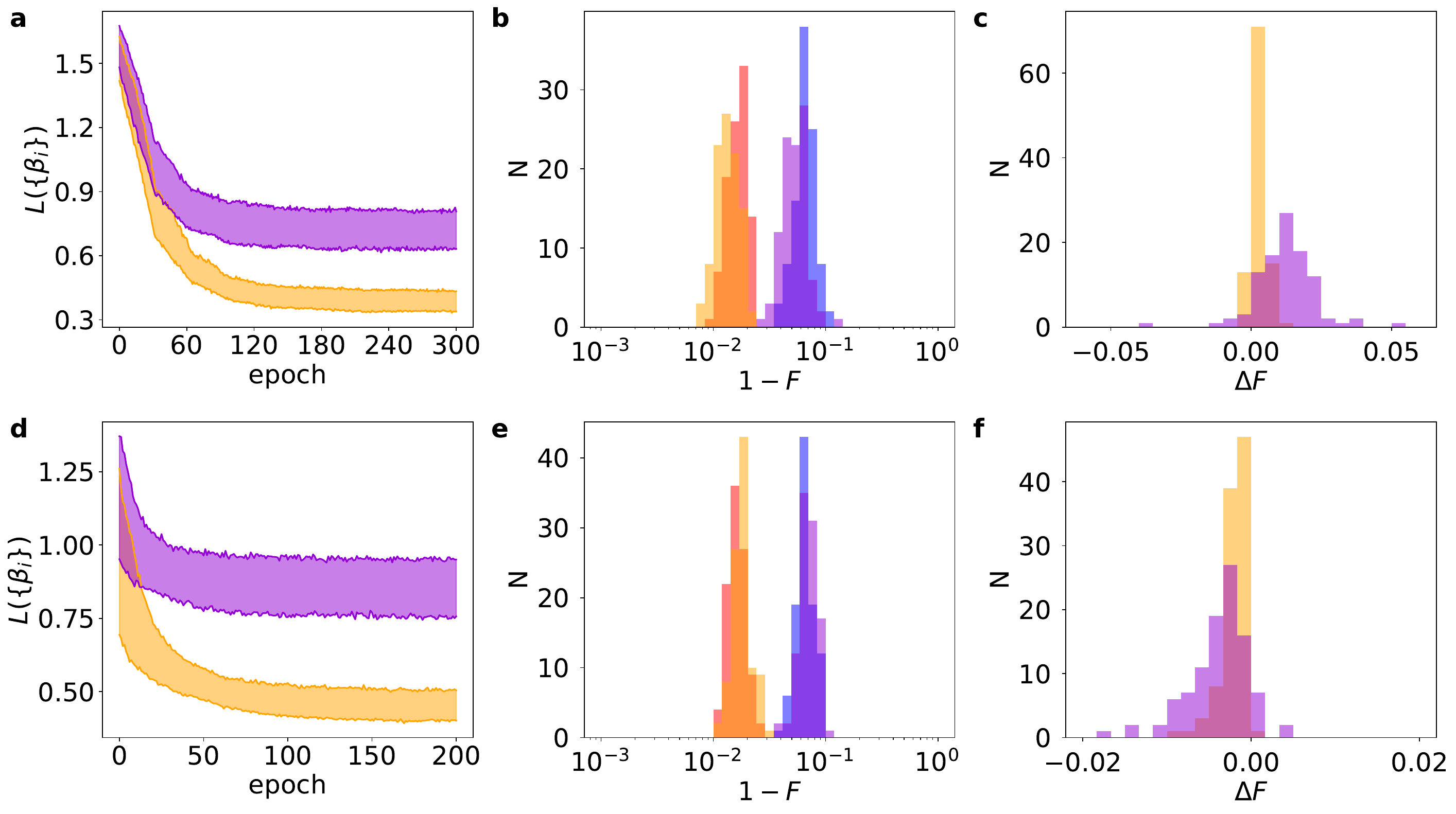}
\centering
\caption{\textbf{Training process for 32-mode, 3D triangular layout interferometers, with different off-diagonal errors.} \textbf{Top row} (\textbf{a}, \textbf{b} \textbf{c}): simulations for the 3D layout with direct control. \textbf{Bottom row} (\textbf{d}, \textbf{e}, \textbf{f}): simulations for the 3D layout with multiple phase control. Orange data sets: $\delta C_{i,j} = 0.01$. Purple data sets: $\delta C_{i,j} = 0.02$. \textbf{a} and \textbf{d}, Loss. For the training, the target unitary is calculated with the correct coupling coefficients $\{ C^{t}_{i,j} \}$ and propagation constants $\{ \beta^{t}_{i} \}$. Panels \textbf{b} and \textbf{e} compare the fidelity ${F}_{\mathrm{base}}$ reached when using the target parameters $\{ \beta^{t}_{i} \}$ (in red and blue) with the fidelity ${F}_{\mathrm{noise}}$ reached when using the final trained parameters $\{ \beta^{f}_{i} \}$ (in orange and purple), in the situation where the shifted coupling coefficients $\{ C^{s}_{i,j} \}$ are used. The x axis is in logarithmic scale. Panels \textbf{c} and \textbf{f} compare the difference in fidelity $\Delta F = {F}_{\mathrm{noise}} - {F}_{\mathrm{base}}$. 100 runs were performed for each configuration. For more information, refer to App. \ref{app:CCOffDiagonal}.}
\label{fig:TriangularOffDiagonal}
\end{figure*}

To complement this analysis taking into account the presence of small errors in the coupling estimations, we also did some simulations inserting shifts in the estimation of the coupling parameters $C_{i,j}$ as explained in the previous section (more details on this simulation procedure can be found in App. \ref{app:CCOffDiagonal}). We did the tests for the 10 mode planar layout, and for the 32 mode triangular layout, in the latter case considering both direct control of the propagation constants and multiple phase control. The results for the triangular layouts are shown in Fig. \ref{fig:TriangularOffDiagonal} and in Tables \ref{tab:TriangularOffDiagonal} and \ref{tab:TriangularOffDiagonal2}. More specifically, we compare the performances in this scenario with the one assuming perfect knowledge on the coupling constants. We observe that, even with a moderate amount of error in the coupling constant knowledge, the training procedure still reaches high fidelity values, close to the ones achieved with $\{ \beta^{t}_{i} \}$. In the case of the 32 mode with direct control of the propagation constants, we can notice that the trained parameters $\{ \beta^{f}_{i} \}$ perform even slightly better than the target parameters $\{ \beta^{t}_{i} \}$, when using the shifted coupling coefficients $\{ C^{s}_{i,j} \}$. This indicates that, for this error regime, the training process within the tested layout was capable to compensate for characterization errors and find a unitary close to the target.

\begin{table}[ht!]
\begin{center}
\begin{tabular}{|c | c | c | c |} 
 \hline
$\delta C_{i,j}$  & $\overline{F}_{\mathrm{base}}$ & $\overline{F}_{\mathrm{noise}}$ & $\overline{\Delta F}$\\
 \hline
 0.01 & 0.9835 & 0.9863 & 0.0028  \\
 \hline
 0.02 & 0.9335 & 0.9456 & 0.0121  \\
 \hline
\end{tabular}
\end{center}
\caption{\textbf{Summary table of the output fidelity for the simulations with the triangular layout with direct control in the case of imperfections in the estimation of $C_{i,j}$}. Average fidelity using the target ($\overline{F}_{\mathrm{base}}$) and trained ($\overline{F}_{\mathrm{noise}}$) parameters while using the shifted values of $C_{i,j}$, as well as their difference $\Delta F = {F}_{\mathrm{noise}} - {F}_{\mathrm{base}}$, for different values of the shift $\delta C_{i,j}$. All results are averaged over $100$ different instances, with a starting fidelity of $F_{\mathrm{in}} \sim 0.7$. For more information, refer to App. \ref{app:CCOffDiagonal}.}
\label{tab:TriangularOffDiagonal}
\end{table}

\begin{table}[ht!]
\begin{center}
\begin{tabular}{|c | c | c | c |} 
 \hline
$\delta C_{i,j}$  & $\overline{F}_{\mathrm{base}}$ & $\overline{F}_{\mathrm{noise}}$ & $\overline{\Delta F}$\\
 \hline
 0.01 & 0.9836 & 0.9816 & -0.0020  \\
 \hline
 0.02 & 0.9332 & 0.9294 & -0.0038  \\
 \hline
\end{tabular}
\end{center}
\caption{\textbf{Summary table of the output fidelity for the simulations with the triangular layout with multiple phase control in the case of imperfections in the estimation of $C_{i,j}$}. Average fidelity using the target ($\overline{F}_{\mathrm{base}}$) and trained ($\overline{F}_{\mathrm{noise}}$) parameters while using the shifted values of $C_{i,j}$, as well as their difference $\Delta F = {F}_{\mathrm{noise}} - {F}_{\mathrm{base}}$, for different values of the shift $\delta C_{i,j}$. All results are averaged over $100$ different instances, with starting propagation constants $\beta^{(0)}_{i}$ chosen randomly. For more information, refer to App. \ref{app:CCOffDiagonal}.}
\label{tab:TriangularOffDiagonal2}
\end{table}

\section{Identifying the optimization starting point}
\label{sec:LocalToGlobal}

The variational method to train the integrated interferometers described above has been tested by considering a starting point which, in most of the cases, requires an initial knowledge on the circuit programming. One question that then arises is how one can find such a starting position. In some cases, where the optical chip can reproduce a limited range of unitaries, a random starting point can be in principle used since any programmed unitary will have a non-negligible fidelity with the target. However, when the interferometer spans a significant portion of the unitary space, it is necessary to develop additional methods that can be employed to find a suitable starting position.

\begin{figure}[h!]
\centering
\includegraphics[width=0.49\textwidth]{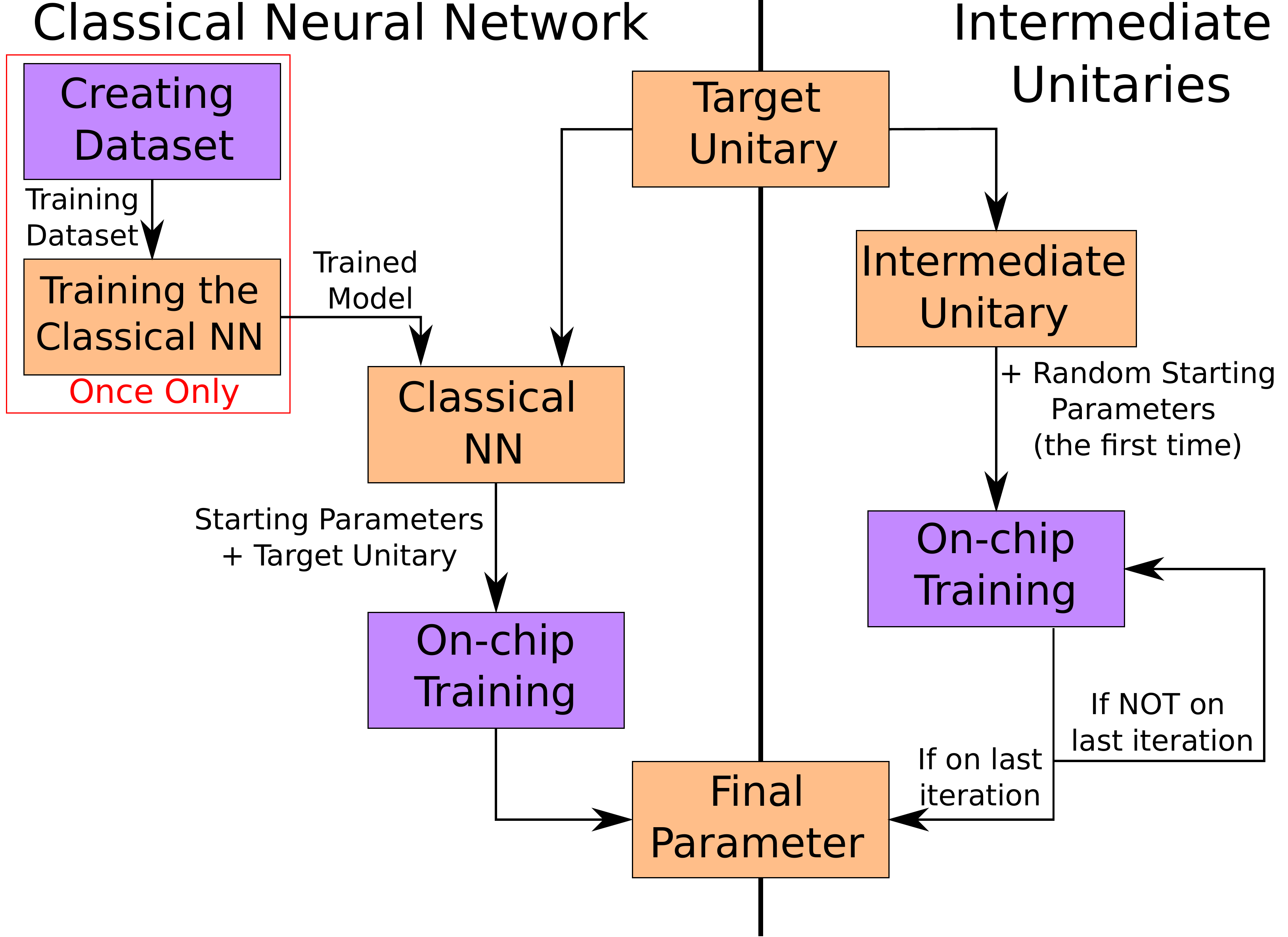}
\caption{\textbf{Schematic of the two different methods to define the starting point for our training algorithm.} Logical sequence of operations to implement the two methods described in the main text. In purple we highlight the parts that are performed on the optical chip, while in orange are shown the parts performed on classical hardware.}
\label{fig:SchematicGlobal}
\end{figure}

Here, we consider the analysis of two specific approaches. The first one is based on training a classical machine learning model to obtain the starting parameters, by giving it as input the target unitary. The second approach relies on using the variational method to progressively search for a sequence of unitaries, starting from a random one, where the unitary at each step $k+1$ is progressively closer to the target one. A schematic view of the working principles for two approaches is shown in Fig. \ref{fig:SchematicGlobal}. We tested these techniques on the 10 and 20 mode planar layouts. Both methods are shown to have their benefits and challenges, that we discuss below in this section. 

\subsection{Classical Machine Learning approach}

We start by describing the classical machine learning method. In this approach, one constructs a neural network (NN) that takes as input a unitary, and gives as output a set of parameters that is as close as possible to those used to generate said unitary. As said, this method is implemented to provide the starting point for the variational local optimization algorithm described above. For this reason, for this initial step we consider the training to be successful if the fidelity between the input unitary and the one corresponding to the network output parameter satisfies $F > 0.69$. This bound is chosen since this fidelity threshold is found to be sufficient to provide a reasonable starting point for the local optimization. 

In all of the trainings of the classical neural network, the input set is obtained by generating a single set of coupling coefficients $\{C_{i,j}\}$, which are common for all generated unitaries for that specific training set, and then selecting a random set of propagation constants $\{ \beta_{i}\}$, different for each generated unitary. The input unitary is then computed from this set of parameters. Unless otherwise stated, the train-test split was 0.8, meaning that 80\% of the input unitaries were used for training, and 20\% for testing. Further details on the structure and training of the neural network are reported in App. \ref{app:CCClassical}. 

As a first analysis, we applied this approach to the 10-mode scenario with planar layout. In particular, we tested the performances as a function of the number of unitaries shown to the NN during the training process. The results of this analysis are reported in Tab. \ref{tab:10ModeNN}. The performances of the trained network are tested by retrieving the circuit parameters for a distinct set of $1000$ unitaries, employed as the test set and common to all trainings. We then calculated the success rate, defined as the frequency where a fidelity $F > 0.69$ between the target unitary and the one obtained using the network output. We observed that, for a training process with $10000$ unitaries, a success rate of $98\%$ is obtained on the test set. Then, we selected a subset of $100$ unitaries, and used the output of the network as a starting point for the local optimization algorithm described in the previous section. We observe that $97\%$ of these unitaries reach a fidelity $F > 0.98$ after the complete training process.

As a further test for this classical neural network method, we also tested its resilience to noise. This was performed by using in the training set noisy unitaries, thus reflecting the scenario where a finite number of measurements is performed to characterize the unitaries of the training set. Robustness to this kind of noisy also suggests that less measurements need to be performed for the unitary characterization in this initial stage, thus lowering the resource requirements. Noisy unitaries are simulated by starting from the set of $10000$ noiseless unitaries used previously, and applying to each unitary $U_{i}$ in the set a random noise term obtained as:
\begin{equation}
U^{\mathrm{noise}} = e^{\imath \epsilon H},
\end{equation}
where $\epsilon$ is a parameter quantifying the noise strength, and $H$ is a random Hermitian matrix. The noisy unitary is found as $U_{i}^{\mathrm{noisy}} = U_{i}U^{\mathrm{noise}}$. In our case, the noise parameter was set to $\epsilon = 0.1$, corresponding to an average fidelity $\overline{F} \sim 0.95$ between the noisy and noiseless unitaries. After training the network with the noisy unitary, we used a test sample with noiseless unitaries to check its performance, and obtained a success rate of $96\%$, which shows only a limited decrease in performance of the trained model (see Tab. \ref{tab:10ModeNN}).

\begin{table}[h!]
\begin{center}
\begin{tabular}{|c | c | c |} 
 \hline
\# unitaries in the training set & Success rate & $\overline{F}$ \\
 \hline
 100 & 8.9\% & 0.4049  \\
 \hline
 1000 & 66.7\% & 0.7246  \\
 \hline
  10000 & 98.3\% & 0.9350  \\
 \hline
  10000 (noisy) & 96.7\% & 0.9060  \\
 \hline
\end{tabular}
\caption{\textbf{Summary table of the success rate of the trained neural network model for the 10-mode planar layout.}  Success rate and average fidelity of the trained neural network model for the 10 mode planar layout with respect to the number and type of unitaries used during training. The values are obtained by testing on 1000 random unitaries, which are the same for all configurations tested.}
\label{tab:10ModeNN}
\end{center}
\end{table}

We have then repeated a similar analysis for the 20-mode planar layout, corresponding to an increased size system. In this case, a training set composed of $40000$ input unitaries leads to a success rate of $\sim 75\%$, while increasing the size of the training set to $400000$ provides a success rate of $\sim 98\%$.

\subsection{Training through intermediate unitaries}

\begin{figure}[h!]
\includegraphics[width=0.49\textwidth]{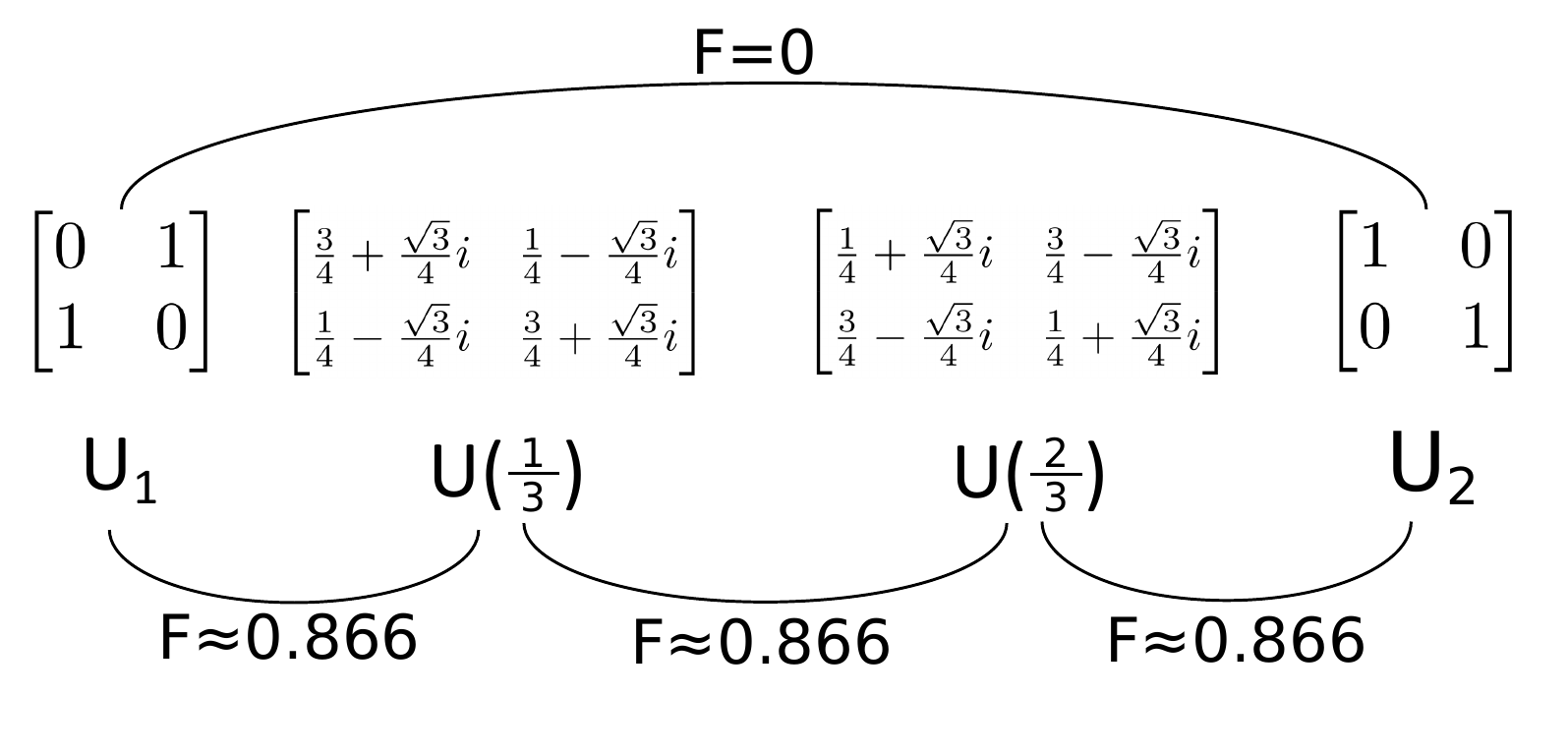}
\centering
\caption{\textbf{Example of the intermediate unitaries method applied on a 2x2 unitary matrix}. The fidelity $F$ between the starting unitary $U_1$ and the target one $U_2$ is small, thus leading to reduced performance of the training algorithm. This can be circumvented by generating intermediate unitaries, that permits us to divide the training process in steps with targets that have a sufficient value of the fidelity for the method to be effective.}
\label{fig:IntermediateUnitaries}
\end{figure}

As a second approach to obtain the starting point for the local optimization, we considered a different approach based on dividing the optimization in multiple steps. The idea is to perform a progressive training through multiple steps having as intermediate targets a set of unitaries which are progressively closer to the target one, chosen according to an appropriate metric. This set of intermediate unitaries can be generated by using the geodesic \cite{GeodesicUnitary}. More specifically, let us call $U_1$ and $U_2$ the starting and target unitaries, respectively. Then, we can define:
\begin{equation}
    \Gamma = U_1^{\dagger} U_2 
\end{equation}
and obtain:
\begin{equation}
U(t) = U_1 e^{t \log{\Gamma}}
\end{equation}
The parameter $t$ identifies the position of a given unitary $U(t)$ in the sequence, with extremal points the unitaries $U_1$ for $t = 0$ and $U_2$ for $t = 1$, and between them for values of $t$ between 0 and 1. A schematic view of this approach applied on a 2x2 unitary is shown in Fig. \ref{fig:IntermediateUnitaries}.

For the intermediate unitaries method, in the case of the 10 mode planar layout we used 3 steps, that is, 2 intermediate unitaries before final target unitary. We found that, using the same 100 sample set that we used to test the classical machine learning method, only around $45\%$ of the trainings resulted in a final fidelity $F > 0.98$. The reason for this behaviour can be found in that  the intermediate unitaries can end up being outside of the range of unitaries that the actual device can reach, given the specific architecture. This would lead to the training process for that step failing, and as a result impacting the whole training procedure. We argue that the efficacy of this method can improve as larger sets of unitaries can be obtained within the circuit architecture. Considering this issue, in the case of the 20 mode layout the performances of the approach worsen, as even using 5 steps the training failed to converge in the majority of the cases.

\subsection{Comparison of the two approaches}

Both methods presented have their advantages and disadvantages. In the case of the method which uses classical machine learning to obtain the starting parameters, one of the main disadvantages is that it requires an overhead of starting unitaries associated to the respective parameters to perform the training. This means that, on a physical device, a measurement overhead to perform the on-chip static characterization of the unitaries is present. However, once trained the neural network is capable of providing a reliable set of starting parameters in short computational time. On the other hand, the intermediate unitaries method does not require significant overheads. However, this approach requires a longer training process, since it effectively requires training the chip multiple times with different targets. Furthermore, the intermediate unitaries has a significant failure rate for those architectures where the set of unitaries achievable by the device is limited.

\section{Discussion}
\label{sec:Discussion}

In this work, we have presented a machine learning based approach to program multi-mode integrated interferometers based on continuously-coupled waveguides. One of the main features of this method is the capability of performing the desired task without requiring extensive characterization of the integrated device. Furthermore, this approach removes the need of performing an accurate modeling of the circuit operations, since the interferometers are treated as black boxes. This makes this technique flexible, and capable to be used with circuits of different layouts. We have shown via numerical simulations the effectiveness of this approach for different circuit layouts, and its robustness to experimental errors. The results of the simulations support the capability to program integrated devices to implement a desired unitary transformation via the developed approach, provided that the target unitary can be reached by tuning the circuit parameters. Finally, to complement this approach we have also presented possible methods to find the starting point of the optimization process.

Using the method described above, an estimate of the time required for the training process is of approximately $\sim$10 hours for the 20-mode planar layout, and $\sim$15 hours for the 32-mode 3D configuration, assuming a rate of $10^5$ two-fold coincidences per minute, using $N_{s} = 10^{4}$ and using 10 two-photon input states at each epoch. This has to be compared with the significantly higher cost required to perform the calibration with other approaches involving the reconstruction of the unitary transformations for several parameter values. It must also be noted that this time estimate is a baseline, as further improvements to the training process can be made, such as using a smaller number of shots per input $N_{s}$ during the beginning of the training, and then increasing the value of $N_{s}$ as training progresses, which can lead to additional speedups.

A currently open problem in quantum machine learning is how to efficiently train large quantum circuits. Our method provides a step forward in that direction, providing several optimizations to reduce the time required to train the circuits, leading to a training time that is expected to scale polynomially. Room for improvement can be found for instance in the presence of a reduced success rate when the starting fidelity is below a certain threshold, which we were able to partially mitigate with the use of classical machine learning. Thus, future research directions can be found in devising optimizations of the method to more efficiently train the circuits when their size increases. These improvements include finding ways to optimize multiple parameters simultaneously, and implementing solutions to support the training process when starting from an initial set of parameters that is far from the target parameters.

The obtained results show that the devised method could be useful in a scenario where an accurate modeling of the devices is not available, while it is also applicable in all cases where a comprehensive characterization via standard means requires an exceedingly amount of resources. Hence, this approach is expected to be used as an effective and viable tool to program and operate reconfigurable integrated interferometers with various layouts.

\appendix

\section{Simulating the continuously coupled chip}
\label{app:continuously-coupled}

In this appendix we provide more details on the methodology used to simulate the evolution in multi-mode interferometers based on continuously-coupled waveguides.

\textit{General structure of the simulation process. --} As discussed in the main text, the overall structure to simulate the multi-mode interferometer is composed of different steps. The starting point is the definition of the circuit structure and thus of the Hamiltonian, by specifying the propagation constants $\beta_i$ and the coupling coefficients $C_{i,j}$ between the modes, and of the interaction length $z$. Then, by using the Qutip \cite{Qutip} library we retrieve the unitary matrix $U$ corresponding to the overall evolution of the interferometer. Such a unitary is then used to simulate samples of $N_{s}$ events obtained at the output of the interferometer by using the Perceval \cite{Perceval} library. Regarding the definition of the Hamiltonian and of the corresponding unitary matrix, they are defined by the number of modes, by the internal structure of the circuit and by the interaction length. 

After we generate the starting parameters of the simulated chips, we take a copy of the tunable parameters $\{ \beta^t_{i} \}$, and shift their values randomly from a uniform distribution. More specifically, the size of this shift depends on the size and mesh of the circuit, and on the required starting fidelity. For instance, in the case of the 10-mode planar layout and of the 32-mode 3D layout, we use a random shift between $\pm 0.1$ mm$^{-1}$ on each parameter. With this process we obtain the starting point for the trainable parameters $\{ \beta^{(0)}_{i} \}$. This process is iterated until the starting fidelity is obtained in the required range.

We detail below the different cases analyzed in the main text.

\textit{Integrated interferometers with planar layout. --}  For the circuits with planar layout, the length of the circuit has been set as being $2 m$ mm, with $m$ being the number of modes. For the propagation constants, we generate a number of parameters equal to the number of modes, each one corresponding to a propagation constant $\beta_i$, and we generate them by sampling from a random uniform distribution between 0.7 mm$^{-1}$ and 1.3 mm$^{-1}$. Then we generate a set of coupling coefficients, with a number of parameters defined by the design and size of the chip, generating them from a random distribution between 0.1 mm$^{-1}$ and 0.3 mm$^{-1}$. This was chosen according to the values reported in recent implementations \cite{3DChipLab}, where the coupling coefficients were chosen around 0.2 mm$^{-1}$.

\textit{3D interferometers with direct control. --} For the circuits with the 3D layout we set the length of the circuit to be $36$ mm, following the specification of the 3D circuits implemented in Ref. \cite{3DChipLab}. The main differences when compared to the planar layout are that the Hamiltonian has a different structure, and that the coupling constants correspond to a different structure for the neighbor modes. Thus, the simulation process has been adapted to take into account these two differences with respect to the planar layout.

\textit{3D interferometers with multiple control. --}  For the circuits with the 3D layout we set the length of the circuit to be $36$ mm, following the specification of the 3D circuits implemented in Ref. \cite{3DChipLab}. The simulation process of this layout is modified with respect the other cases to reproduce the behaviour of this kind of architecture.

As already stated in the main text, the first modification requiring considering a layout with multiple connected segments, which can be controlled independently. Then, we changed the way in which the parameters are controlled, no longer being able to control the propagation constants individually, but instead including the effect of thermal cross-talks. The latter effect correspond to having multiple propagation constants affected simultaneously by changing the current in a single resistor, which is due to heat propagation mechanism in the glass.

To reproduce the structure of the interferometers with this layout, we divide the simulated chip into 18 separate segments, each of length 2 mm. In 8 of these segments, thermal phase shifters are inserted to modulate the propagation constants. More specifically, we alternate segments without active control and segments with phase shifters. For each segment where active control is available, we train only 2 parameters, which are associated with the currents $I_{k}$ of the 2 available resistors. Each of these parameters then influences the propagation constants $\{ \beta_{i} \}$. In our specific case, we start with a fixed base value for the propagation constants, chosen to be 0.7 mm$^{-1}$. Then, for each propagation constant we add an amount equal to $\log(d_{x_i}) x_i$, where $x_i$ is the value of the parameter, and $d_{x_i}$ is the lattice distance between the mode and the resistor associated with such parameter. We implement this as a fixed multiplicative mesh, so that it can be altered to suit different chip structures. Regarding the value of $x_i$, we choose its value to be restricted between 0 mm$^{-1}$ and 0.6 mm$^{-1}$. In such a way, the final value of the propagation constants of the modes closest to the resistors is between 0.7 mm$^{-1}$ and 1.3 mm$^{-1}$, thus reproducing the range used for the previously considered architectures.

\section{Loss landscape}
\label{app:CCLossLandscape}

To get some additional insights on the training process, we analyzed the loss landscape when tuning only a single parameter.

This analysis is performed by choosing a single propagation constant, identified by a given index $k$. Then, we set $\beta^{(0)}_{k}$ and $\beta^{t}_{k}$ in the target and trainable set to the same value, placed in the middle of the range of possible values. This means that such parameter is set to 1.0 mm$^{-1}$ in the case with direct control, and 0.3 mm$^{-1}$ in the case with multiple control. We then evaluate the loss by tuning parameter $\beta^{(0)}_{k}$ in the range $\pm 0.6$ mm$^{-1}$ with respect to the central value, equal to twice the range used in the training process. This analysis provide an intuition of how the loss varies by changing a single parameter, as performed in a single step of the training process, by showing the emergence of eventual local or global minima.

\begin{figure}[h!]
\includegraphics[width=0.45\textwidth]{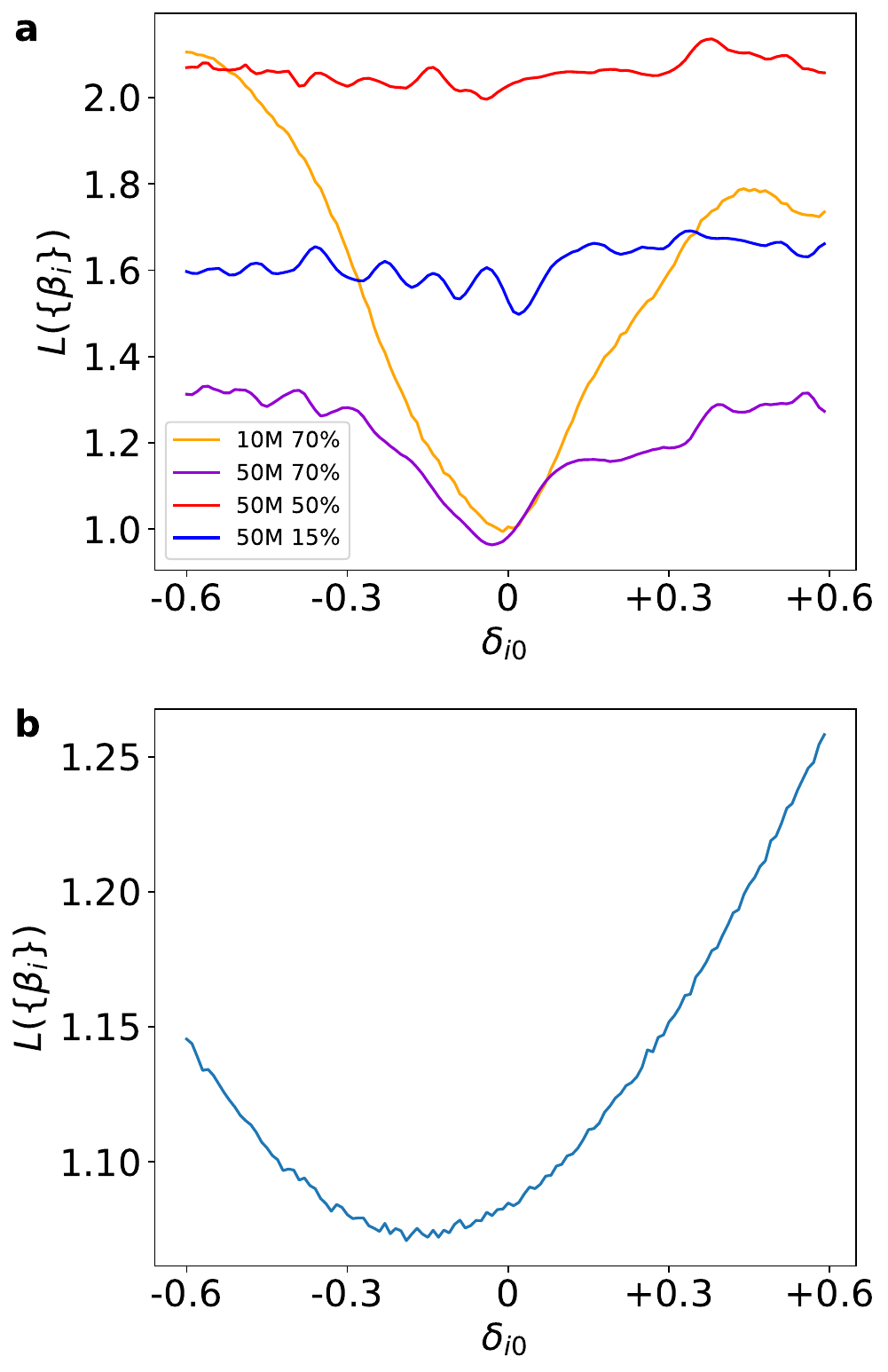}
\centering
\caption{\textbf{Loss landscape when changing a single parameter $\beta_{i0}$ in the training process.} Loss change as a function of the parameter shift with respect to the correct value. \textbf{a}, Loss landscape for the planar structure. Orange curve: $m = 10$ and $F_{in} \sim 0.7$. Purple curve: $m = 50$ and $F_{in} \sim 0.7$. Blue curve: $m = 50$ and $F_{in} \sim 0.5$. Red curve: $m = 50$ and $F_{in} \sim 0.15$. \textbf{b}, Loss landscape for the 32-mode 3D layout according to Ref. \cite{3DChipLab}.}
\label{fig:LossLandscape}
\end{figure} 

As shown in Fig. \ref{fig:LossLandscape}a for specific instances of interferometers with planar layout, we observe two different behaviours depending from the starting point of the optimization. More specifically, when the starting point has an initial fidelity $F_{\mathrm{in}} \sim 0.7$, we observe the emergence of a global minimum of the loss when the parameter is equal to the target $\beta^{t}_{k}$. Conversely, for lower starting fidelities the loss landscape shows the presence of different local minima, and thus the training process is characterized by higher failure probability. This highlights the need of choosing appropriately the starting fidelity for the training process, as discussed in Sec. \ref{sec:Results}.

In Fig. \ref{fig:LossLandscape}b we report a similar analysis for the 3D interferometers with multiple phase control. In this case, the loss landscape of the single parameter has its global minima at a different value than the target $\beta^{t}_{k}$. This is likely due to the feature that, with this layout, multiple configurations can generate the same unitary. This also provides an intuition on the possibility of starting the training of interferometers with this layout from a random set of parameters, since it is likely that from any starting position there is a close configuration that leads to the desired unitary. We can also notice that the curve of the loss shows one well defined minimum, thus helping the training process.

\section{Details on the approximation of the gradient and on the training}
\label{app:CCTrainingDetails}

In this subsection we will give some additional details on our modified FDSA and overall training method. 

The choice of the parameter to be trained is performed in the following way. The parameters are organized in an ordered list, and we proceed changing the parameter at each epoch in sequence. Then, once the full set of parameters have been used, the list is randomly permuted. Then we proceed in sequence changing the parameter following the shuffled list. This appraoch is then iterated until the training reaches convergence. This method, while retaining a degree of randomness in the choice, ensures at the same time that no parameter is left untouched for too long. 

The training is performed then by shifting the value of the chosen parameter by a certain amount $\delta$, and by evaluating how the loss changes with the shifted value. The choice of $\delta$ is performed as a compromise between two different trends. More specifically, reducing the value of $\delta$ permits to obtain a better approximation of the gradient, while at the same time it increases vulnerability to sampling noise as well as other forms of noise. The value of $\delta$ changes based on the size and on the interferometer layout. In our case, we tested various values and found out the one that simultaneously minimizes both the loss and the probability of the training getting stuck in a local minimum. Furthermore, the used value of $\delta$ is reduced as the loss decreases. Adapting this values permits to obtain faster convergence rate around the global minima at the beginning of the training. As an added advantage, we found that this approach reduces the risk of getting stuck in a narrow local minima during the training process. 

Given a specific choice of $\delta$, the training step involves shifting the value by $\pm \delta$, comparing the obtained losses with the one obtained when the parameter is left unchanged, and then moving in the direction where the loss is lower. If neither direction has lower loss, we do not change the value of that parameter. This takes 3 total evaluations per epoch, instead of the usual 2 needed by FDSA, while reducing the risk of obtaining an increase in the loss during that step of the process.

Once the correct direction for the parameter change has been identified, one then needs to choose the effective strength of the parameter shift to move to the next training step. To this end, we tried two approaches. The first method involves adding a change in the parameter in the direction where the loss decreases, by an amount proportional to the approximated gradient. The second approach corresponds to using the value of $\pm \delta$, already evaluated previously to identify the direction, corresponding to the lower value of the loss. This second method also allows the use of only 2 evaluations per epoch, instead of 3, since we already know the loss of the unaltered value from the previous epoch. In general, we found out that the first method leads to larger reduction rate of the loss when far away from the target point. Conversely, the second method permits to converge to a lower loss value when the parameter is close to the target, and removed the risk of increasing the loss from one epoch to the next. Given that we start from $F_{\mathrm{in}} \sim 0.7$, we privileged the use of the second method. Further  optimization could be possible by using the first method for the initial part of training, and then switching to the second method for the second part of training. Considering the involved parameter range, we expect that in the investigated scenarios this would lead to a marginal improvement in the algorithm efficiency. 

Finally, the complete training process consists of multiple epochs that follow the procedure described above. The specific number of epochs changes depending on the structure and on the number of modes of the optical interferometer.

As a further direction for improvement to be used for interferometers with a large number of parameters and/or a lower starting fidelity, a possible approach involves performing the initial part of the training with SPSA, to obtain a first approximation of the target, and then use our modified FDSA to perform the second part fo the optimization process.

\section{Details on the simulation with shifts in the coupling parameters}
\label{app:CCOffDiagonal}

In this subsection we will provide a more technical explanation on the fidelities used for the trainings with shifts in the estimation of the coupling coefficients $\{ C_{i,j} \}$. 

In the absence of shifts in the estimation of the coupling coefficients, we start from a set of propagation constants $\{ \beta^{t}_{i} \}$, and a set of coupling coefficients $\{ C^{t}_{i,j} \}$. We then create $\{ \beta^{(0)}_{i} \}$ by taking the values of $\{ \beta^{t}_{i} \}$ and shifting each of them by a random amount. The strength of the shift depends on the target starting fidelity ($F_{\mathrm{in}} \sim 0.7$) and on the structure of the circuit. We then train those propagation constants $\{ \beta^{(0)}_{i} \}$ as explained in the main text, using as target outputs those generated by simulating the circuit using $\{ \beta^{t}_{i} \}$ as the propagation constants. In such a way we obtain the trained version, which we call $\{ \beta^{f}_{i} \}$.

When inserting the shifts in the coupling coefficients, we generate a new set of coupling coefficients called $\{ C^{s}_{i,j} \}$, which can be obtained by applying a small shift $\delta C_{i,j}$ to $\{ C^{t}_{i,j} \}$. In this situation, the parameters $\{ C^{t}_{i,j} \}$ can be considered to represent the knowledge on the coupling coefficients of the device, obtained either via pre-calibration or from the circuit fabrication, while $\{ C^{s}_{i,j} \}$ are the true coupling coefficients of the device.

For notation on the fidelity comparisons, we will use: \textbf{$F$($\{ \beta_{i} \}_1$, $\{ C_{i,j} \}_1$, $\{ \beta_{i} \}_2$, $\{ C_{i,j} \}_2$)} to indicate the fidelity between two unitaries, where $\{ \beta_{i} \}_k$ and $\{ C_{i,j} \}_k$ indicate the propagation constants and the coupling coefficients of the $k$-th unitary.

As a first step, we get the fidelity \textbf{ $F_{\mathrm{base}}$ =  $F$($\{ \beta^{t}_{i} \}$, $\{ C^{s}_{i,j} \}$, $\{ \beta^{t}_{i} \}$, $\{ C^{t}_{i,j} \}$)} to compare the distance from the target transformation induced by the shift. Then, we use $\{ \beta^{t}_{i} \}$ with $\{ C^{t}_{i,j} \}$ to get the target output values. We then get $\{ \beta^{(0)}_{i} \}$ as previously describe to obtain the $F_{\mathrm{in}} \sim 0.7$ starting point, and perform the training using those target values. In the training loop we use $\{ \beta^{(0)}_{i} \}$ for the propagation constants, while we use $\{ C^{s}_{i,j} \}$ for the coupling coefficients. At the end of the training process we obtain $\{ \beta^{f}_{i} \}$. From these parameters, we can calculate the fidelity $F_{\mathrm{noise}}$ =  $F$($\{ \beta^{f}_{i} \}$, $\{ C^{s}_{i,j} \}$, $\{ \beta^{t}_{i} \}$, $\{ C^{t}_{i,j} \}$) between the final trained circuit and the target obtained from the initial knowledge on the coupling constants.

\section{Classical machine learning model}
\label{app:CCClassical}

In this subsection we will give some additional details on the structure and training of the classical neural network we used in Sec. \ref{sec:LocalToGlobal} to find the starting parameters for a given target unitary.

Since the values of the unitary are complex, we first need to split every complex value into 2 real ones for simplicity. This choice is performed considering that, although complex-valued neural networks have been defined \cite{ComplexNN}, they do present some unique challenges. Such kind of neural networks are still partially unexplored, and thus fewer libraries optimized for their training are available. For our network we use a multilayer perceptron with dense layers. If $m$ is the number of modes of the interferometer, we get that the input layer of the network has size $2m^2$. The output layer, on the other hand, has the same size as the number of trainable parameters of the chosen optical interferometer. Conversely, the number and size of the hidden layers varies with the size of the interferometer. For instance, with the 10 mode chip we use 4 hidden layers, with width going from 256 to 64 nodes per layer. In the case of the 20 mode chip, we add at the beginning an additional layer with 1024 nodes.

The inputs parameters of the network are normalized. Conversely, we use a linear transformation to map the output to the range of values $[0,1]$, and use a sigmoid activation function on the last layer. Then, the output values are reverted back in the original range. For all the other layers, we use the Exponential Linear Units (ELU) activation function \cite{ELUActivation}. The used loss function is the Mean Squared Error  $ \mathrm{MSE}  = \frac{1}{n} \sum_{i=1}^{n} (y_{i} - \hat{y}_{i})^2$, being $\hat{y}_{i}$ the output parameters of the network, and $y_{i}$ the target parameters, corresponding to those of the input unitary. For the optimizer, we use Adam \cite{AdamOptimizer}.

We use dropout \cite{Dropout2014} to minimize the risk of overfitting, and improve stability of the trained model. As a further refinement, during the first part of training we increased the batch size instead of decreasing the learning rate. This approach leads to a similar behaviour in training, but allows for greater parallelism, and thus shorter training times \cite{BatchSizeIncrease}. Then, once the batch size is big enough, we start to decrease the learning rate.

\begin{acknowledgments}
This work is supported by the European Union via project
QLASS (“Quantum Glass-based Photonic Integrated Circuits” - Grant Agreement No. 101135876) and by the ERC Advanced Grant QU-BOSS (QUantum advantage via non-linear BOSon Sampling, Grant Agreement No. 884676). D.S. acknowledges Thales Alenia Space Italia for supporting the PhD fellowship.
\end{acknowledgments}

%\bibliography{main}

\begin{thebibliography}{64}%
\makeatletter
\providecommand \@ifxundefined [1]{%
 \@ifx{#1\undefined}
}%
\providecommand \@ifnum [1]{%
 \ifnum #1\expandafter \@firstoftwo
 \else \expandafter \@secondoftwo
 \fi
}%
\providecommand \@ifx [1]{%
 \ifx #1\expandafter \@firstoftwo
 \else \expandafter \@secondoftwo
 \fi
}%
\providecommand \natexlab [1]{#1}%
\providecommand \enquote  [1]{``#1''}%
\providecommand \bibnamefont  [1]{#1}%
\providecommand \bibfnamefont [1]{#1}%
\providecommand \citenamefont [1]{#1}%
\providecommand \href@noop [0]{\@secondoftwo}%
\providecommand \href [0]{\begingroup \@sanitize@url \@href}%
\providecommand \@href[1]{\@@startlink{#1}\@@href}%
\providecommand \@@href[1]{\endgroup#1\@@endlink}%
\providecommand \@sanitize@url [0]{\catcode `\\12\catcode `\$12\catcode
  `\&12\catcode `\#12\catcode `\^12\catcode `\_12\catcode `\%12\relax}%
\providecommand \@@startlink[1]{}%
\providecommand \@@endlink[0]{}%
\providecommand \url  [0]{\begingroup\@sanitize@url \@url }%
\providecommand \@url [1]{\endgroup\@href {#1}{\urlprefix }}%
\providecommand \urlprefix  [0]{URL }%
\providecommand \Eprint [0]{\href }%
\providecommand \doibase [0]{https://doi.org/}%
\providecommand \selectlanguage [0]{\@gobble}%
\providecommand \bibinfo  [0]{\@secondoftwo}%
\providecommand \bibfield  [0]{\@secondoftwo}%
\providecommand \translation [1]{[#1]}%
\providecommand \BibitemOpen [0]{}%
\providecommand \bibitemStop [0]{}%
\providecommand \bibitemNoStop [0]{.\EOS\space}%
\providecommand \EOS [0]{\spacefactor3000\relax}%
\providecommand \BibitemShut  [1]{\csname bibitem#1\endcsname}%
\let\auto@bib@innerbib\@empty
%</preamble>
\bibitem [{\citenamefont {Arute}\ \emph {et~al.}(2019)\citenamefont {Arute},
  \citenamefont {Arya}, \citenamefont {Babbush}, \citenamefont {Bacon},
  \citenamefont {Bardin}, \citenamefont {Barends}, \citenamefont {Biswas},
  \citenamefont {Boixo}, \citenamefont {Brandao}, \citenamefont {Buell},
  \citenamefont {Burkett}, \citenamefont {Chen}, \citenamefont {Chen},
  \citenamefont {Chiaro}, \citenamefont {Collins}, \citenamefont {Courtney},
  \citenamefont {Dunsworth}, \citenamefont {Farhi}, \citenamefont {Foxen},
  \citenamefont {Fowler}, \citenamefont {Gidney}, \citenamefont {Giustina},
  \citenamefont {Graff}, \citenamefont {Guerin}, \citenamefont {Habegger},
  \citenamefont {Harrigan}, \citenamefont {Hartmann}, \citenamefont {Ho},
  \citenamefont {Hoffman}, \citenamefont {Huang}, \citenamefont {Humble},
  \citenamefont {Isakov}, \citenamefont {Jeffrey}, \citenamefont {Jiang},
  \citenamefont {Kafri}, \citenamefont {Kechedzhi}, \citenamefont {Kelly},
  \citenamefont {Klimov}, \citenamefont {Knysh}, \citenamefont {Korotov},
  \citenamefont {Kostritsa}, \citenamefont {Landhuis}, \citenamefont
  {Lindmark}, \citenamefont {Lucero}, \citenamefont {Lyakh}, \citenamefont
  {Mandr\`a}, \citenamefont {McClean}, \citenamefont {McEwen}, \citenamefont
  {Megrant}, \citenamefont {Mi}, \citenamefont {Michielsen}, \citenamefont
  {Mohseni}, \citenamefont {Motus}, \citenamefont {Naaman}, \citenamefont
  {Neeley}, \citenamefont {Neill}, \citenamefont {{Yuezhen Niu}}, \citenamefont
  {Ostby}, \citenamefont {Petukhov}, \citenamefont {Platt}, \citenamefont
  {Quintana}, \citenamefont {Rieffel}, \citenamefont {Rousham}, \citenamefont
  {Rubin}, \citenamefont {Sank}, \citenamefont {Satzinger}, \citenamefont
  {Smelyanskiy}, \citenamefont {Sung}, \citenamefont {Trevithick},
  \citenamefont {Vainsencher}, \citenamefont {Villalonga}, \citenamefont
  {White}, \citenamefont {Yao}, \citenamefont {Yeh}, \citenamefont {Zalcman},
  \citenamefont {Neven},\ and\ \citenamefont {Martinis}}]{QuantumSupGoogle}%
  \BibitemOpen
  \bibfield  {author} {\bibinfo {author} {\bibfnamefont {F.}~\bibnamefont
  {Arute}}, \bibinfo {author} {\bibfnamefont {K.}~\bibnamefont {Arya}},
  \bibinfo {author} {\bibfnamefont {R.}~\bibnamefont {Babbush}}, \bibinfo
  {author} {\bibfnamefont {D.}~\bibnamefont {Bacon}}, \bibinfo {author}
  {\bibfnamefont {J.~C.}\ \bibnamefont {Bardin}}, \bibinfo {author}
  {\bibfnamefont {R.}~\bibnamefont {Barends}}, \bibinfo {author} {\bibfnamefont
  {R.}~\bibnamefont {Biswas}}, \bibinfo {author} {\bibfnamefont
  {S.}~\bibnamefont {Boixo}}, \bibinfo {author} {\bibfnamefont {F.~G. S.~L.}\
  \bibnamefont {Brandao}}, \bibinfo {author} {\bibfnamefont {D.~A.}\
  \bibnamefont {Buell}}, \bibinfo {author} {\bibfnamefont {B.}~\bibnamefont
  {Burkett}}, \bibinfo {author} {\bibfnamefont {Y.}~\bibnamefont {Chen}},
  \bibinfo {author} {\bibfnamefont {Z.}~\bibnamefont {Chen}}, \bibinfo {author}
  {\bibfnamefont {B.}~\bibnamefont {Chiaro}}, \bibinfo {author} {\bibfnamefont
  {R.}~\bibnamefont {Collins}}, \bibinfo {author} {\bibfnamefont
  {W.}~\bibnamefont {Courtney}}, \bibinfo {author} {\bibfnamefont
  {A.}~\bibnamefont {Dunsworth}}, \bibinfo {author} {\bibfnamefont
  {E.}~\bibnamefont {Farhi}}, \bibinfo {author} {\bibfnamefont
  {B.}~\bibnamefont {Foxen}}, \bibinfo {author} {\bibfnamefont
  {A.}~\bibnamefont {Fowler}}, \bibinfo {author} {\bibfnamefont
  {C.}~\bibnamefont {Gidney}}, \bibinfo {author} {\bibfnamefont
  {M.}~\bibnamefont {Giustina}}, \bibinfo {author} {\bibfnamefont
  {R.}~\bibnamefont {Graff}}, \bibinfo {author} {\bibfnamefont
  {K.}~\bibnamefont {Guerin}}, \bibinfo {author} {\bibfnamefont
  {S.}~\bibnamefont {Habegger}}, \bibinfo {author} {\bibfnamefont {M.~P.}\
  \bibnamefont {Harrigan}}, \bibinfo {author} {\bibfnamefont {M.~J.}\
  \bibnamefont {Hartmann}}, \bibinfo {author} {\bibfnamefont {A.}~\bibnamefont
  {Ho}}, \bibinfo {author} {\bibfnamefont {M.}~\bibnamefont {Hoffman}},
  \bibinfo {author} {\bibfnamefont {T.}~\bibnamefont {Huang}}, \bibinfo
  {author} {\bibfnamefont {T.~S.}\ \bibnamefont {Humble}}, \bibinfo {author}
  {\bibfnamefont {S.~V.}\ \bibnamefont {Isakov}}, \bibinfo {author}
  {\bibfnamefont {E.}~\bibnamefont {Jeffrey}}, \bibinfo {author} {\bibfnamefont
  {Z.}~\bibnamefont {Jiang}}, \bibinfo {author} {\bibfnamefont
  {D.}~\bibnamefont {Kafri}}, \bibinfo {author} {\bibfnamefont
  {K.}~\bibnamefont {Kechedzhi}}, \bibinfo {author} {\bibfnamefont
  {J.}~\bibnamefont {Kelly}}, \bibinfo {author} {\bibfnamefont {P.~V.}\
  \bibnamefont {Klimov}}, \bibinfo {author} {\bibfnamefont {S.}~\bibnamefont
  {Knysh}}, \bibinfo {author} {\bibfnamefont {A.}~\bibnamefont {Korotov}},
  \bibinfo {author} {\bibfnamefont {F.}~\bibnamefont {Kostritsa}}, \bibinfo
  {author} {\bibfnamefont {D.}~\bibnamefont {Landhuis}}, \bibinfo {author}
  {\bibfnamefont {M.}~\bibnamefont {Lindmark}}, \bibinfo {author}
  {\bibfnamefont {E.}~\bibnamefont {Lucero}}, \bibinfo {author} {\bibfnamefont
  {D.}~\bibnamefont {Lyakh}}, \bibinfo {author} {\bibfnamefont
  {S.}~\bibnamefont {Mandr\`a}}, \bibinfo {author} {\bibfnamefont {J.~R.}\
  \bibnamefont {McClean}}, \bibinfo {author} {\bibfnamefont {M.}~\bibnamefont
  {McEwen}}, \bibinfo {author} {\bibfnamefont {A.}~\bibnamefont {Megrant}},
  \bibinfo {author} {\bibfnamefont {X.}~\bibnamefont {Mi}}, \bibinfo {author}
  {\bibfnamefont {K.}~\bibnamefont {Michielsen}}, \bibinfo {author}
  {\bibfnamefont {M.}~\bibnamefont {Mohseni}}, \bibinfo {author} {\bibfnamefont
  {J.}~\bibnamefont {Motus}}, \bibinfo {author} {\bibfnamefont
  {O.}~\bibnamefont {Naaman}}, \bibinfo {author} {\bibfnamefont
  {M.}~\bibnamefont {Neeley}}, \bibinfo {author} {\bibfnamefont
  {C.}~\bibnamefont {Neill}}, \bibinfo {author} {\bibfnamefont
  {M.}~\bibnamefont {{Yuezhen Niu}}}, \bibinfo {author} {\bibfnamefont
  {E.}~\bibnamefont {Ostby}}, \bibinfo {author} {\bibfnamefont
  {A.}~\bibnamefont {Petukhov}}, \bibinfo {author} {\bibfnamefont {J.~C.}\
  \bibnamefont {Platt}}, \bibinfo {author} {\bibfnamefont {C.}~\bibnamefont
  {Quintana}}, \bibinfo {author} {\bibfnamefont {E.~G.}\ \bibnamefont
  {Rieffel}}, \bibinfo {author} {\bibfnamefont {P.}~\bibnamefont {Rousham}},
  \bibinfo {author} {\bibfnamefont {N.~C.}\ \bibnamefont {Rubin}}, \bibinfo
  {author} {\bibfnamefont {D.}~\bibnamefont {Sank}}, \bibinfo {author}
  {\bibfnamefont {K.~J.}\ \bibnamefont {Satzinger}}, \bibinfo {author}
  {\bibfnamefont {V.}~\bibnamefont {Smelyanskiy}}, \bibinfo {author}
  {\bibfnamefont {K.~J.}\ \bibnamefont {Sung}}, \bibinfo {author}
  {\bibfnamefont {M.~D.}\ \bibnamefont {Trevithick}}, \bibinfo {author}
  {\bibfnamefont {A.}~\bibnamefont {Vainsencher}}, \bibinfo {author}
  {\bibfnamefont {B.}~\bibnamefont {Villalonga}}, \bibinfo {author}
  {\bibfnamefont {T.}~\bibnamefont {White}}, \bibinfo {author} {\bibfnamefont
  {Z.~J.}\ \bibnamefont {Yao}}, \bibinfo {author} {\bibfnamefont
  {P.}~\bibnamefont {Yeh}}, \bibinfo {author} {\bibfnamefont {A.}~\bibnamefont
  {Zalcman}}, \bibinfo {author} {\bibfnamefont {H.}~\bibnamefont {Neven}},\
  and\ \bibinfo {author} {\bibfnamefont {J.~M.}\ \bibnamefont {Martinis}},\
  }\bibfield  {title} {\bibinfo {title} {Quantum supremacy using a programmable
  superconducting processor},\ }\href
  {https://doi.org/10.1038/s41586-019-1666-5} {\bibfield  {journal} {\bibinfo
  {journal} {Nature}\ }\textbf {\bibinfo {volume} {574}},\ \bibinfo {pages}
  {505–510} (\bibinfo {year} {2019})}\BibitemShut {NoStop}%
\bibitem [{\citenamefont {Zhong}\ \emph {et~al.}(2020)\citenamefont {Zhong},
  \citenamefont {Wang}, \citenamefont {Deng}, \citenamefont {Chen},
  \citenamefont {Peng}, \citenamefont {Luo}, \citenamefont {Qin}, \citenamefont
  {Wu}, \citenamefont {Ding}, \citenamefont {Hu}, \citenamefont {Hu},
  \citenamefont {Yang}, \citenamefont {Zhang}, \citenamefont {Li},
  \citenamefont {Li}, \citenamefont {Jiang}, \citenamefont {Gan}, \citenamefont
  {Yang}, \citenamefont {You}, \citenamefont {Wang}, \citenamefont {Li},
  \citenamefont {Liu}, \citenamefont {Lu},\ and\ \citenamefont
  {Pan}}]{QuantumSupChina}%
  \BibitemOpen
  \bibfield  {author} {\bibinfo {author} {\bibfnamefont {H.-S.}\ \bibnamefont
  {Zhong}}, \bibinfo {author} {\bibfnamefont {H.}~\bibnamefont {Wang}},
  \bibinfo {author} {\bibfnamefont {Y.-H.}\ \bibnamefont {Deng}}, \bibinfo
  {author} {\bibfnamefont {M.-C.}\ \bibnamefont {Chen}}, \bibinfo {author}
  {\bibfnamefont {L.-C.}\ \bibnamefont {Peng}}, \bibinfo {author}
  {\bibfnamefont {Y.-H.}\ \bibnamefont {Luo}}, \bibinfo {author} {\bibfnamefont
  {J.}~\bibnamefont {Qin}}, \bibinfo {author} {\bibfnamefont {D.}~\bibnamefont
  {Wu}}, \bibinfo {author} {\bibfnamefont {X.}~\bibnamefont {Ding}}, \bibinfo
  {author} {\bibfnamefont {Y.}~\bibnamefont {Hu}}, \bibinfo {author}
  {\bibfnamefont {P.}~\bibnamefont {Hu}}, \bibinfo {author} {\bibfnamefont
  {X.-Y.}\ \bibnamefont {Yang}}, \bibinfo {author} {\bibfnamefont {W.-J.}\
  \bibnamefont {Zhang}}, \bibinfo {author} {\bibfnamefont {H.}~\bibnamefont
  {Li}}, \bibinfo {author} {\bibfnamefont {Y.}~\bibnamefont {Li}}, \bibinfo
  {author} {\bibfnamefont {X.}~\bibnamefont {Jiang}}, \bibinfo {author}
  {\bibfnamefont {L.}~\bibnamefont {Gan}}, \bibinfo {author} {\bibfnamefont
  {G.}~\bibnamefont {Yang}}, \bibinfo {author} {\bibfnamefont {L.}~\bibnamefont
  {You}}, \bibinfo {author} {\bibfnamefont {Z.}~\bibnamefont {Wang}}, \bibinfo
  {author} {\bibfnamefont {L.}~\bibnamefont {Li}}, \bibinfo {author}
  {\bibfnamefont {N.-L.}\ \bibnamefont {Liu}}, \bibinfo {author} {\bibfnamefont
  {C.-Y.}\ \bibnamefont {Lu}},\ and\ \bibinfo {author} {\bibfnamefont {J.-W.}\
  \bibnamefont {Pan}},\ }\bibfield  {title} {\bibinfo {title} {Quantum
  computational advantage using photons},\ }\href
  {https://doi.org/10.1126/science.abe8770} {\bibfield  {journal} {\bibinfo
  {journal} {Science}\ }\textbf {\bibinfo {volume} {370}},\ \bibinfo {pages}
  {1460} (\bibinfo {year} {2020})}\BibitemShut {NoStop}%
\bibitem [{\citenamefont {Zhong}\ \emph {et~al.}(2021)\citenamefont {Zhong},
  \citenamefont {Deng}, \citenamefont {Qin}, \citenamefont {Wang},
  \citenamefont {Chen}, \citenamefont {Peng}, \citenamefont {Luo},
  \citenamefont {Wu}, \citenamefont {Gong}, \citenamefont {Su}, \citenamefont
  {Hu}, \citenamefont {Hu}, \citenamefont {Yang}, \citenamefont {Zhang},
  \citenamefont {Li}, \citenamefont {Li}, \citenamefont {Jiang}, \citenamefont
  {Gan}, \citenamefont {Yang}, \citenamefont {You}, \citenamefont {Wang},
  \citenamefont {Li}, \citenamefont {Liu}, \citenamefont {Renema},
  \citenamefont {Lu},\ and\ \citenamefont {Pan}}]{QuantumSupGBSCHN}%
  \BibitemOpen
  \bibfield  {author} {\bibinfo {author} {\bibfnamefont {H.-S.}\ \bibnamefont
  {Zhong}}, \bibinfo {author} {\bibfnamefont {Y.-H.}\ \bibnamefont {Deng}},
  \bibinfo {author} {\bibfnamefont {J.}~\bibnamefont {Qin}}, \bibinfo {author}
  {\bibfnamefont {H.}~\bibnamefont {Wang}}, \bibinfo {author} {\bibfnamefont
  {M.-C.}\ \bibnamefont {Chen}}, \bibinfo {author} {\bibfnamefont {L.-C.}\
  \bibnamefont {Peng}}, \bibinfo {author} {\bibfnamefont {Y.-H.}\ \bibnamefont
  {Luo}}, \bibinfo {author} {\bibfnamefont {D.}~\bibnamefont {Wu}}, \bibinfo
  {author} {\bibfnamefont {S.-Q.}\ \bibnamefont {Gong}}, \bibinfo {author}
  {\bibfnamefont {H.}~\bibnamefont {Su}}, \bibinfo {author} {\bibfnamefont
  {Y.}~\bibnamefont {Hu}}, \bibinfo {author} {\bibfnamefont {P.}~\bibnamefont
  {Hu}}, \bibinfo {author} {\bibfnamefont {X.-Y.}\ \bibnamefont {Yang}},
  \bibinfo {author} {\bibfnamefont {W.-J.}\ \bibnamefont {Zhang}}, \bibinfo
  {author} {\bibfnamefont {H.}~\bibnamefont {Li}}, \bibinfo {author}
  {\bibfnamefont {Y.}~\bibnamefont {Li}}, \bibinfo {author} {\bibfnamefont
  {X.}~\bibnamefont {Jiang}}, \bibinfo {author} {\bibfnamefont
  {L.}~\bibnamefont {Gan}}, \bibinfo {author} {\bibfnamefont {G.}~\bibnamefont
  {Yang}}, \bibinfo {author} {\bibfnamefont {L.}~\bibnamefont {You}}, \bibinfo
  {author} {\bibfnamefont {Z.}~\bibnamefont {Wang}}, \bibinfo {author}
  {\bibfnamefont {L.}~\bibnamefont {Li}}, \bibinfo {author} {\bibfnamefont
  {N.-L.}\ \bibnamefont {Liu}}, \bibinfo {author} {\bibfnamefont {J.~J.}\
  \bibnamefont {Renema}}, \bibinfo {author} {\bibfnamefont {C.-Y.}\
  \bibnamefont {Lu}},\ and\ \bibinfo {author} {\bibfnamefont {J.-W.}\
  \bibnamefont {Pan}},\ }\bibfield  {title} {\bibinfo {title}
  {Phase-programmable gaussian boson sampling using stimulated squeezed
  light},\ }\href {https://doi.org/10.1103/PhysRevLett.127.180502} {\bibfield
  {journal} {\bibinfo  {journal} {Phys. Rev. Lett.}\ }\textbf {\bibinfo
  {volume} {127}},\ \bibinfo {pages} {180502} (\bibinfo {year}
  {2021})}\BibitemShut {NoStop}%
\bibitem [{\citenamefont {Wu}\ \emph {et~al.}(2021)\citenamefont {Wu},
  \citenamefont {Bao}, \citenamefont {Cao}, \citenamefont {Chen}, \citenamefont
  {Chen}, \citenamefont {Chen}, \citenamefont {Chung}, \citenamefont {Deng},
  \citenamefont {Du}, \citenamefont {Fan}, \citenamefont {Gong}, \citenamefont
  {Guo}, \citenamefont {Guo}, \citenamefont {Guo}, \citenamefont {Han},
  \citenamefont {Hong}, \citenamefont {Huang}, \citenamefont {Huo},
  \citenamefont {Li}, \citenamefont {Li}, \citenamefont {Li}, \citenamefont
  {Li}, \citenamefont {Liang}, \citenamefont {Lin}, \citenamefont {Lin},
  \citenamefont {Qian}, \citenamefont {Qiao}, \citenamefont {Rong},
  \citenamefont {Su}, \citenamefont {Sun}, \citenamefont {Wang}, \citenamefont
  {Wang}, \citenamefont {Wu}, \citenamefont {Xu}, \citenamefont {Yan},
  \citenamefont {Yang}, \citenamefont {Yang}, \citenamefont {Ye}, \citenamefont
  {Yin}, \citenamefont {Ying}, \citenamefont {Yu}, \citenamefont {Zha},
  \citenamefont {Zhang}, \citenamefont {Zhang}, \citenamefont {Zhang},
  \citenamefont {Zhang}, \citenamefont {Zhao}, \citenamefont {Zhao},
  \citenamefont {Zhou}, \citenamefont {Zhu}, \citenamefont {Lu}, \citenamefont
  {Peng}, \citenamefont {Zhu},\ and\ \citenamefont {Pan}}]{QuantumSupSuCHN}%
  \BibitemOpen
  \bibfield  {author} {\bibinfo {author} {\bibfnamefont {Y.}~\bibnamefont
  {Wu}}, \bibinfo {author} {\bibfnamefont {W.-S.}\ \bibnamefont {Bao}},
  \bibinfo {author} {\bibfnamefont {S.}~\bibnamefont {Cao}}, \bibinfo {author}
  {\bibfnamefont {F.}~\bibnamefont {Chen}}, \bibinfo {author} {\bibfnamefont
  {M.-C.}\ \bibnamefont {Chen}}, \bibinfo {author} {\bibfnamefont
  {X.}~\bibnamefont {Chen}}, \bibinfo {author} {\bibfnamefont {T.-H.}\
  \bibnamefont {Chung}}, \bibinfo {author} {\bibfnamefont {H.}~\bibnamefont
  {Deng}}, \bibinfo {author} {\bibfnamefont {Y.}~\bibnamefont {Du}}, \bibinfo
  {author} {\bibfnamefont {D.}~\bibnamefont {Fan}}, \bibinfo {author}
  {\bibfnamefont {M.}~\bibnamefont {Gong}}, \bibinfo {author} {\bibfnamefont
  {C.}~\bibnamefont {Guo}}, \bibinfo {author} {\bibfnamefont {C.}~\bibnamefont
  {Guo}}, \bibinfo {author} {\bibfnamefont {S.}~\bibnamefont {Guo}}, \bibinfo
  {author} {\bibfnamefont {L.}~\bibnamefont {Han}}, \bibinfo {author}
  {\bibfnamefont {L.}~\bibnamefont {Hong}}, \bibinfo {author} {\bibfnamefont
  {H.-L.}\ \bibnamefont {Huang}}, \bibinfo {author} {\bibfnamefont {Y.-H.}\
  \bibnamefont {Huo}}, \bibinfo {author} {\bibfnamefont {L.}~\bibnamefont
  {Li}}, \bibinfo {author} {\bibfnamefont {N.}~\bibnamefont {Li}}, \bibinfo
  {author} {\bibfnamefont {S.}~\bibnamefont {Li}}, \bibinfo {author}
  {\bibfnamefont {Y.}~\bibnamefont {Li}}, \bibinfo {author} {\bibfnamefont
  {F.}~\bibnamefont {Liang}}, \bibinfo {author} {\bibfnamefont
  {C.}~\bibnamefont {Lin}}, \bibinfo {author} {\bibfnamefont {J.}~\bibnamefont
  {Lin}}, \bibinfo {author} {\bibfnamefont {H.}~\bibnamefont {Qian}}, \bibinfo
  {author} {\bibfnamefont {D.}~\bibnamefont {Qiao}}, \bibinfo {author}
  {\bibfnamefont {H.}~\bibnamefont {Rong}}, \bibinfo {author} {\bibfnamefont
  {H.}~\bibnamefont {Su}}, \bibinfo {author} {\bibfnamefont {L.}~\bibnamefont
  {Sun}}, \bibinfo {author} {\bibfnamefont {L.}~\bibnamefont {Wang}}, \bibinfo
  {author} {\bibfnamefont {S.}~\bibnamefont {Wang}}, \bibinfo {author}
  {\bibfnamefont {D.}~\bibnamefont {Wu}}, \bibinfo {author} {\bibfnamefont
  {Y.}~\bibnamefont {Xu}}, \bibinfo {author} {\bibfnamefont {K.}~\bibnamefont
  {Yan}}, \bibinfo {author} {\bibfnamefont {W.}~\bibnamefont {Yang}}, \bibinfo
  {author} {\bibfnamefont {Y.}~\bibnamefont {Yang}}, \bibinfo {author}
  {\bibfnamefont {Y.}~\bibnamefont {Ye}}, \bibinfo {author} {\bibfnamefont
  {J.}~\bibnamefont {Yin}}, \bibinfo {author} {\bibfnamefont {C.}~\bibnamefont
  {Ying}}, \bibinfo {author} {\bibfnamefont {J.}~\bibnamefont {Yu}}, \bibinfo
  {author} {\bibfnamefont {C.}~\bibnamefont {Zha}}, \bibinfo {author}
  {\bibfnamefont {C.}~\bibnamefont {Zhang}}, \bibinfo {author} {\bibfnamefont
  {H.}~\bibnamefont {Zhang}}, \bibinfo {author} {\bibfnamefont
  {K.}~\bibnamefont {Zhang}}, \bibinfo {author} {\bibfnamefont
  {Y.}~\bibnamefont {Zhang}}, \bibinfo {author} {\bibfnamefont
  {H.}~\bibnamefont {Zhao}}, \bibinfo {author} {\bibfnamefont {Y.}~\bibnamefont
  {Zhao}}, \bibinfo {author} {\bibfnamefont {L.}~\bibnamefont {Zhou}}, \bibinfo
  {author} {\bibfnamefont {Q.}~\bibnamefont {Zhu}}, \bibinfo {author}
  {\bibfnamefont {C.-Y.}\ \bibnamefont {Lu}}, \bibinfo {author} {\bibfnamefont
  {C.-Z.}\ \bibnamefont {Peng}}, \bibinfo {author} {\bibfnamefont
  {X.}~\bibnamefont {Zhu}},\ and\ \bibinfo {author} {\bibfnamefont {J.-W.}\
  \bibnamefont {Pan}},\ }\bibfield  {title} {\bibinfo {title} {Strong quantum
  computational advantage using a superconducting quantum processor},\ }\href
  {https://doi.org/10.1103/PhysRevLett.127.180501} {\bibfield  {journal}
  {\bibinfo  {journal} {Phys. Rev. Lett.}\ }\textbf {\bibinfo {volume} {127}},\
  \bibinfo {pages} {180501} (\bibinfo {year} {2021})}\BibitemShut {NoStop}%
\bibitem [{\citenamefont {Madsen}\ \emph {et~al.}(2022)\citenamefont {Madsen},
  \citenamefont {Laudenbach}, \citenamefont {Falamarzi.Askarani}, \citenamefont
  {Rortais}, \citenamefont {Vincent}, \citenamefont {Bulmer}, \citenamefont
  {Miatto}, \citenamefont {Neuhaus}, \citenamefont {Helt}, \citenamefont
  {Collins}, \citenamefont {Lita}, \citenamefont {Gerrits}, \citenamefont
  {Nam}, \citenamefont {Vaidya}, \citenamefont {Menotti}, \citenamefont
  {Dhand}, \citenamefont {Vernon}, \citenamefont {Quesada},\ and\ \citenamefont
  {Lavoie}}]{XanaduQAdv}%
  \BibitemOpen
  \bibfield  {author} {\bibinfo {author} {\bibfnamefont {L.~S.}\ \bibnamefont
  {Madsen}}, \bibinfo {author} {\bibfnamefont {F.}~\bibnamefont {Laudenbach}},
  \bibinfo {author} {\bibfnamefont {M.}~\bibnamefont {Falamarzi.Askarani}},
  \bibinfo {author} {\bibfnamefont {F.}~\bibnamefont {Rortais}}, \bibinfo
  {author} {\bibfnamefont {T.}~\bibnamefont {Vincent}}, \bibinfo {author}
  {\bibfnamefont {J.~F.~F.}\ \bibnamefont {Bulmer}}, \bibinfo {author}
  {\bibfnamefont {F.~M.}\ \bibnamefont {Miatto}}, \bibinfo {author}
  {\bibfnamefont {L.}~\bibnamefont {Neuhaus}}, \bibinfo {author} {\bibfnamefont
  {L.~G.}\ \bibnamefont {Helt}}, \bibinfo {author} {\bibfnamefont {M.~J.}\
  \bibnamefont {Collins}}, \bibinfo {author} {\bibfnamefont {A.~E.}\
  \bibnamefont {Lita}}, \bibinfo {author} {\bibfnamefont {T.}~\bibnamefont
  {Gerrits}}, \bibinfo {author} {\bibfnamefont {S.~W.}\ \bibnamefont {Nam}},
  \bibinfo {author} {\bibfnamefont {V.~D.}\ \bibnamefont {Vaidya}}, \bibinfo
  {author} {\bibfnamefont {M.}~\bibnamefont {Menotti}}, \bibinfo {author}
  {\bibfnamefont {I.}~\bibnamefont {Dhand}}, \bibinfo {author} {\bibfnamefont
  {Z.}~\bibnamefont {Vernon}}, \bibinfo {author} {\bibfnamefont
  {N.}~\bibnamefont {Quesada}},\ and\ \bibinfo {author} {\bibfnamefont
  {J.}~\bibnamefont {Lavoie}},\ }\bibfield  {title} {\bibinfo {title} {Quantum
  computational advantage with a programmable photonic processor},\ }\href
  {https://doi.org/10.1038/s41586-022-04725-x} {\bibfield  {journal} {\bibinfo
  {journal} {Nature}\ }\textbf {\bibinfo {volume} {606}},\ \bibinfo {pages}
  {75} (\bibinfo {year} {2022})}\BibitemShut {NoStop}%
\bibitem [{\citenamefont {Deng}\ \emph {et~al.}(2023)\citenamefont {Deng},
  \citenamefont {Gu}, \citenamefont {Liu}, \citenamefont {Gong}, \citenamefont
  {Su}, \citenamefont {Zhang}, \citenamefont {Tang}, \citenamefont {Jia},
  \citenamefont {Xu}, \citenamefont {Chen}, \citenamefont {Zhong},
  \citenamefont {Qin}, \citenamefont {Wang}, \citenamefont {Peng},
  \citenamefont {Yan}, \citenamefont {Hu}, \citenamefont {Huang}, \citenamefont
  {Li}, \citenamefont {Li}, \citenamefont {Chen}, \citenamefont {Jiang},
  \citenamefont {Gan}, \citenamefont {Yang}, \citenamefont {You}, \citenamefont
  {Li}, \citenamefont {Liu}, \citenamefont {Renema}, \citenamefont {Lu},\ and\
  \citenamefont {Pan}}]{Jiuzhang3}%
  \BibitemOpen
  \bibfield  {author} {\bibinfo {author} {\bibfnamefont {Y.-H.}\ \bibnamefont
  {Deng}}, \bibinfo {author} {\bibfnamefont {Y.-C.}\ \bibnamefont {Gu}},
  \bibinfo {author} {\bibfnamefont {H.-L.}\ \bibnamefont {Liu}}, \bibinfo
  {author} {\bibfnamefont {S.-Q.}\ \bibnamefont {Gong}}, \bibinfo {author}
  {\bibfnamefont {H.}~\bibnamefont {Su}}, \bibinfo {author} {\bibfnamefont
  {Z.-J.}\ \bibnamefont {Zhang}}, \bibinfo {author} {\bibfnamefont {H.-Y.}\
  \bibnamefont {Tang}}, \bibinfo {author} {\bibfnamefont {M.-H.}\ \bibnamefont
  {Jia}}, \bibinfo {author} {\bibfnamefont {J.-M.}\ \bibnamefont {Xu}},
  \bibinfo {author} {\bibfnamefont {M.-C.}\ \bibnamefont {Chen}}, \bibinfo
  {author} {\bibfnamefont {H.-S.}\ \bibnamefont {Zhong}}, \bibinfo {author}
  {\bibfnamefont {J.}~\bibnamefont {Qin}}, \bibinfo {author} {\bibfnamefont
  {H.}~\bibnamefont {Wang}}, \bibinfo {author} {\bibfnamefont {L.-C.}\
  \bibnamefont {Peng}}, \bibinfo {author} {\bibfnamefont {J.}~\bibnamefont
  {Yan}}, \bibinfo {author} {\bibfnamefont {Y.}~\bibnamefont {Hu}}, \bibinfo
  {author} {\bibfnamefont {J.}~\bibnamefont {Huang}}, \bibinfo {author}
  {\bibfnamefont {H.}~\bibnamefont {Li}}, \bibinfo {author} {\bibfnamefont
  {Y.}~\bibnamefont {Li}}, \bibinfo {author} {\bibfnamefont {Y.}~\bibnamefont
  {Chen}}, \bibinfo {author} {\bibfnamefont {X.}~\bibnamefont {Jiang}},
  \bibinfo {author} {\bibfnamefont {L.}~\bibnamefont {Gan}}, \bibinfo {author}
  {\bibfnamefont {G.}~\bibnamefont {Yang}}, \bibinfo {author} {\bibfnamefont
  {L.}~\bibnamefont {You}}, \bibinfo {author} {\bibfnamefont {L.}~\bibnamefont
  {Li}}, \bibinfo {author} {\bibfnamefont {N.-L.}\ \bibnamefont {Liu}},
  \bibinfo {author} {\bibfnamefont {J.~J.}\ \bibnamefont {Renema}}, \bibinfo
  {author} {\bibfnamefont {C.-Y.}\ \bibnamefont {Lu}},\ and\ \bibinfo {author}
  {\bibfnamefont {J.-W.}\ \bibnamefont {Pan}},\ }\bibfield  {title} {\bibinfo
  {title} {Gaussian boson sampling with pseudo-photon-number resolving
  detectors and quantum computational advantage},\ }\href
  {https://doi.org/10.1103/PhysRevLett.131.150601} {\bibfield  {journal}
  {\bibinfo  {journal} {Phys. Rev. Lett.}\ }\textbf {\bibinfo {volume} {131}},\
  \bibinfo {pages} {150601} (\bibinfo {year} {2023})}\BibitemShut {NoStop}%
\bibitem [{\citenamefont {Kim}\ \emph {et~al.}(2023)\citenamefont {Kim},
  \citenamefont {Eddins}, \citenamefont {Anand}, \citenamefont {Wei},
  \citenamefont {van~den Berg}, \citenamefont {Rosenblatt}, \citenamefont
  {Nayfeh}, \citenamefont {Wu}, \citenamefont {Zaletel}, \citenamefont
  {Temme},\ and\ \citenamefont {Kandala}}]{QuantumSupIBM}%
  \BibitemOpen
  \bibfield  {author} {\bibinfo {author} {\bibfnamefont {Y.}~\bibnamefont
  {Kim}}, \bibinfo {author} {\bibfnamefont {A.}~\bibnamefont {Eddins}},
  \bibinfo {author} {\bibfnamefont {S.}~\bibnamefont {Anand}}, \bibinfo
  {author} {\bibfnamefont {K.~X.}\ \bibnamefont {Wei}}, \bibinfo {author}
  {\bibfnamefont {E.}~\bibnamefont {van~den Berg}}, \bibinfo {author}
  {\bibfnamefont {S.}~\bibnamefont {Rosenblatt}}, \bibinfo {author}
  {\bibfnamefont {H.}~\bibnamefont {Nayfeh}}, \bibinfo {author} {\bibfnamefont
  {Y.}~\bibnamefont {Wu}}, \bibinfo {author} {\bibfnamefont {M.}~\bibnamefont
  {Zaletel}}, \bibinfo {author} {\bibfnamefont {K.}~\bibnamefont {Temme}},\
  and\ \bibinfo {author} {\bibfnamefont {A.}~\bibnamefont {Kandala}},\
  }\bibfield  {title} {\bibinfo {title} {Evidence for the utility of quantum
  computing before fault tolerance},\ }\href
  {https://doi.org/10.1038/s41586-023-06096-3} {\bibfield  {journal} {\bibinfo
  {journal} {Nature}\ }\textbf {\bibinfo {volume} {618}},\ \bibinfo {pages}
  {500} (\bibinfo {year} {2023})}\BibitemShut {NoStop}%
\bibitem [{\citenamefont {Acharya}\ \emph {et~al.}(2024)\citenamefont
  {Acharya}, \citenamefont {Abanin}, \citenamefont {Aghababaie-Beni},
  \citenamefont {Aleiner}, \citenamefont {Andersen}, \citenamefont {Ansmann},
  \citenamefont {Arute}, \citenamefont {Arya}, \citenamefont {Asfaw},
  \citenamefont {Astrakhantsev}, \citenamefont {Atalaya}, \citenamefont
  {Babbush}, \citenamefont {Bacon}, \citenamefont {Ballard}, \citenamefont
  {Bardin}, \citenamefont {Bausch}, \citenamefont {Bengtsson}, \citenamefont
  {Bilmes}, \citenamefont {Blackwell}, \citenamefont {Boixo}, \citenamefont
  {Bortoli}, \citenamefont {Bourassa}, \citenamefont {Bovaird}, \citenamefont
  {Brill}, \citenamefont {Broughton}, \citenamefont {Browne}, \citenamefont
  {Buchea}, \citenamefont {Buckley}, \citenamefont {Buell}, \citenamefont
  {Burger}, \citenamefont {Burkett}, \citenamefont {Bushnell}, \citenamefont
  {Cabrera}, \citenamefont {Campero}, \citenamefont {Chang}, \citenamefont
  {Chen}, \citenamefont {Chen}, \citenamefont {Chiaro}, \citenamefont {Chik},
  \citenamefont {Chou}, \citenamefont {Claes}, \citenamefont {Cleland},
  \citenamefont {Cogan}, \citenamefont {Collins}, \citenamefont {Conner},
  \citenamefont {Courtney}, \citenamefont {Crook}, \citenamefont {Curtin},
  \citenamefont {Das}, \citenamefont {Davies}, \citenamefont {De~Lorenzo},
  \citenamefont {Debroy}, \citenamefont {Demura}, \citenamefont {Devoret},
  \citenamefont {Di~Paolo}, \citenamefont {Donohoe}, \citenamefont {Drozdov},
  \citenamefont {Dunsworth}, \citenamefont {Earle}, \citenamefont {Edlich},
  \citenamefont {Eickbusch}, \citenamefont {Elbag}, \citenamefont {Elzouka},
  \citenamefont {Erickson}, \citenamefont {Faoro}, \citenamefont {Farhi},
  \citenamefont {Ferreira}, \citenamefont {Burgos}, \citenamefont {Forati},
  \citenamefont {Fowler}, \citenamefont {Foxen}, \citenamefont {Ganjam},
  \citenamefont {Garcia}, \citenamefont {Gasca}, \citenamefont {Genois},
  \citenamefont {Giang}, \citenamefont {Gidney}, \citenamefont {Gilboa},
  \citenamefont {Gosula}, \citenamefont {Dau}, \citenamefont {Graumann},
  \citenamefont {Greene}, \citenamefont {Gross}, \citenamefont {Habegger},
  \citenamefont {Hall}, \citenamefont {Hamilton}, \citenamefont {Hansen},
  \citenamefont {Harrigan}, \citenamefont {Harrington}, \citenamefont {Heras},
  \citenamefont {Heslin}, \citenamefont {Heu}, \citenamefont {Higgott},
  \citenamefont {Hill}, \citenamefont {Hilton}, \citenamefont {Holland},
  \citenamefont {Hong}, \citenamefont {Huang}, \citenamefont {Huff},
  \citenamefont {Huggins}, \citenamefont {Ioffe}, \citenamefont {Isakov},
  \citenamefont {Iveland}, \citenamefont {Jeffrey}, \citenamefont {Jiang},
  \citenamefont {Jones}, \citenamefont {Jordan}, \citenamefont {Joshi},
  \citenamefont {Juhas}, \citenamefont {Kafri}, \citenamefont {Kang},
  \citenamefont {Karamlou}, \citenamefont {Kechedzhi}, \citenamefont {Kelly},
  \citenamefont {Khaire}, \citenamefont {Khattar}, \citenamefont {Khezri},
  \citenamefont {Kim}, \citenamefont {Klimov}, \citenamefont {Klots},
  \citenamefont {Kobrin}, \citenamefont {Kohli}, \citenamefont {Korotkov},
  \citenamefont {Kostritsa}, \citenamefont {Kothari}, \citenamefont
  {Kozlovskii}, \citenamefont {Kreikebaum}, \citenamefont {Kurilovich},
  \citenamefont {Lacroix}, \citenamefont {Landhuis}, \citenamefont {Lange-Dei},
  \citenamefont {Langley}, \citenamefont {Laptev}, \citenamefont {Lau},
  \citenamefont {Le~Guevel}, \citenamefont {Ledford}, \citenamefont {Lee},
  \citenamefont {Lee}, \citenamefont {Lensky}, \citenamefont {Leon},
  \citenamefont {Lester}, \citenamefont {Li}, \citenamefont {Li}, \citenamefont
  {Lill}, \citenamefont {Liu}, \citenamefont {Livingston}, \citenamefont
  {Locharla}, \citenamefont {Lucero}, \citenamefont {Lundahl}, \citenamefont
  {Lunt}, \citenamefont {Madhuk}, \citenamefont {Malone}, \citenamefont
  {Maloney}, \citenamefont {Mandrà}, \citenamefont {Manyika}, \citenamefont
  {Martin}, \citenamefont {Martin}, \citenamefont {Martin}, \citenamefont
  {Maxfield}, \citenamefont {McClean}, \citenamefont {McEwen}, \citenamefont
  {Meeks}, \citenamefont {Megrant}, \citenamefont {Mi}, \citenamefont {Miao},
  \citenamefont {Mieszala}, \citenamefont {Molavi}, \citenamefont {Molina},
  \citenamefont {Montazeri}, \citenamefont {Morvan}, \citenamefont {Movassagh},
  \citenamefont {Mruczkiewicz}, \citenamefont {Naaman}, \citenamefont {Neeley},
  \citenamefont {Neill}, \citenamefont {Nersisyan}, \citenamefont {Neven},
  \citenamefont {Newman}, \citenamefont {Ng}, \citenamefont {Nguyen},
  \citenamefont {Nguyen}, \citenamefont {Ni}, \citenamefont {Niu},
  \citenamefont {O’Brien}, \citenamefont {Oliver}, \citenamefont {Opremcak},
  \citenamefont {Ottosson}, \citenamefont {Petukhov}, \citenamefont {Pizzuto},
  \citenamefont {Platt}, \citenamefont {Potter}, \citenamefont {Pritchard},
  \citenamefont {Pryadko}, \citenamefont {Quintana}, \citenamefont
  {Ramachandran}, \citenamefont {Reagor}, \citenamefont {Redding},
  \citenamefont {Rhodes}, \citenamefont {Roberts}, \citenamefont {Rosenberg},
  \citenamefont {Rosenfeld}, \citenamefont {Roushan}, \citenamefont {Rubin},
  \citenamefont {Saei}, \citenamefont {Sank}, \citenamefont {Sankaragomathi},
  \citenamefont {Satzinger}, \citenamefont {Schurkus}, \citenamefont
  {Schuster}, \citenamefont {Senior}, \citenamefont {Shearn}, \citenamefont
  {Shorter}, \citenamefont {Shutty}, \citenamefont {Shvarts}, \citenamefont
  {Singh}, \citenamefont {Sivak}, \citenamefont {Skruzny}, \citenamefont
  {Small}, \citenamefont {Smelyanskiy}, \citenamefont {Smith}, \citenamefont
  {Somma}, \citenamefont {Springer}, \citenamefont {Sterling}, \citenamefont
  {Strain}, \citenamefont {Suchard}, \citenamefont {Szasz}, \citenamefont
  {Sztein}, \citenamefont {Thor}, \citenamefont {Torres}, \citenamefont
  {Torunbalci}, \citenamefont {Vaishnav}, \citenamefont {Vargas}, \citenamefont
  {Vdovichev}, \citenamefont {Vidal}, \citenamefont {Villalonga}, \citenamefont
  {Heidweiller}, \citenamefont {Waltman}, \citenamefont {Wang}, \citenamefont
  {Ware}, \citenamefont {Weber}, \citenamefont {Weidel}, \citenamefont {White},
  \citenamefont {Wong}, \citenamefont {Woo}, \citenamefont {Xing},
  \citenamefont {Yao}, \citenamefont {Yeh}, \citenamefont {Ying}, \citenamefont
  {Yoo}, \citenamefont {Yosri}, \citenamefont {Young}, \citenamefont {Zalcman},
  \citenamefont {Zhang}, \citenamefont {Zhu},\ and\ \citenamefont
  {Zobrist}}]{GoogleWillow}%
  \BibitemOpen
  \bibfield  {author} {\bibinfo {author} {\bibfnamefont {R.}~\bibnamefont
  {Acharya}}, \bibinfo {author} {\bibfnamefont {D.~A.}\ \bibnamefont {Abanin}},
  \bibinfo {author} {\bibfnamefont {L.}~\bibnamefont {Aghababaie-Beni}},
  \bibinfo {author} {\bibfnamefont {I.}~\bibnamefont {Aleiner}}, \bibinfo
  {author} {\bibfnamefont {T.~I.}\ \bibnamefont {Andersen}}, \bibinfo {author}
  {\bibfnamefont {M.}~\bibnamefont {Ansmann}}, \bibinfo {author} {\bibfnamefont
  {F.}~\bibnamefont {Arute}}, \bibinfo {author} {\bibfnamefont
  {K.}~\bibnamefont {Arya}}, \bibinfo {author} {\bibfnamefont {A.}~\bibnamefont
  {Asfaw}}, \bibinfo {author} {\bibfnamefont {N.}~\bibnamefont
  {Astrakhantsev}}, \bibinfo {author} {\bibfnamefont {J.}~\bibnamefont
  {Atalaya}}, \bibinfo {author} {\bibfnamefont {R.}~\bibnamefont {Babbush}},
  \bibinfo {author} {\bibfnamefont {D.}~\bibnamefont {Bacon}}, \bibinfo
  {author} {\bibfnamefont {B.}~\bibnamefont {Ballard}}, \bibinfo {author}
  {\bibfnamefont {J.~C.}\ \bibnamefont {Bardin}}, \bibinfo {author}
  {\bibfnamefont {J.}~\bibnamefont {Bausch}}, \bibinfo {author} {\bibfnamefont
  {A.}~\bibnamefont {Bengtsson}}, \bibinfo {author} {\bibfnamefont
  {A.}~\bibnamefont {Bilmes}}, \bibinfo {author} {\bibfnamefont
  {S.}~\bibnamefont {Blackwell}}, \bibinfo {author} {\bibfnamefont
  {S.}~\bibnamefont {Boixo}}, \bibinfo {author} {\bibfnamefont
  {G.}~\bibnamefont {Bortoli}}, \bibinfo {author} {\bibfnamefont
  {A.}~\bibnamefont {Bourassa}}, \bibinfo {author} {\bibfnamefont
  {J.}~\bibnamefont {Bovaird}}, \bibinfo {author} {\bibfnamefont
  {L.}~\bibnamefont {Brill}}, \bibinfo {author} {\bibfnamefont
  {M.}~\bibnamefont {Broughton}}, \bibinfo {author} {\bibfnamefont {D.~A.}\
  \bibnamefont {Browne}}, \bibinfo {author} {\bibfnamefont {B.}~\bibnamefont
  {Buchea}}, \bibinfo {author} {\bibfnamefont {B.~B.}\ \bibnamefont {Buckley}},
  \bibinfo {author} {\bibfnamefont {D.~A.}\ \bibnamefont {Buell}}, \bibinfo
  {author} {\bibfnamefont {T.}~\bibnamefont {Burger}}, \bibinfo {author}
  {\bibfnamefont {B.}~\bibnamefont {Burkett}}, \bibinfo {author} {\bibfnamefont
  {N.}~\bibnamefont {Bushnell}}, \bibinfo {author} {\bibfnamefont
  {A.}~\bibnamefont {Cabrera}}, \bibinfo {author} {\bibfnamefont
  {J.}~\bibnamefont {Campero}}, \bibinfo {author} {\bibfnamefont {H.-S.}\
  \bibnamefont {Chang}}, \bibinfo {author} {\bibfnamefont {Y.}~\bibnamefont
  {Chen}}, \bibinfo {author} {\bibfnamefont {Z.}~\bibnamefont {Chen}}, \bibinfo
  {author} {\bibfnamefont {B.}~\bibnamefont {Chiaro}}, \bibinfo {author}
  {\bibfnamefont {D.}~\bibnamefont {Chik}}, \bibinfo {author} {\bibfnamefont
  {C.}~\bibnamefont {Chou}}, \bibinfo {author} {\bibfnamefont {J.}~\bibnamefont
  {Claes}}, \bibinfo {author} {\bibfnamefont {A.~Y.}\ \bibnamefont {Cleland}},
  \bibinfo {author} {\bibfnamefont {J.}~\bibnamefont {Cogan}}, \bibinfo
  {author} {\bibfnamefont {R.}~\bibnamefont {Collins}}, \bibinfo {author}
  {\bibfnamefont {P.}~\bibnamefont {Conner}}, \bibinfo {author} {\bibfnamefont
  {W.}~\bibnamefont {Courtney}}, \bibinfo {author} {\bibfnamefont {A.~L.}\
  \bibnamefont {Crook}}, \bibinfo {author} {\bibfnamefont {B.}~\bibnamefont
  {Curtin}}, \bibinfo {author} {\bibfnamefont {S.}~\bibnamefont {Das}},
  \bibinfo {author} {\bibfnamefont {A.}~\bibnamefont {Davies}}, \bibinfo
  {author} {\bibfnamefont {L.}~\bibnamefont {De~Lorenzo}}, \bibinfo {author}
  {\bibfnamefont {D.~M.}\ \bibnamefont {Debroy}}, \bibinfo {author}
  {\bibfnamefont {S.}~\bibnamefont {Demura}}, \bibinfo {author} {\bibfnamefont
  {M.}~\bibnamefont {Devoret}}, \bibinfo {author} {\bibfnamefont
  {A.}~\bibnamefont {Di~Paolo}}, \bibinfo {author} {\bibfnamefont
  {P.}~\bibnamefont {Donohoe}}, \bibinfo {author} {\bibfnamefont
  {I.}~\bibnamefont {Drozdov}}, \bibinfo {author} {\bibfnamefont
  {A.}~\bibnamefont {Dunsworth}}, \bibinfo {author} {\bibfnamefont
  {C.}~\bibnamefont {Earle}}, \bibinfo {author} {\bibfnamefont
  {T.}~\bibnamefont {Edlich}}, \bibinfo {author} {\bibfnamefont
  {A.}~\bibnamefont {Eickbusch}}, \bibinfo {author} {\bibfnamefont {A.~M.}\
  \bibnamefont {Elbag}}, \bibinfo {author} {\bibfnamefont {M.}~\bibnamefont
  {Elzouka}}, \bibinfo {author} {\bibfnamefont {C.}~\bibnamefont {Erickson}},
  \bibinfo {author} {\bibfnamefont {L.}~\bibnamefont {Faoro}}, \bibinfo
  {author} {\bibfnamefont {E.}~\bibnamefont {Farhi}}, \bibinfo {author}
  {\bibfnamefont {V.~S.}\ \bibnamefont {Ferreira}}, \bibinfo {author}
  {\bibfnamefont {L.~F.}\ \bibnamefont {Burgos}}, \bibinfo {author}
  {\bibfnamefont {E.}~\bibnamefont {Forati}}, \bibinfo {author} {\bibfnamefont
  {A.~G.}\ \bibnamefont {Fowler}}, \bibinfo {author} {\bibfnamefont
  {B.}~\bibnamefont {Foxen}}, \bibinfo {author} {\bibfnamefont
  {S.}~\bibnamefont {Ganjam}}, \bibinfo {author} {\bibfnamefont
  {G.}~\bibnamefont {Garcia}}, \bibinfo {author} {\bibfnamefont
  {R.}~\bibnamefont {Gasca}}, \bibinfo {author} {\bibfnamefont
  {E.}~\bibnamefont {Genois}}, \bibinfo {author} {\bibfnamefont
  {W.}~\bibnamefont {Giang}}, \bibinfo {author} {\bibfnamefont
  {C.}~\bibnamefont {Gidney}}, \bibinfo {author} {\bibfnamefont
  {D.}~\bibnamefont {Gilboa}}, \bibinfo {author} {\bibfnamefont
  {R.}~\bibnamefont {Gosula}}, \bibinfo {author} {\bibfnamefont {A.~G.}\
  \bibnamefont {Dau}}, \bibinfo {author} {\bibfnamefont {D.}~\bibnamefont
  {Graumann}}, \bibinfo {author} {\bibfnamefont {A.}~\bibnamefont {Greene}},
  \bibinfo {author} {\bibfnamefont {J.~A.}\ \bibnamefont {Gross}}, \bibinfo
  {author} {\bibfnamefont {S.}~\bibnamefont {Habegger}}, \bibinfo {author}
  {\bibfnamefont {J.}~\bibnamefont {Hall}}, \bibinfo {author} {\bibfnamefont
  {M.~C.}\ \bibnamefont {Hamilton}}, \bibinfo {author} {\bibfnamefont
  {M.}~\bibnamefont {Hansen}}, \bibinfo {author} {\bibfnamefont {M.~P.}\
  \bibnamefont {Harrigan}}, \bibinfo {author} {\bibfnamefont {S.~D.}\
  \bibnamefont {Harrington}}, \bibinfo {author} {\bibfnamefont {F.~J.~H.}\
  \bibnamefont {Heras}}, \bibinfo {author} {\bibfnamefont {S.}~\bibnamefont
  {Heslin}}, \bibinfo {author} {\bibfnamefont {P.}~\bibnamefont {Heu}},
  \bibinfo {author} {\bibfnamefont {O.}~\bibnamefont {Higgott}}, \bibinfo
  {author} {\bibfnamefont {G.}~\bibnamefont {Hill}}, \bibinfo {author}
  {\bibfnamefont {J.}~\bibnamefont {Hilton}}, \bibinfo {author} {\bibfnamefont
  {G.}~\bibnamefont {Holland}}, \bibinfo {author} {\bibfnamefont
  {S.}~\bibnamefont {Hong}}, \bibinfo {author} {\bibfnamefont {H.-Y.}\
  \bibnamefont {Huang}}, \bibinfo {author} {\bibfnamefont {A.}~\bibnamefont
  {Huff}}, \bibinfo {author} {\bibfnamefont {W.~J.}\ \bibnamefont {Huggins}},
  \bibinfo {author} {\bibfnamefont {L.~B.}\ \bibnamefont {Ioffe}}, \bibinfo
  {author} {\bibfnamefont {S.~V.}\ \bibnamefont {Isakov}}, \bibinfo {author}
  {\bibfnamefont {J.}~\bibnamefont {Iveland}}, \bibinfo {author} {\bibfnamefont
  {E.}~\bibnamefont {Jeffrey}}, \bibinfo {author} {\bibfnamefont
  {Z.}~\bibnamefont {Jiang}}, \bibinfo {author} {\bibfnamefont
  {C.}~\bibnamefont {Jones}}, \bibinfo {author} {\bibfnamefont
  {S.}~\bibnamefont {Jordan}}, \bibinfo {author} {\bibfnamefont
  {C.}~\bibnamefont {Joshi}}, \bibinfo {author} {\bibfnamefont
  {P.}~\bibnamefont {Juhas}}, \bibinfo {author} {\bibfnamefont
  {D.}~\bibnamefont {Kafri}}, \bibinfo {author} {\bibfnamefont
  {H.}~\bibnamefont {Kang}}, \bibinfo {author} {\bibfnamefont {A.~H.}\
  \bibnamefont {Karamlou}}, \bibinfo {author} {\bibfnamefont {K.}~\bibnamefont
  {Kechedzhi}}, \bibinfo {author} {\bibfnamefont {J.}~\bibnamefont {Kelly}},
  \bibinfo {author} {\bibfnamefont {T.}~\bibnamefont {Khaire}}, \bibinfo
  {author} {\bibfnamefont {T.}~\bibnamefont {Khattar}}, \bibinfo {author}
  {\bibfnamefont {M.}~\bibnamefont {Khezri}}, \bibinfo {author} {\bibfnamefont
  {S.}~\bibnamefont {Kim}}, \bibinfo {author} {\bibfnamefont {P.~V.}\
  \bibnamefont {Klimov}}, \bibinfo {author} {\bibfnamefont {A.~R.}\
  \bibnamefont {Klots}}, \bibinfo {author} {\bibfnamefont {B.}~\bibnamefont
  {Kobrin}}, \bibinfo {author} {\bibfnamefont {P.}~\bibnamefont {Kohli}},
  \bibinfo {author} {\bibfnamefont {A.~N.}\ \bibnamefont {Korotkov}}, \bibinfo
  {author} {\bibfnamefont {F.}~\bibnamefont {Kostritsa}}, \bibinfo {author}
  {\bibfnamefont {R.}~\bibnamefont {Kothari}}, \bibinfo {author} {\bibfnamefont
  {B.}~\bibnamefont {Kozlovskii}}, \bibinfo {author} {\bibfnamefont {J.~M.}\
  \bibnamefont {Kreikebaum}}, \bibinfo {author} {\bibfnamefont {V.~D.}\
  \bibnamefont {Kurilovich}}, \bibinfo {author} {\bibfnamefont
  {N.}~\bibnamefont {Lacroix}}, \bibinfo {author} {\bibfnamefont
  {D.}~\bibnamefont {Landhuis}}, \bibinfo {author} {\bibfnamefont
  {T.}~\bibnamefont {Lange-Dei}}, \bibinfo {author} {\bibfnamefont {B.~W.}\
  \bibnamefont {Langley}}, \bibinfo {author} {\bibfnamefont {P.}~\bibnamefont
  {Laptev}}, \bibinfo {author} {\bibfnamefont {K.-M.}\ \bibnamefont {Lau}},
  \bibinfo {author} {\bibfnamefont {L.}~\bibnamefont {Le~Guevel}}, \bibinfo
  {author} {\bibfnamefont {J.}~\bibnamefont {Ledford}}, \bibinfo {author}
  {\bibfnamefont {J.}~\bibnamefont {Lee}}, \bibinfo {author} {\bibfnamefont
  {K.}~\bibnamefont {Lee}}, \bibinfo {author} {\bibfnamefont {Y.~D.}\
  \bibnamefont {Lensky}}, \bibinfo {author} {\bibfnamefont {S.}~\bibnamefont
  {Leon}}, \bibinfo {author} {\bibfnamefont {B.~J.}\ \bibnamefont {Lester}},
  \bibinfo {author} {\bibfnamefont {W.~Y.}\ \bibnamefont {Li}}, \bibinfo
  {author} {\bibfnamefont {Y.}~\bibnamefont {Li}}, \bibinfo {author}
  {\bibfnamefont {A.~T.}\ \bibnamefont {Lill}}, \bibinfo {author}
  {\bibfnamefont {W.}~\bibnamefont {Liu}}, \bibinfo {author} {\bibfnamefont
  {W.~P.}\ \bibnamefont {Livingston}}, \bibinfo {author} {\bibfnamefont
  {A.}~\bibnamefont {Locharla}}, \bibinfo {author} {\bibfnamefont
  {E.}~\bibnamefont {Lucero}}, \bibinfo {author} {\bibfnamefont
  {D.}~\bibnamefont {Lundahl}}, \bibinfo {author} {\bibfnamefont
  {A.}~\bibnamefont {Lunt}}, \bibinfo {author} {\bibfnamefont {S.}~\bibnamefont
  {Madhuk}}, \bibinfo {author} {\bibfnamefont {F.~D.}\ \bibnamefont {Malone}},
  \bibinfo {author} {\bibfnamefont {A.}~\bibnamefont {Maloney}}, \bibinfo
  {author} {\bibfnamefont {S.}~\bibnamefont {Mandrà}}, \bibinfo {author}
  {\bibfnamefont {J.}~\bibnamefont {Manyika}}, \bibinfo {author} {\bibfnamefont
  {L.~S.}\ \bibnamefont {Martin}}, \bibinfo {author} {\bibfnamefont
  {O.}~\bibnamefont {Martin}}, \bibinfo {author} {\bibfnamefont
  {S.}~\bibnamefont {Martin}}, \bibinfo {author} {\bibfnamefont
  {C.}~\bibnamefont {Maxfield}}, \bibinfo {author} {\bibfnamefont {J.~R.}\
  \bibnamefont {McClean}}, \bibinfo {author} {\bibfnamefont {M.}~\bibnamefont
  {McEwen}}, \bibinfo {author} {\bibfnamefont {S.}~\bibnamefont {Meeks}},
  \bibinfo {author} {\bibfnamefont {A.}~\bibnamefont {Megrant}}, \bibinfo
  {author} {\bibfnamefont {X.}~\bibnamefont {Mi}}, \bibinfo {author}
  {\bibfnamefont {K.~C.}\ \bibnamefont {Miao}}, \bibinfo {author}
  {\bibfnamefont {A.}~\bibnamefont {Mieszala}}, \bibinfo {author}
  {\bibfnamefont {R.}~\bibnamefont {Molavi}}, \bibinfo {author} {\bibfnamefont
  {S.}~\bibnamefont {Molina}}, \bibinfo {author} {\bibfnamefont
  {S.}~\bibnamefont {Montazeri}}, \bibinfo {author} {\bibfnamefont
  {A.}~\bibnamefont {Morvan}}, \bibinfo {author} {\bibfnamefont
  {R.}~\bibnamefont {Movassagh}}, \bibinfo {author} {\bibfnamefont
  {W.}~\bibnamefont {Mruczkiewicz}}, \bibinfo {author} {\bibfnamefont
  {O.}~\bibnamefont {Naaman}}, \bibinfo {author} {\bibfnamefont
  {M.}~\bibnamefont {Neeley}}, \bibinfo {author} {\bibfnamefont
  {C.}~\bibnamefont {Neill}}, \bibinfo {author} {\bibfnamefont
  {A.}~\bibnamefont {Nersisyan}}, \bibinfo {author} {\bibfnamefont
  {H.}~\bibnamefont {Neven}}, \bibinfo {author} {\bibfnamefont
  {M.}~\bibnamefont {Newman}}, \bibinfo {author} {\bibfnamefont {J.~H.}\
  \bibnamefont {Ng}}, \bibinfo {author} {\bibfnamefont {A.}~\bibnamefont
  {Nguyen}}, \bibinfo {author} {\bibfnamefont {M.}~\bibnamefont {Nguyen}},
  \bibinfo {author} {\bibfnamefont {C.-H.}\ \bibnamefont {Ni}}, \bibinfo
  {author} {\bibfnamefont {M.~Y.}\ \bibnamefont {Niu}}, \bibinfo {author}
  {\bibfnamefont {T.~E.}\ \bibnamefont {O’Brien}}, \bibinfo {author}
  {\bibfnamefont {W.~D.}\ \bibnamefont {Oliver}}, \bibinfo {author}
  {\bibfnamefont {A.}~\bibnamefont {Opremcak}}, \bibinfo {author}
  {\bibfnamefont {K.}~\bibnamefont {Ottosson}}, \bibinfo {author}
  {\bibfnamefont {A.}~\bibnamefont {Petukhov}}, \bibinfo {author}
  {\bibfnamefont {A.}~\bibnamefont {Pizzuto}}, \bibinfo {author} {\bibfnamefont
  {J.}~\bibnamefont {Platt}}, \bibinfo {author} {\bibfnamefont
  {R.}~\bibnamefont {Potter}}, \bibinfo {author} {\bibfnamefont
  {O.}~\bibnamefont {Pritchard}}, \bibinfo {author} {\bibfnamefont {L.~P.}\
  \bibnamefont {Pryadko}}, \bibinfo {author} {\bibfnamefont {C.}~\bibnamefont
  {Quintana}}, \bibinfo {author} {\bibfnamefont {G.}~\bibnamefont
  {Ramachandran}}, \bibinfo {author} {\bibfnamefont {M.~J.}\ \bibnamefont
  {Reagor}}, \bibinfo {author} {\bibfnamefont {J.}~\bibnamefont {Redding}},
  \bibinfo {author} {\bibfnamefont {D.~M.}\ \bibnamefont {Rhodes}}, \bibinfo
  {author} {\bibfnamefont {G.}~\bibnamefont {Roberts}}, \bibinfo {author}
  {\bibfnamefont {E.}~\bibnamefont {Rosenberg}}, \bibinfo {author}
  {\bibfnamefont {E.}~\bibnamefont {Rosenfeld}}, \bibinfo {author}
  {\bibfnamefont {P.}~\bibnamefont {Roushan}}, \bibinfo {author} {\bibfnamefont
  {N.~C.}\ \bibnamefont {Rubin}}, \bibinfo {author} {\bibfnamefont
  {N.}~\bibnamefont {Saei}}, \bibinfo {author} {\bibfnamefont {D.}~\bibnamefont
  {Sank}}, \bibinfo {author} {\bibfnamefont {K.}~\bibnamefont
  {Sankaragomathi}}, \bibinfo {author} {\bibfnamefont {K.~J.}\ \bibnamefont
  {Satzinger}}, \bibinfo {author} {\bibfnamefont {H.~F.}\ \bibnamefont
  {Schurkus}}, \bibinfo {author} {\bibfnamefont {C.}~\bibnamefont {Schuster}},
  \bibinfo {author} {\bibfnamefont {A.~W.}\ \bibnamefont {Senior}}, \bibinfo
  {author} {\bibfnamefont {M.~J.}\ \bibnamefont {Shearn}}, \bibinfo {author}
  {\bibfnamefont {A.}~\bibnamefont {Shorter}}, \bibinfo {author} {\bibfnamefont
  {N.}~\bibnamefont {Shutty}}, \bibinfo {author} {\bibfnamefont
  {V.}~\bibnamefont {Shvarts}}, \bibinfo {author} {\bibfnamefont
  {S.}~\bibnamefont {Singh}}, \bibinfo {author} {\bibfnamefont
  {V.}~\bibnamefont {Sivak}}, \bibinfo {author} {\bibfnamefont
  {J.}~\bibnamefont {Skruzny}}, \bibinfo {author} {\bibfnamefont
  {S.}~\bibnamefont {Small}}, \bibinfo {author} {\bibfnamefont
  {V.}~\bibnamefont {Smelyanskiy}}, \bibinfo {author} {\bibfnamefont {W.~C.}\
  \bibnamefont {Smith}}, \bibinfo {author} {\bibfnamefont {R.~D.}\ \bibnamefont
  {Somma}}, \bibinfo {author} {\bibfnamefont {S.}~\bibnamefont {Springer}},
  \bibinfo {author} {\bibfnamefont {G.}~\bibnamefont {Sterling}}, \bibinfo
  {author} {\bibfnamefont {D.}~\bibnamefont {Strain}}, \bibinfo {author}
  {\bibfnamefont {J.}~\bibnamefont {Suchard}}, \bibinfo {author} {\bibfnamefont
  {A.}~\bibnamefont {Szasz}}, \bibinfo {author} {\bibfnamefont
  {A.}~\bibnamefont {Sztein}}, \bibinfo {author} {\bibfnamefont
  {D.}~\bibnamefont {Thor}}, \bibinfo {author} {\bibfnamefont {A.}~\bibnamefont
  {Torres}}, \bibinfo {author} {\bibfnamefont {M.~M.}\ \bibnamefont
  {Torunbalci}}, \bibinfo {author} {\bibfnamefont {A.}~\bibnamefont
  {Vaishnav}}, \bibinfo {author} {\bibfnamefont {J.}~\bibnamefont {Vargas}},
  \bibinfo {author} {\bibfnamefont {S.}~\bibnamefont {Vdovichev}}, \bibinfo
  {author} {\bibfnamefont {G.}~\bibnamefont {Vidal}}, \bibinfo {author}
  {\bibfnamefont {B.}~\bibnamefont {Villalonga}}, \bibinfo {author}
  {\bibfnamefont {C.~V.}\ \bibnamefont {Heidweiller}}, \bibinfo {author}
  {\bibfnamefont {S.}~\bibnamefont {Waltman}}, \bibinfo {author} {\bibfnamefont
  {S.~X.}\ \bibnamefont {Wang}}, \bibinfo {author} {\bibfnamefont
  {B.}~\bibnamefont {Ware}}, \bibinfo {author} {\bibfnamefont {K.}~\bibnamefont
  {Weber}}, \bibinfo {author} {\bibfnamefont {T.}~\bibnamefont {Weidel}},
  \bibinfo {author} {\bibfnamefont {T.}~\bibnamefont {White}}, \bibinfo
  {author} {\bibfnamefont {K.}~\bibnamefont {Wong}}, \bibinfo {author}
  {\bibfnamefont {B.~W.~K.}\ \bibnamefont {Woo}}, \bibinfo {author}
  {\bibfnamefont {C.}~\bibnamefont {Xing}}, \bibinfo {author} {\bibfnamefont
  {Z.~J.}\ \bibnamefont {Yao}}, \bibinfo {author} {\bibfnamefont
  {P.}~\bibnamefont {Yeh}}, \bibinfo {author} {\bibfnamefont {B.}~\bibnamefont
  {Ying}}, \bibinfo {author} {\bibfnamefont {J.}~\bibnamefont {Yoo}}, \bibinfo
  {author} {\bibfnamefont {N.}~\bibnamefont {Yosri}}, \bibinfo {author}
  {\bibfnamefont {G.}~\bibnamefont {Young}}, \bibinfo {author} {\bibfnamefont
  {A.}~\bibnamefont {Zalcman}}, \bibinfo {author} {\bibfnamefont
  {Y.}~\bibnamefont {Zhang}}, \bibinfo {author} {\bibfnamefont
  {N.}~\bibnamefont {Zhu}},\ and\ \bibinfo {author} {\bibfnamefont
  {N.}~\bibnamefont {Zobrist}},\ }\bibfield  {title} {\bibinfo {title} {Quantum
  error correction below the surface code threshold},\ }\href
  {https://doi.org/10.1038/s41586-024-08449-y} {\bibfield  {journal} {\bibinfo
  {journal} {Nature}\ }\textbf {\bibinfo {volume} {638}},\ \bibinfo {pages}
  {920–926} (\bibinfo {year} {2024})}\BibitemShut {NoStop}%
\bibitem [{\citenamefont {Wang}\ \emph {et~al.}(2019)\citenamefont {Wang},
  \citenamefont {Sciarrino}, \citenamefont {Laing},\ and\ \citenamefont
  {Thompson}}]{PhotonicTech}%
  \BibitemOpen
  \bibfield  {author} {\bibinfo {author} {\bibfnamefont {J.}~\bibnamefont
  {Wang}}, \bibinfo {author} {\bibfnamefont {F.}~\bibnamefont {Sciarrino}},
  \bibinfo {author} {\bibfnamefont {A.}~\bibnamefont {Laing}},\ and\ \bibinfo
  {author} {\bibfnamefont {M.~G.}\ \bibnamefont {Thompson}},\ }\bibfield
  {title} {\bibinfo {title} {Integrated photonic quantum technologies},\ }\href
  {https://doi.org/10.1038/s41566-019-0532-1} {\bibfield  {journal} {\bibinfo
  {journal} {Nat. Photon.}\ }\textbf {\bibinfo {volume} {14}},\ \bibinfo
  {pages} {273–284} (\bibinfo {year} {2019})}\BibitemShut {NoStop}%
\bibitem [{\citenamefont {Rohde}(2015)}]{LoopLinear}%
  \BibitemOpen
  \bibfield  {author} {\bibinfo {author} {\bibfnamefont {P.~P.}\ \bibnamefont
  {Rohde}},\ }\bibfield  {title} {\bibinfo {title} {Simple scheme for universal
  linear-optics quantum computing with constant experimental complexity using
  fiber loops},\ }\href {https://doi.org/10.1103/PhysRevA.91.012306} {\bibfield
   {journal} {\bibinfo  {journal} {Phys. Rev. A}\ }\textbf {\bibinfo {volume}
  {91}},\ \bibinfo {pages} {012306} (\bibinfo {year} {2015})}\BibitemShut
  {NoStop}%
\bibitem [{\citenamefont {Bartlett}\ \emph {et~al.}(2021)\citenamefont
  {Bartlett}, \citenamefont {Dutt},\ and\ \citenamefont {Fan}}]{BartlettLoop}%
  \BibitemOpen
  \bibfield  {author} {\bibinfo {author} {\bibfnamefont {B.}~\bibnamefont
  {Bartlett}}, \bibinfo {author} {\bibfnamefont {A.}~\bibnamefont {Dutt}},\
  and\ \bibinfo {author} {\bibfnamefont {S.}~\bibnamefont {Fan}},\ }\bibfield
  {title} {\bibinfo {title} {Deterministic photonic quantum computation in a
  synthetic time dimension},\ }\href {https://doi.org/10.1364/OPTICA.424258}
  {\bibfield  {journal} {\bibinfo  {journal} {Optica}\ }\textbf {\bibinfo
  {volume} {8}},\ \bibinfo {pages} {1515} (\bibinfo {year} {2021})}\BibitemShut
  {NoStop}%
\bibitem [{\citenamefont {Humphreys}\ \emph {et~al.}(2013)\citenamefont
  {Humphreys}, \citenamefont {Metcalf}, \citenamefont {Spring}, \citenamefont
  {Moore}, \citenamefont {Jin}, \citenamefont {Barbieri}, \citenamefont
  {Kolthammer},\ and\ \citenamefont {Walmsley}}]{TimeBinCPHASE}%
  \BibitemOpen
  \bibfield  {author} {\bibinfo {author} {\bibfnamefont {P.~C.}\ \bibnamefont
  {Humphreys}}, \bibinfo {author} {\bibfnamefont {B.~J.}\ \bibnamefont
  {Metcalf}}, \bibinfo {author} {\bibfnamefont {J.~B.}\ \bibnamefont {Spring}},
  \bibinfo {author} {\bibfnamefont {M.}~\bibnamefont {Moore}}, \bibinfo
  {author} {\bibfnamefont {X.-M.}\ \bibnamefont {Jin}}, \bibinfo {author}
  {\bibfnamefont {M.}~\bibnamefont {Barbieri}}, \bibinfo {author}
  {\bibfnamefont {W.~S.}\ \bibnamefont {Kolthammer}},\ and\ \bibinfo {author}
  {\bibfnamefont {I.~A.}\ \bibnamefont {Walmsley}},\ }\bibfield  {title}
  {\bibinfo {title} {Linear optical quantum computing in a single spatial
  mode},\ }\href {https://doi.org/10.1103/PhysRevLett.111.150501} {\bibfield
  {journal} {\bibinfo  {journal} {Phys. Rev. Lett.}\ }\textbf {\bibinfo
  {volume} {111}},\ \bibinfo {pages} {150501} (\bibinfo {year}
  {2013})}\BibitemShut {NoStop}%
\bibitem [{\citenamefont {Carosini}\ \emph {et~al.}(2024)\citenamefont
  {Carosini}, \citenamefont {Oddi}, \citenamefont {Giorgino}, \citenamefont
  {Hansen}, \citenamefont {Seron}, \citenamefont {Piacentini}, \citenamefont
  {Guggemos}, \citenamefont {Agresti}, \citenamefont {Loredo},\ and\
  \citenamefont {Walther}}]{TimeBinCarosini}%
  \BibitemOpen
  \bibfield  {author} {\bibinfo {author} {\bibfnamefont {L.}~\bibnamefont
  {Carosini}}, \bibinfo {author} {\bibfnamefont {V.}~\bibnamefont {Oddi}},
  \bibinfo {author} {\bibfnamefont {F.}~\bibnamefont {Giorgino}}, \bibinfo
  {author} {\bibfnamefont {L.~M.}\ \bibnamefont {Hansen}}, \bibinfo {author}
  {\bibfnamefont {B.}~\bibnamefont {Seron}}, \bibinfo {author} {\bibfnamefont
  {S.}~\bibnamefont {Piacentini}}, \bibinfo {author} {\bibfnamefont
  {T.}~\bibnamefont {Guggemos}}, \bibinfo {author} {\bibfnamefont
  {I.}~\bibnamefont {Agresti}}, \bibinfo {author} {\bibfnamefont {J.~C.}\
  \bibnamefont {Loredo}},\ and\ \bibinfo {author} {\bibfnamefont
  {P.}~\bibnamefont {Walther}},\ }\bibfield  {title} {\bibinfo {title}
  {Programmable multiphoton quantum interference in a single spatial mode},\
  }\href {https://doi.org/10.1126/sciadv.adj0993} {\bibfield  {journal}
  {\bibinfo  {journal} {Sci. Adv.}\ }\textbf {\bibinfo {volume} {10}},\
  \bibinfo {pages} {eadj0993} (\bibinfo {year} {2024})}\BibitemShut {NoStop}%
\bibitem [{\citenamefont {Flamini}\ \emph {et~al.}(2018)\citenamefont
  {Flamini}, \citenamefont {Spagnolo},\ and\ \citenamefont
  {Sciarrino}}]{flamini18}%
  \BibitemOpen
  \bibfield  {author} {\bibinfo {author} {\bibfnamefont {F.}~\bibnamefont
  {Flamini}}, \bibinfo {author} {\bibfnamefont {N.}~\bibnamefont {Spagnolo}},\
  and\ \bibinfo {author} {\bibfnamefont {F.}~\bibnamefont {Sciarrino}},\
  }\bibfield  {title} {\bibinfo {title} {Photonic quantum information
  processing: a review},\ }\href {https://doi.org/10.1088/1361-6633/aad5b2}
  {\bibfield  {journal} {\bibinfo  {journal} {Rep. Prog. Phys.}\ }\textbf
  {\bibinfo {volume} {82}},\ \bibinfo {pages} {016001} (\bibinfo {year}
  {2018})}\BibitemShut {NoStop}%
\bibitem [{\citenamefont {Fukui}\ and\ \citenamefont
  {Takeda}(2022)}]{QuantuCompCV}%
  \BibitemOpen
  \bibfield  {author} {\bibinfo {author} {\bibfnamefont {K.}~\bibnamefont
  {Fukui}}\ and\ \bibinfo {author} {\bibfnamefont {S.}~\bibnamefont {Takeda}},\
  }\bibfield  {title} {\bibinfo {title} {Building a large-scale quantum
  computer with continuous-variable optical technologies},\ }\href
  {https://doi.org/10.1088/1361-6455/ac489c} {\bibfield  {journal} {\bibinfo
  {journal} {J. Phys. B: At. Mol. Opt. Phys.}\ }\textbf {\bibinfo {volume}
  {55}},\ \bibinfo {pages} {012001} (\bibinfo {year} {2022})}\BibitemShut
  {NoStop}%
\bibitem [{\citenamefont {Gottesman}\ \emph {et~al.}(2001)\citenamefont
  {Gottesman}, \citenamefont {Kitaev},\ and\ \citenamefont
  {Preskill}}]{GKP_Orig}%
  \BibitemOpen
  \bibfield  {author} {\bibinfo {author} {\bibfnamefont {D.}~\bibnamefont
  {Gottesman}}, \bibinfo {author} {\bibfnamefont {A.}~\bibnamefont {Kitaev}},\
  and\ \bibinfo {author} {\bibfnamefont {J.}~\bibnamefont {Preskill}},\
  }\bibfield  {title} {\bibinfo {title} {Encoding a qubit in an oscillator},\
  }\href {https://doi.org/10.1103/PhysRevA.64.012310} {\bibfield  {journal}
  {\bibinfo  {journal} {Phys. Rev. A}\ }\textbf {\bibinfo {volume} {64}},\
  \bibinfo {pages} {012310} (\bibinfo {year} {2001})}\BibitemShut {NoStop}%
\bibitem [{\citenamefont {Liu}\ and\ \citenamefont
  {Wei}(2023)}]{LinearOpticalUniversalQuantumGate}%
  \BibitemOpen
  \bibfield  {author} {\bibinfo {author} {\bibfnamefont {W.-Q.}\ \bibnamefont
  {Liu}}\ and\ \bibinfo {author} {\bibfnamefont {H.-R.}\ \bibnamefont {Wei}},\
  }\bibfield  {title} {\bibinfo {title} {Linear optical universal quantum gates
  with higher success probabilities},\ }\href
  {https://doi.org/https://doi.org/10.1002/qute.202300009} {\bibfield
  {journal} {\bibinfo  {journal} {Adv. Quantum Technol.}\ }\textbf {\bibinfo
  {volume} {6}},\ \bibinfo {pages} {2300009} (\bibinfo {year}
  {2023})}\BibitemShut {NoStop}%
\bibitem [{\citenamefont {Bartolucci}\ \emph {et~al.}(2023)\citenamefont
  {Bartolucci}, \citenamefont {Birchall}, \citenamefont {Bomb{\'{\i}}n},
  \citenamefont {Cable}, \citenamefont {Dawson}, \citenamefont
  {Gimeno-Segovia}, \citenamefont {Johnston}, \citenamefont {Kieling},
  \citenamefont {Nickerson}, \citenamefont {Pant}, \citenamefont {Pastawski},
  \citenamefont {Rudolph},\ and\ \citenamefont {Sparrow}}]{PsiquantumFusion}%
  \BibitemOpen
  \bibfield  {author} {\bibinfo {author} {\bibfnamefont {S.}~\bibnamefont
  {Bartolucci}}, \bibinfo {author} {\bibfnamefont {P.}~\bibnamefont
  {Birchall}}, \bibinfo {author} {\bibfnamefont {H.}~\bibnamefont
  {Bomb{\'{\i}}n}}, \bibinfo {author} {\bibfnamefont {H.}~\bibnamefont
  {Cable}}, \bibinfo {author} {\bibfnamefont {C.}~\bibnamefont {Dawson}},
  \bibinfo {author} {\bibfnamefont {M.}~\bibnamefont {Gimeno-Segovia}},
  \bibinfo {author} {\bibfnamefont {E.}~\bibnamefont {Johnston}}, \bibinfo
  {author} {\bibfnamefont {K.}~\bibnamefont {Kieling}}, \bibinfo {author}
  {\bibfnamefont {N.}~\bibnamefont {Nickerson}}, \bibinfo {author}
  {\bibfnamefont {M.}~\bibnamefont {Pant}}, \bibinfo {author} {\bibfnamefont
  {F.}~\bibnamefont {Pastawski}}, \bibinfo {author} {\bibfnamefont
  {T.}~\bibnamefont {Rudolph}},\ and\ \bibinfo {author} {\bibfnamefont
  {C.}~\bibnamefont {Sparrow}},\ }\bibfield  {title} {\bibinfo {title}
  {Fusion-based quantum computation},\ }\href
  {https://doi.org/10.1038/s41467-023-36493-1} {\bibfield  {journal} {\bibinfo
  {journal} {Nat. Commun.}\ }\textbf {\bibinfo {volume} {14}},\ \bibinfo
  {pages} {912} (\bibinfo {year} {2023})}\BibitemShut {NoStop}%
\bibitem [{\citenamefont {Zhan}\ \emph {et~al.}(2024)\citenamefont {Zhan},
  \citenamefont {Zhang}, \citenamefont {Erbanni}, \citenamefont {Burger},
  \citenamefont {Wan}, \citenamefont {Jiang}, \citenamefont {Chae},
  \citenamefont {Liu}, \citenamefont {Poletti},\ and\ \citenamefont
  {Kwek}}]{LoopLinearMultiTime}%
  \BibitemOpen
  \bibfield  {author} {\bibinfo {author} {\bibfnamefont {Y.}~\bibnamefont
  {Zhan}}, \bibinfo {author} {\bibfnamefont {H.}~\bibnamefont {Zhang}},
  \bibinfo {author} {\bibfnamefont {R.}~\bibnamefont {Erbanni}}, \bibinfo
  {author} {\bibfnamefont {A.}~\bibnamefont {Burger}}, \bibinfo {author}
  {\bibfnamefont {L.}~\bibnamefont {Wan}}, \bibinfo {author} {\bibfnamefont
  {X.}~\bibnamefont {Jiang}}, \bibinfo {author} {\bibfnamefont
  {S.}~\bibnamefont {Chae}}, \bibinfo {author} {\bibfnamefont {A.}~\bibnamefont
  {Liu}}, \bibinfo {author} {\bibfnamefont {D.}~\bibnamefont {Poletti}},\ and\
  \bibinfo {author} {\bibfnamefont {L.~C.}\ \bibnamefont {Kwek}},\ }\href@noop
  {} {\bibinfo {title} {Loop quantum photonic chip for coherent multi-time-step
  evolution}} (\bibinfo {year} {2024}),\ \Eprint
  {https://arxiv.org/abs/2411.11307} {arXiv:2411.11307 [quant-ph]} \BibitemShut
  {NoStop}%
\bibitem [{\citenamefont {Paesani}\ \emph {et~al.}(2019)\citenamefont
  {Paesani}, \citenamefont {Ding}, \citenamefont {Santagati}, \citenamefont
  {Chakhmakhchyan}, \citenamefont {Vigliar}, \citenamefont {Rottwitt},
  \citenamefont {Oxenløwe}, \citenamefont {Wang}, \citenamefont {Thompson},\
  and\ \citenamefont {Laing}}]{BosonSamplingPaesani}%
  \BibitemOpen
  \bibfield  {author} {\bibinfo {author} {\bibfnamefont {S.}~\bibnamefont
  {Paesani}}, \bibinfo {author} {\bibfnamefont {Y.}~\bibnamefont {Ding}},
  \bibinfo {author} {\bibfnamefont {R.}~\bibnamefont {Santagati}}, \bibinfo
  {author} {\bibfnamefont {L.}~\bibnamefont {Chakhmakhchyan}}, \bibinfo
  {author} {\bibfnamefont {C.}~\bibnamefont {Vigliar}}, \bibinfo {author}
  {\bibfnamefont {K.}~\bibnamefont {Rottwitt}}, \bibinfo {author}
  {\bibfnamefont {L.~K.}\ \bibnamefont {Oxenløwe}}, \bibinfo {author}
  {\bibfnamefont {J.}~\bibnamefont {Wang}}, \bibinfo {author} {\bibfnamefont
  {M.~G.}\ \bibnamefont {Thompson}},\ and\ \bibinfo {author} {\bibfnamefont
  {A.}~\bibnamefont {Laing}},\ }\bibfield  {title} {\bibinfo {title}
  {Generation and sampling of quantum states of light in a silicon chip},\
  }\href {https://doi.org/10.1038/s41567-019-0567-8} {\bibfield  {journal}
  {\bibinfo  {journal} {Nat. Phys.}\ }\textbf {\bibinfo {volume} {15}},\
  \bibinfo {pages} {925–929} (\bibinfo {year} {2019})}\BibitemShut {NoStop}%
\bibitem [{\citenamefont {Reck}\ \emph {et~al.}(1994)\citenamefont {Reck},
  \citenamefont {Zeilinger}, \citenamefont {Bernstein},\ and\ \citenamefont
  {Bertani}}]{ReckDecomposition}%
  \BibitemOpen
  \bibfield  {author} {\bibinfo {author} {\bibfnamefont {M.}~\bibnamefont
  {Reck}}, \bibinfo {author} {\bibfnamefont {A.}~\bibnamefont {Zeilinger}},
  \bibinfo {author} {\bibfnamefont {H.~J.}\ \bibnamefont {Bernstein}},\ and\
  \bibinfo {author} {\bibfnamefont {P.}~\bibnamefont {Bertani}},\ }\bibfield
  {title} {\bibinfo {title} {Experimental realization of any discrete unitary
  operator},\ }\href {https://doi.org/10.1103/PhysRevLett.73.58} {\bibfield
  {journal} {\bibinfo  {journal} {Phys. Rev. Lett.}\ }\textbf {\bibinfo
  {volume} {73}},\ \bibinfo {pages} {58} (\bibinfo {year} {1994})}\BibitemShut
  {NoStop}%
\bibitem [{\citenamefont {Clements}\ \emph {et~al.}(2016)\citenamefont
  {Clements}, \citenamefont {Humphreys}, \citenamefont {Metcalf}, \citenamefont
  {Kolthammer},\ and\ \citenamefont {Walmsley}}]{ClementsDecomposition}%
  \BibitemOpen
  \bibfield  {author} {\bibinfo {author} {\bibfnamefont {W.~R.}\ \bibnamefont
  {Clements}}, \bibinfo {author} {\bibfnamefont {P.~C.}\ \bibnamefont
  {Humphreys}}, \bibinfo {author} {\bibfnamefont {B.~J.}\ \bibnamefont
  {Metcalf}}, \bibinfo {author} {\bibfnamefont {W.~S.}\ \bibnamefont
  {Kolthammer}},\ and\ \bibinfo {author} {\bibfnamefont {I.~A.}\ \bibnamefont
  {Walmsley}},\ }\bibfield  {title} {\bibinfo {title} {Optimal design for
  universal multiport interferometers},\ }\href
  {https://doi.org/10.1364/OPTICA.3.001460} {\bibfield  {journal} {\bibinfo
  {journal} {Optica}\ }\textbf {\bibinfo {volume} {3}},\ \bibinfo {pages}
  {1460} (\bibinfo {year} {2016})}\BibitemShut {NoStop}%
\bibitem [{\citenamefont {Giordani}\ \emph {et~al.}(2023)\citenamefont
  {Giordani}, \citenamefont {Hoch}, \citenamefont {Carvacho}, \citenamefont
  {Spagnolo},\ and\ \citenamefont {Sciarrino}}]{IntegratedReview}%
  \BibitemOpen
  \bibfield  {author} {\bibinfo {author} {\bibfnamefont {T.}~\bibnamefont
  {Giordani}}, \bibinfo {author} {\bibfnamefont {F.}~\bibnamefont {Hoch}},
  \bibinfo {author} {\bibfnamefont {G.}~\bibnamefont {Carvacho}}, \bibinfo
  {author} {\bibfnamefont {N.}~\bibnamefont {Spagnolo}},\ and\ \bibinfo
  {author} {\bibfnamefont {F.}~\bibnamefont {Sciarrino}},\ }\bibfield  {title}
  {\bibinfo {title} {Integrated photonics in quantum technologies},\ }\href
  {https://doi.org/10.1007/s40766-023-00040-x} {\bibfield  {journal} {\bibinfo
  {journal} {Riv. Nuovo Cim.}\ }\textbf {\bibinfo {volume} {46}},\ \bibinfo
  {pages} {71} (\bibinfo {year} {2023})}\BibitemShut {NoStop}%
\bibitem [{\citenamefont {Peruzzo}\ \emph {et~al.}(2010)\citenamefont
  {Peruzzo}, \citenamefont {Lobino}, \citenamefont {Matthews}, \citenamefont
  {Matsuda}, \citenamefont {Politi}, \citenamefont {Poulios}, \citenamefont
  {Zhou}, \citenamefont {Lahini}, \citenamefont {Ismail}, \citenamefont
  {W\"orhoff}, \citenamefont {Bromberg}, \citenamefont {Silberberg},
  \citenamefont {Thompson},\ and\ \citenamefont {O'Brien}}]{peruzzo2010}%
  \BibitemOpen
  \bibfield  {author} {\bibinfo {author} {\bibfnamefont {A.}~\bibnamefont
  {Peruzzo}}, \bibinfo {author} {\bibfnamefont {M.}~\bibnamefont {Lobino}},
  \bibinfo {author} {\bibfnamefont {J.~C.~F.}\ \bibnamefont {Matthews}},
  \bibinfo {author} {\bibfnamefont {N.}~\bibnamefont {Matsuda}}, \bibinfo
  {author} {\bibfnamefont {A.}~\bibnamefont {Politi}}, \bibinfo {author}
  {\bibfnamefont {K.}~\bibnamefont {Poulios}}, \bibinfo {author} {\bibfnamefont
  {X.-Q.}\ \bibnamefont {Zhou}}, \bibinfo {author} {\bibfnamefont
  {Y.}~\bibnamefont {Lahini}}, \bibinfo {author} {\bibfnamefont
  {N.}~\bibnamefont {Ismail}}, \bibinfo {author} {\bibfnamefont
  {K.}~\bibnamefont {W\"orhoff}}, \bibinfo {author} {\bibfnamefont
  {Y.}~\bibnamefont {Bromberg}}, \bibinfo {author} {\bibfnamefont
  {Y.}~\bibnamefont {Silberberg}}, \bibinfo {author} {\bibfnamefont {M.~G.}\
  \bibnamefont {Thompson}},\ and\ \bibinfo {author} {\bibfnamefont {J.~L.}\
  \bibnamefont {O'Brien}},\ }\bibfield  {title} {\bibinfo {title} {Quantum
  walks of correlated photons},\ }\href {https://doi.org/10.1126/science.1193}
  {\bibfield  {journal} {\bibinfo  {journal} {Science}\ }\textbf {\bibinfo
  {volume} {329}},\ \bibinfo {pages} {1500} (\bibinfo {year}
  {2010})}\BibitemShut {NoStop}%
\bibitem [{\citenamefont {Hoch}\ \emph {et~al.}(2022)\citenamefont {Hoch},
  \citenamefont {Piacentini}, \citenamefont {Giordani}, \citenamefont {Tian},
  \citenamefont {Iuliano}, \citenamefont {Esposito}, \citenamefont {Camillini},
  \citenamefont {Carvacho}, \citenamefont {Ceccarelli}, \citenamefont
  {Spagnolo}, \citenamefont {Crespi}, \citenamefont {Sciarrino},\ and\
  \citenamefont {Osellame}}]{3DChipLab}%
  \BibitemOpen
  \bibfield  {author} {\bibinfo {author} {\bibfnamefont {F.}~\bibnamefont
  {Hoch}}, \bibinfo {author} {\bibfnamefont {S.}~\bibnamefont {Piacentini}},
  \bibinfo {author} {\bibfnamefont {T.}~\bibnamefont {Giordani}}, \bibinfo
  {author} {\bibfnamefont {Z.-N.}\ \bibnamefont {Tian}}, \bibinfo {author}
  {\bibfnamefont {M.}~\bibnamefont {Iuliano}}, \bibinfo {author} {\bibfnamefont
  {C.}~\bibnamefont {Esposito}}, \bibinfo {author} {\bibfnamefont
  {A.}~\bibnamefont {Camillini}}, \bibinfo {author} {\bibfnamefont
  {G.}~\bibnamefont {Carvacho}}, \bibinfo {author} {\bibfnamefont
  {F.}~\bibnamefont {Ceccarelli}}, \bibinfo {author} {\bibfnamefont
  {N.}~\bibnamefont {Spagnolo}}, \bibinfo {author} {\bibfnamefont
  {A.}~\bibnamefont {Crespi}}, \bibinfo {author} {\bibfnamefont
  {F.}~\bibnamefont {Sciarrino}},\ and\ \bibinfo {author} {\bibfnamefont
  {R.}~\bibnamefont {Osellame}},\ }\bibfield  {title} {\bibinfo {title}
  {Reconfigurable continuously-coupled 3d photonic circuit for boson sampling
  experiments},\ }\href {https://doi.org/10.1038/s41534-022-00568-6} {\bibfield
   {journal} {\bibinfo  {journal} {npj Quantum Inf.}\ }\textbf {\bibinfo
  {volume} {8}},\ \bibinfo {pages} {35} (\bibinfo {year} {2022})}\BibitemShut
  {NoStop}%
\bibitem [{\citenamefont {Yang}\ \emph {et~al.}(2024)\citenamefont {Yang},
  \citenamefont {Chapman}, \citenamefont {Haylock}, \citenamefont {Lenzini},
  \citenamefont {Joglekar}, \citenamefont {Lobino},\ and\ \citenamefont
  {Peruzzo}}]{CCElectro1}%
  \BibitemOpen
  \bibfield  {author} {\bibinfo {author} {\bibfnamefont {Y.}~\bibnamefont
  {Yang}}, \bibinfo {author} {\bibfnamefont {R.~J.}\ \bibnamefont {Chapman}},
  \bibinfo {author} {\bibfnamefont {B.}~\bibnamefont {Haylock}}, \bibinfo
  {author} {\bibfnamefont {F.}~\bibnamefont {Lenzini}}, \bibinfo {author}
  {\bibfnamefont {Y.~N.}\ \bibnamefont {Joglekar}}, \bibinfo {author}
  {\bibfnamefont {M.}~\bibnamefont {Lobino}},\ and\ \bibinfo {author}
  {\bibfnamefont {A.}~\bibnamefont {Peruzzo}},\ }\bibfield  {title} {\bibinfo
  {title} {Programmable high-dimensional hamiltonian in a photonic waveguide
  array},\ }\href {https://doi.org/10.1038/s41467-023-44185-z} {\bibfield
  {journal} {\bibinfo  {journal} {Nat. Commun.}\ }\textbf {\bibinfo {volume}
  {15}},\ \bibinfo {pages} {50} (\bibinfo {year} {2024})}\BibitemShut {NoStop}%
\bibitem [{\citenamefont {Yang}\ \emph {et~al.}(2025)\citenamefont {Yang},
  \citenamefont {Chapman}, \citenamefont {Youssry}, \citenamefont {Haylock},
  \citenamefont {Lenzini}, \citenamefont {Lobino},\ and\ \citenamefont
  {Peruzzo}}]{CCElectro2}%
  \BibitemOpen
  \bibfield  {author} {\bibinfo {author} {\bibfnamefont {Y.}~\bibnamefont
  {Yang}}, \bibinfo {author} {\bibfnamefont {R.~J.}\ \bibnamefont {Chapman}},
  \bibinfo {author} {\bibfnamefont {A.}~\bibnamefont {Youssry}}, \bibinfo
  {author} {\bibfnamefont {B.}~\bibnamefont {Haylock}}, \bibinfo {author}
  {\bibfnamefont {F.}~\bibnamefont {Lenzini}}, \bibinfo {author} {\bibfnamefont
  {M.}~\bibnamefont {Lobino}},\ and\ \bibinfo {author} {\bibfnamefont
  {A.}~\bibnamefont {Peruzzo}},\ }\bibfield  {title} {\bibinfo {title}
  {Programmable quantum circuits in a large-scale photonic waveguide array},\
  }\href {https://doi.org/10.1038/s41534-024-00934-6} {\bibfield  {journal}
  {\bibinfo  {journal} {npj Quantum Inf.}\ }\textbf {\bibinfo {volume} {11}},\
  \bibinfo {pages} {19} (\bibinfo {year} {2025})}\BibitemShut {NoStop}%
\bibitem [{\citenamefont {Zhou}\ \emph {et~al.}(2024)\citenamefont {Zhou},
  \citenamefont {Wang}, \citenamefont {Ren}, \citenamefont {Fu}, \citenamefont
  {Chang}, \citenamefont {Xu}, \citenamefont {Tang},\ and\ \citenamefont
  {Jin}}]{CC3D1}%
  \BibitemOpen
  \bibfield  {author} {\bibinfo {author} {\bibfnamefont {W.-H.}\ \bibnamefont
  {Zhou}}, \bibinfo {author} {\bibfnamefont {X.-W.}\ \bibnamefont {Wang}},
  \bibinfo {author} {\bibfnamefont {R.-J.}\ \bibnamefont {Ren}}, \bibinfo
  {author} {\bibfnamefont {Y.-X.}\ \bibnamefont {Fu}}, \bibinfo {author}
  {\bibfnamefont {Y.-J.}\ \bibnamefont {Chang}}, \bibinfo {author}
  {\bibfnamefont {X.-Y.}\ \bibnamefont {Xu}}, \bibinfo {author} {\bibfnamefont
  {H.}~\bibnamefont {Tang}},\ and\ \bibinfo {author} {\bibfnamefont {X.-M.}\
  \bibnamefont {Jin}},\ }\bibfield  {title} {\bibinfo {title} {Multi-particle
  quantum walks on 3d integrated photonic chip},\ }\href
  {https://doi.org/10.1038/s41377-024-01627-7} {\bibfield  {journal} {\bibinfo
  {journal} {Light: Sci. Appl.}\ }\textbf {\bibinfo {volume} {13}},\ \bibinfo
  {pages} {296} (\bibinfo {year} {2024})}\BibitemShut {NoStop}%
\bibitem [{\citenamefont {Jiao}\ \emph {et~al.}(2021)\citenamefont {Jiao},
  \citenamefont {Gao}, \citenamefont {Zhou}, \citenamefont {Wang},
  \citenamefont {Ren}, \citenamefont {Xu}, \citenamefont {Qiao}, \citenamefont
  {Wang},\ and\ \citenamefont {Jin}}]{CC3D2}%
  \BibitemOpen
  \bibfield  {author} {\bibinfo {author} {\bibfnamefont {Z.-Q.}\ \bibnamefont
  {Jiao}}, \bibinfo {author} {\bibfnamefont {J.}~\bibnamefont {Gao}}, \bibinfo
  {author} {\bibfnamefont {W.-H.}\ \bibnamefont {Zhou}}, \bibinfo {author}
  {\bibfnamefont {X.-W.}\ \bibnamefont {Wang}}, \bibinfo {author}
  {\bibfnamefont {R.-J.}\ \bibnamefont {Ren}}, \bibinfo {author} {\bibfnamefont
  {X.-Y.}\ \bibnamefont {Xu}}, \bibinfo {author} {\bibfnamefont {L.-F.}\
  \bibnamefont {Qiao}}, \bibinfo {author} {\bibfnamefont {Y.}~\bibnamefont
  {Wang}},\ and\ \bibinfo {author} {\bibfnamefont {X.-M.}\ \bibnamefont
  {Jin}},\ }\bibfield  {title} {\bibinfo {title} {Two-dimensional quantum walks
  of correlated photons},\ }\href {https://doi.org/10.1364/OPTICA.425879}
  {\bibfield  {journal} {\bibinfo  {journal} {Optica}\ }\textbf {\bibinfo
  {volume} {8}},\ \bibinfo {pages} {1129} (\bibinfo {year} {2021})}\BibitemShut
  {NoStop}%
\bibitem [{\citenamefont {Poulios}\ \emph {et~al.}(2014)\citenamefont
  {Poulios}, \citenamefont {Keil}, \citenamefont {Fry}, \citenamefont
  {Meinecke}, \citenamefont {Matthews}, \citenamefont {Politi}, \citenamefont
  {Lobino}, \citenamefont {Gr\"{a}fe}, \citenamefont {Heinrich}, \citenamefont
  {Nolte}, \citenamefont {Szameit},\ and\ \citenamefont {O'Brien}}]{CC3D3}%
  \BibitemOpen
  \bibfield  {author} {\bibinfo {author} {\bibfnamefont {K.}~\bibnamefont
  {Poulios}}, \bibinfo {author} {\bibfnamefont {R.}~\bibnamefont {Keil}},
  \bibinfo {author} {\bibfnamefont {D.}~\bibnamefont {Fry}}, \bibinfo {author}
  {\bibfnamefont {J.~D.~A.}\ \bibnamefont {Meinecke}}, \bibinfo {author}
  {\bibfnamefont {J.~C.~F.}\ \bibnamefont {Matthews}}, \bibinfo {author}
  {\bibfnamefont {A.}~\bibnamefont {Politi}}, \bibinfo {author} {\bibfnamefont
  {M.}~\bibnamefont {Lobino}}, \bibinfo {author} {\bibfnamefont
  {M.}~\bibnamefont {Gr\"{a}fe}}, \bibinfo {author} {\bibfnamefont
  {M.}~\bibnamefont {Heinrich}}, \bibinfo {author} {\bibfnamefont
  {S.}~\bibnamefont {Nolte}}, \bibinfo {author} {\bibfnamefont
  {A.}~\bibnamefont {Szameit}},\ and\ \bibinfo {author} {\bibfnamefont {J.~L.}\
  \bibnamefont {O'Brien}},\ }\bibfield  {title} {\bibinfo {title} {Quantum
  walks of correlated photon pairs in two-dimensional waveguide arrays},\
  }\href {https://doi.org/10.1103/PhysRevLett.112.143604} {\bibfield  {journal}
  {\bibinfo  {journal} {Phys. Rev. Lett.}\ }\textbf {\bibinfo {volume} {112}},\
  \bibinfo {pages} {143604} (\bibinfo {year} {2014})}\BibitemShut {NoStop}%
\bibitem [{\citenamefont {Biggerstaff}\ \emph {et~al.}(2016)\citenamefont
  {Biggerstaff}, \citenamefont {Heilmann}, \citenamefont {Zecevik},
  \citenamefont {Gr\"{a}fe}, \citenamefont {Broome}, \citenamefont {Fedrizzi},
  \citenamefont {Nolte}, \citenamefont {Szameit}, \citenamefont {White},\ and\
  \citenamefont {Kassal}}]{CC3D4}%
  \BibitemOpen
  \bibfield  {author} {\bibinfo {author} {\bibfnamefont {D.~N.}\ \bibnamefont
  {Biggerstaff}}, \bibinfo {author} {\bibfnamefont {R.}~\bibnamefont
  {Heilmann}}, \bibinfo {author} {\bibfnamefont {A.~A.}\ \bibnamefont
  {Zecevik}}, \bibinfo {author} {\bibfnamefont {M.}~\bibnamefont {Gr\"{a}fe}},
  \bibinfo {author} {\bibfnamefont {M.~A.}\ \bibnamefont {Broome}}, \bibinfo
  {author} {\bibfnamefont {A.}~\bibnamefont {Fedrizzi}}, \bibinfo {author}
  {\bibfnamefont {S.}~\bibnamefont {Nolte}}, \bibinfo {author} {\bibfnamefont
  {A.}~\bibnamefont {Szameit}}, \bibinfo {author} {\bibfnamefont {A.~G.}\
  \bibnamefont {White}},\ and\ \bibinfo {author} {\bibfnamefont
  {I.}~\bibnamefont {Kassal}},\ }\bibfield  {title} {\bibinfo {title}
  {Enhancing coherent transport in a photonic network using controllable
  decoherence},\ }\href {https://doi.org/10.1038/ncomms11282} {\bibfield
  {journal} {\bibinfo  {journal} {Nat. Commun.}\ }\textbf {\bibinfo {volume}
  {7}},\ \bibinfo {pages} {11282} (\bibinfo {year} {2016})}\BibitemShut
  {NoStop}%
\bibitem [{\citenamefont {Caruso}\ \emph {et~al.}(2016)\citenamefont {Caruso},
  \citenamefont {Crespi}, \citenamefont {Ciriolo}, \citenamefont {Sciarrino},\
  and\ \citenamefont {Osellame}}]{CC3D5}%
  \BibitemOpen
  \bibfield  {author} {\bibinfo {author} {\bibfnamefont {F.}~\bibnamefont
  {Caruso}}, \bibinfo {author} {\bibfnamefont {A.}~\bibnamefont {Crespi}},
  \bibinfo {author} {\bibfnamefont {A.~G.}\ \bibnamefont {Ciriolo}}, \bibinfo
  {author} {\bibfnamefont {F.}~\bibnamefont {Sciarrino}},\ and\ \bibinfo
  {author} {\bibfnamefont {R.}~\bibnamefont {Osellame}},\ }\bibfield  {title}
  {\bibinfo {title} {Fast escape of a quantum walker from an integrated
  photonic maze},\ }\href {https://doi.org/10.1038/ncomms11682} {\bibfield
  {journal} {\bibinfo  {journal} {Nat. Commun.}\ }\textbf {\bibinfo {volume}
  {7}},\ \bibinfo {pages} {11682} (\bibinfo {year} {2016})}\BibitemShut
  {NoStop}%
\bibitem [{\citenamefont {Albertsson}\ \emph {et~al.}(2018)\citenamefont
  {Albertsson}, \citenamefont {Altoe}, \citenamefont {Anderson}, \citenamefont
  {Andrews}, \citenamefont {Araque~Espinosa}, \citenamefont {Aurisano},
  \citenamefont {Basara}, \citenamefont {Bevan}, \citenamefont {Bhimji},
  \citenamefont {Bonacorsi}, \citenamefont {Calafiura}, \citenamefont
  {Campanelli}, \citenamefont {Capps}, \citenamefont {Carminati}, \citenamefont
  {Carrazza}, \citenamefont {Childers}, \citenamefont {Coniavitis},
  \citenamefont {Cranmer}, \citenamefont {David}, \citenamefont {Davis},
  \citenamefont {Duarte}, \citenamefont {Erdmann}, \citenamefont {Eschle},
  \citenamefont {Farbin}, \citenamefont {Feickert}, \citenamefont {Castro},
  \citenamefont {Fitzpatrick}, \citenamefont {Floris}, \citenamefont {Forti},
  \citenamefont {Garra-Tico}, \citenamefont {Gemmler}, \citenamefont {Girone},
  \citenamefont {Glaysher}, \citenamefont {Gleyzer}, \citenamefont {Gligorov},
  \citenamefont {Golling}, \citenamefont {Graw}, \citenamefont {Gray},
  \citenamefont {Greenwood}, \citenamefont {Hacker}, \citenamefont {Harvey},
  \citenamefont {Hegner}, \citenamefont {Heinrich}, \citenamefont {Hooberman},
  \citenamefont {Junggeburth}, \citenamefont {Kagan}, \citenamefont {Kane},
  \citenamefont {Kanishchev}, \citenamefont {Karpiński}, \citenamefont
  {Kassabov}, \citenamefont {Kaul}, \citenamefont {Kcira}, \citenamefont
  {Keck}, \citenamefont {Klimentov}, \citenamefont {Kowalkowski}, \citenamefont
  {Kreczko}, \citenamefont {Kurepin}, \citenamefont {Kutschke}, \citenamefont
  {Kuznetsov}, \citenamefont {K\"{o}hler}, \citenamefont {Lakomov},
  \citenamefont {Lannon}, \citenamefont {Lassnig}, \citenamefont {Limosani},
  \citenamefont {Louppe}, \citenamefont {Mangu}, \citenamefont {Mato},
  \citenamefont {Meinhard}, \citenamefont {Menasce}, \citenamefont {Moneta},
  \citenamefont {Moortgat}, \citenamefont {Narain}, \citenamefont {Neubauer},
  \citenamefont {Newman}, \citenamefont {Pabst}, \citenamefont {Paganini},
  \citenamefont {Paulini}, \citenamefont {Perdue}, \citenamefont {Perez},
  \citenamefont {Picazio}, \citenamefont {Pivarski}, \citenamefont {Prosper},
  \citenamefont {Psihas}, \citenamefont {Radovic}, \citenamefont {Reece},
  \citenamefont {Rinkevicius}, \citenamefont {Rodrigues}, \citenamefont
  {Rorie}, \citenamefont {Rousseau}, \citenamefont {Sauers}, \citenamefont
  {Schramm}, \citenamefont {Schwartzman}, \citenamefont {Severini},
  \citenamefont {Seyfert}, \citenamefont {Siroky}, \citenamefont {Skazytkin},
  \citenamefont {Sokoloff}, \citenamefont {Stewart}, \citenamefont {Stienen},
  \citenamefont {Stockdale}, \citenamefont {Strong}, \citenamefont {Thais},
  \citenamefont {Tomko}, \citenamefont {Upfal}, \citenamefont {Usai},
  \citenamefont {Ustyuzhanin}, \citenamefont {Vala}, \citenamefont
  {Vallecorsa}, \citenamefont {Vasel}, \citenamefont {Verzetti}, \citenamefont
  {Vilasís-Cardona}, \citenamefont {Vlimant}, \citenamefont {Vukotic},
  \citenamefont {Wang}, \citenamefont {Watts}, \citenamefont {Williams},
  \citenamefont {Wu}, \citenamefont {Wunsch},\ and\ \citenamefont
  {Zapata}}]{MLPhysics}%
  \BibitemOpen
  \bibfield  {author} {\bibinfo {author} {\bibfnamefont {K.}~\bibnamefont
  {Albertsson}}, \bibinfo {author} {\bibfnamefont {P.}~\bibnamefont {Altoe}},
  \bibinfo {author} {\bibfnamefont {D.}~\bibnamefont {Anderson}}, \bibinfo
  {author} {\bibfnamefont {M.}~\bibnamefont {Andrews}}, \bibinfo {author}
  {\bibfnamefont {J.~P.}\ \bibnamefont {Araque~Espinosa}}, \bibinfo {author}
  {\bibfnamefont {A.}~\bibnamefont {Aurisano}}, \bibinfo {author}
  {\bibfnamefont {L.}~\bibnamefont {Basara}}, \bibinfo {author} {\bibfnamefont
  {A.}~\bibnamefont {Bevan}}, \bibinfo {author} {\bibfnamefont
  {W.}~\bibnamefont {Bhimji}}, \bibinfo {author} {\bibfnamefont
  {D.}~\bibnamefont {Bonacorsi}}, \bibinfo {author} {\bibfnamefont
  {P.}~\bibnamefont {Calafiura}}, \bibinfo {author} {\bibfnamefont
  {M.}~\bibnamefont {Campanelli}}, \bibinfo {author} {\bibfnamefont
  {L.}~\bibnamefont {Capps}}, \bibinfo {author} {\bibfnamefont
  {F.}~\bibnamefont {Carminati}}, \bibinfo {author} {\bibfnamefont
  {S.}~\bibnamefont {Carrazza}}, \bibinfo {author} {\bibfnamefont
  {T.}~\bibnamefont {Childers}}, \bibinfo {author} {\bibfnamefont
  {E.}~\bibnamefont {Coniavitis}}, \bibinfo {author} {\bibfnamefont
  {K.}~\bibnamefont {Cranmer}}, \bibinfo {author} {\bibfnamefont
  {C.}~\bibnamefont {David}}, \bibinfo {author} {\bibfnamefont
  {D.}~\bibnamefont {Davis}}, \bibinfo {author} {\bibfnamefont
  {J.}~\bibnamefont {Duarte}}, \bibinfo {author} {\bibfnamefont
  {M.}~\bibnamefont {Erdmann}}, \bibinfo {author} {\bibfnamefont
  {J.}~\bibnamefont {Eschle}}, \bibinfo {author} {\bibfnamefont
  {A.}~\bibnamefont {Farbin}}, \bibinfo {author} {\bibfnamefont
  {M.}~\bibnamefont {Feickert}}, \bibinfo {author} {\bibfnamefont {N.~F.}\
  \bibnamefont {Castro}}, \bibinfo {author} {\bibfnamefont {C.}~\bibnamefont
  {Fitzpatrick}}, \bibinfo {author} {\bibfnamefont {M.}~\bibnamefont {Floris}},
  \bibinfo {author} {\bibfnamefont {A.}~\bibnamefont {Forti}}, \bibinfo
  {author} {\bibfnamefont {J.}~\bibnamefont {Garra-Tico}}, \bibinfo {author}
  {\bibfnamefont {J.}~\bibnamefont {Gemmler}}, \bibinfo {author} {\bibfnamefont
  {M.}~\bibnamefont {Girone}}, \bibinfo {author} {\bibfnamefont
  {P.}~\bibnamefont {Glaysher}}, \bibinfo {author} {\bibfnamefont
  {S.}~\bibnamefont {Gleyzer}}, \bibinfo {author} {\bibfnamefont
  {V.}~\bibnamefont {Gligorov}}, \bibinfo {author} {\bibfnamefont
  {T.}~\bibnamefont {Golling}}, \bibinfo {author} {\bibfnamefont
  {J.}~\bibnamefont {Graw}}, \bibinfo {author} {\bibfnamefont {L.}~\bibnamefont
  {Gray}}, \bibinfo {author} {\bibfnamefont {D.}~\bibnamefont {Greenwood}},
  \bibinfo {author} {\bibfnamefont {T.}~\bibnamefont {Hacker}}, \bibinfo
  {author} {\bibfnamefont {J.}~\bibnamefont {Harvey}}, \bibinfo {author}
  {\bibfnamefont {B.}~\bibnamefont {Hegner}}, \bibinfo {author} {\bibfnamefont
  {L.}~\bibnamefont {Heinrich}}, \bibinfo {author} {\bibfnamefont
  {B.}~\bibnamefont {Hooberman}}, \bibinfo {author} {\bibfnamefont
  {J.}~\bibnamefont {Junggeburth}}, \bibinfo {author} {\bibfnamefont
  {M.}~\bibnamefont {Kagan}}, \bibinfo {author} {\bibfnamefont
  {M.}~\bibnamefont {Kane}}, \bibinfo {author} {\bibfnamefont {K.}~\bibnamefont
  {Kanishchev}}, \bibinfo {author} {\bibfnamefont {P.}~\bibnamefont
  {Karpiński}}, \bibinfo {author} {\bibfnamefont {Z.}~\bibnamefont
  {Kassabov}}, \bibinfo {author} {\bibfnamefont {G.}~\bibnamefont {Kaul}},
  \bibinfo {author} {\bibfnamefont {D.}~\bibnamefont {Kcira}}, \bibinfo
  {author} {\bibfnamefont {T.}~\bibnamefont {Keck}}, \bibinfo {author}
  {\bibfnamefont {A.}~\bibnamefont {Klimentov}}, \bibinfo {author}
  {\bibfnamefont {J.}~\bibnamefont {Kowalkowski}}, \bibinfo {author}
  {\bibfnamefont {L.}~\bibnamefont {Kreczko}}, \bibinfo {author} {\bibfnamefont
  {A.}~\bibnamefont {Kurepin}}, \bibinfo {author} {\bibfnamefont
  {R.}~\bibnamefont {Kutschke}}, \bibinfo {author} {\bibfnamefont
  {V.}~\bibnamefont {Kuznetsov}}, \bibinfo {author} {\bibfnamefont
  {N.}~\bibnamefont {K\"{o}hler}}, \bibinfo {author} {\bibfnamefont
  {I.}~\bibnamefont {Lakomov}}, \bibinfo {author} {\bibfnamefont
  {K.}~\bibnamefont {Lannon}}, \bibinfo {author} {\bibfnamefont
  {M.}~\bibnamefont {Lassnig}}, \bibinfo {author} {\bibfnamefont
  {A.}~\bibnamefont {Limosani}}, \bibinfo {author} {\bibfnamefont
  {G.}~\bibnamefont {Louppe}}, \bibinfo {author} {\bibfnamefont
  {A.}~\bibnamefont {Mangu}}, \bibinfo {author} {\bibfnamefont
  {P.}~\bibnamefont {Mato}}, \bibinfo {author} {\bibfnamefont {H.}~\bibnamefont
  {Meinhard}}, \bibinfo {author} {\bibfnamefont {D.}~\bibnamefont {Menasce}},
  \bibinfo {author} {\bibfnamefont {L.}~\bibnamefont {Moneta}}, \bibinfo
  {author} {\bibfnamefont {S.}~\bibnamefont {Moortgat}}, \bibinfo {author}
  {\bibfnamefont {M.}~\bibnamefont {Narain}}, \bibinfo {author} {\bibfnamefont
  {M.}~\bibnamefont {Neubauer}}, \bibinfo {author} {\bibfnamefont
  {H.}~\bibnamefont {Newman}}, \bibinfo {author} {\bibfnamefont
  {H.}~\bibnamefont {Pabst}}, \bibinfo {author} {\bibfnamefont
  {M.}~\bibnamefont {Paganini}}, \bibinfo {author} {\bibfnamefont
  {M.}~\bibnamefont {Paulini}}, \bibinfo {author} {\bibfnamefont
  {G.}~\bibnamefont {Perdue}}, \bibinfo {author} {\bibfnamefont
  {U.}~\bibnamefont {Perez}}, \bibinfo {author} {\bibfnamefont
  {A.}~\bibnamefont {Picazio}}, \bibinfo {author} {\bibfnamefont
  {J.}~\bibnamefont {Pivarski}}, \bibinfo {author} {\bibfnamefont
  {H.}~\bibnamefont {Prosper}}, \bibinfo {author} {\bibfnamefont
  {F.}~\bibnamefont {Psihas}}, \bibinfo {author} {\bibfnamefont
  {A.}~\bibnamefont {Radovic}}, \bibinfo {author} {\bibfnamefont
  {R.}~\bibnamefont {Reece}}, \bibinfo {author} {\bibfnamefont
  {A.}~\bibnamefont {Rinkevicius}}, \bibinfo {author} {\bibfnamefont
  {E.}~\bibnamefont {Rodrigues}}, \bibinfo {author} {\bibfnamefont
  {J.}~\bibnamefont {Rorie}}, \bibinfo {author} {\bibfnamefont
  {D.}~\bibnamefont {Rousseau}}, \bibinfo {author} {\bibfnamefont
  {A.}~\bibnamefont {Sauers}}, \bibinfo {author} {\bibfnamefont
  {S.}~\bibnamefont {Schramm}}, \bibinfo {author} {\bibfnamefont
  {A.}~\bibnamefont {Schwartzman}}, \bibinfo {author} {\bibfnamefont
  {H.}~\bibnamefont {Severini}}, \bibinfo {author} {\bibfnamefont
  {P.}~\bibnamefont {Seyfert}}, \bibinfo {author} {\bibfnamefont
  {F.}~\bibnamefont {Siroky}}, \bibinfo {author} {\bibfnamefont
  {K.}~\bibnamefont {Skazytkin}}, \bibinfo {author} {\bibfnamefont
  {M.}~\bibnamefont {Sokoloff}}, \bibinfo {author} {\bibfnamefont
  {G.}~\bibnamefont {Stewart}}, \bibinfo {author} {\bibfnamefont
  {B.}~\bibnamefont {Stienen}}, \bibinfo {author} {\bibfnamefont
  {I.}~\bibnamefont {Stockdale}}, \bibinfo {author} {\bibfnamefont
  {G.}~\bibnamefont {Strong}}, \bibinfo {author} {\bibfnamefont
  {S.}~\bibnamefont {Thais}}, \bibinfo {author} {\bibfnamefont
  {K.}~\bibnamefont {Tomko}}, \bibinfo {author} {\bibfnamefont
  {E.}~\bibnamefont {Upfal}}, \bibinfo {author} {\bibfnamefont
  {E.}~\bibnamefont {Usai}}, \bibinfo {author} {\bibfnamefont {A.}~\bibnamefont
  {Ustyuzhanin}}, \bibinfo {author} {\bibfnamefont {M.}~\bibnamefont {Vala}},
  \bibinfo {author} {\bibfnamefont {S.}~\bibnamefont {Vallecorsa}}, \bibinfo
  {author} {\bibfnamefont {J.}~\bibnamefont {Vasel}}, \bibinfo {author}
  {\bibfnamefont {M.}~\bibnamefont {Verzetti}}, \bibinfo {author}
  {\bibfnamefont {X.}~\bibnamefont {Vilasís-Cardona}}, \bibinfo {author}
  {\bibfnamefont {J.-R.}\ \bibnamefont {Vlimant}}, \bibinfo {author}
  {\bibfnamefont {I.}~\bibnamefont {Vukotic}}, \bibinfo {author} {\bibfnamefont
  {S.-J.}\ \bibnamefont {Wang}}, \bibinfo {author} {\bibfnamefont
  {G.}~\bibnamefont {Watts}}, \bibinfo {author} {\bibfnamefont
  {M.}~\bibnamefont {Williams}}, \bibinfo {author} {\bibfnamefont
  {W.}~\bibnamefont {Wu}}, \bibinfo {author} {\bibfnamefont {S.}~\bibnamefont
  {Wunsch}},\ and\ \bibinfo {author} {\bibfnamefont {O.}~\bibnamefont
  {Zapata}},\ }\bibfield  {title} {\bibinfo {title} {Machine learning in high
  energy physics community white paper},\ }\href
  {https://doi.org/10.1088/1742-6596/1085/2/022008} {\bibfield  {journal}
  {\bibinfo  {journal} {J. Phys. Conf. Ser.}\ }\textbf {\bibinfo {volume}
  {1085}},\ \bibinfo {pages} {022008} (\bibinfo {year} {2018})}\BibitemShut
  {NoStop}%
\bibitem [{\citenamefont {Litjens}\ \emph {et~al.}(2017)\citenamefont
  {Litjens}, \citenamefont {Kooi}, \citenamefont {Bejnordi}, \citenamefont
  {Setio}, \citenamefont {Ciompi}, \citenamefont {Ghafoorian}, \citenamefont
  {van~der Laak}, \citenamefont {van Ginneken},\ and\ \citenamefont
  {Sánchez}}]{MLMed}%
  \BibitemOpen
  \bibfield  {author} {\bibinfo {author} {\bibfnamefont {G.}~\bibnamefont
  {Litjens}}, \bibinfo {author} {\bibfnamefont {T.}~\bibnamefont {Kooi}},
  \bibinfo {author} {\bibfnamefont {B.~E.}\ \bibnamefont {Bejnordi}}, \bibinfo
  {author} {\bibfnamefont {A.~A.~A.}\ \bibnamefont {Setio}}, \bibinfo {author}
  {\bibfnamefont {F.}~\bibnamefont {Ciompi}}, \bibinfo {author} {\bibfnamefont
  {M.}~\bibnamefont {Ghafoorian}}, \bibinfo {author} {\bibfnamefont {J.~A.}\
  \bibnamefont {van~der Laak}}, \bibinfo {author} {\bibfnamefont
  {B.}~\bibnamefont {van Ginneken}},\ and\ \bibinfo {author} {\bibfnamefont
  {C.~I.}\ \bibnamefont {Sánchez}},\ }\bibfield  {title} {\bibinfo {title} {A
  survey on deep learning in medical image analysis},\ }\href
  {https://doi.org/10.1016/j.media.2017.07.005} {\bibfield  {journal} {\bibinfo
   {journal} {Med. Image Anal.}\ }\textbf {\bibinfo {volume} {42}},\ \bibinfo
  {pages} {60–88} (\bibinfo {year} {2017})}\BibitemShut {NoStop}%
\bibitem [{\citenamefont {Vaswani}\ \emph {et~al.}(2017)\citenamefont
  {Vaswani}, \citenamefont {Shazeer}, \citenamefont {Parmar}, \citenamefont
  {Uszkoreit}, \citenamefont {Jones}, \citenamefont {Gomez}, \citenamefont
  {Kaiser},\ and\ \citenamefont {Polosukhin}}]{TransformerNN}%
  \BibitemOpen
  \bibfield  {author} {\bibinfo {author} {\bibfnamefont {A.}~\bibnamefont
  {Vaswani}}, \bibinfo {author} {\bibfnamefont {N.}~\bibnamefont {Shazeer}},
  \bibinfo {author} {\bibfnamefont {N.}~\bibnamefont {Parmar}}, \bibinfo
  {author} {\bibfnamefont {J.}~\bibnamefont {Uszkoreit}}, \bibinfo {author}
  {\bibfnamefont {L.}~\bibnamefont {Jones}}, \bibinfo {author} {\bibfnamefont
  {A.~N.}\ \bibnamefont {Gomez}}, \bibinfo {author} {\bibfnamefont
  {L.}~\bibnamefont {Kaiser}},\ and\ \bibinfo {author} {\bibfnamefont
  {I.}~\bibnamefont {Polosukhin}},\ }\bibfield  {title} {\bibinfo {title}
  {Attention is all you need},\ }in\ \href
  {https://proceedings.neurips.cc/paper_files/paper/2017/file/3f5ee243547dee91fbd053c1c4a845aa-Paper.pdf}
  {\emph {\bibinfo {booktitle} {Advances in Neural Information Processing
  Systems}}},\ Vol.~\bibinfo {volume} {30},\ \bibinfo {editor} {edited by\
  \bibinfo {editor} {\bibfnamefont {I.}~\bibnamefont {Guyon}}, \bibinfo
  {editor} {\bibfnamefont {U.~V.}\ \bibnamefont {Luxburg}}, \bibinfo {editor}
  {\bibfnamefont {S.}~\bibnamefont {Bengio}}, \bibinfo {editor} {\bibfnamefont
  {H.}~\bibnamefont {Wallach}}, \bibinfo {editor} {\bibfnamefont
  {R.}~\bibnamefont {Fergus}}, \bibinfo {editor} {\bibfnamefont
  {S.}~\bibnamefont {Vishwanathan}},\ and\ \bibinfo {editor} {\bibfnamefont
  {R.}~\bibnamefont {Garnett}}}\ (\bibinfo  {publisher} {Curran Associates,
  Inc.},\ \bibinfo {year} {2017})\BibitemShut {NoStop}%
\bibitem [{\citenamefont {{DeepSeek-AI: X. Bi}}\ \emph
  {et~al.}(2024)\citenamefont {{DeepSeek-AI: X. Bi}}, \citenamefont {Chen},
  \citenamefont {Chen}, \citenamefont {Chen}, \citenamefont {Dai},
  \citenamefont {Deng}, \citenamefont {Ding}, \citenamefont {Dong},
  \citenamefont {Du}, \citenamefont {Fu}, \citenamefont {Gao}, \citenamefont
  {Gao}, \citenamefont {Gao}, \citenamefont {Ge}, \citenamefont {Guan},
  \citenamefont {Guo}, \citenamefont {Guo}, \citenamefont {Hao}, \citenamefont
  {Hao}, \citenamefont {He}, \citenamefont {Hu}, \citenamefont {Huang},
  \citenamefont {Li}, \citenamefont {Li}, \citenamefont {Li}, \citenamefont
  {Li}, \citenamefont {Li}, \citenamefont {Liang}, \citenamefont {Lin},
  \citenamefont {Liu}, \citenamefont {Liu}, \citenamefont {Liu}, \citenamefont
  {Liu}, \citenamefont {Liu}, \citenamefont {Liu}, \citenamefont {Lu},
  \citenamefont {Lu}, \citenamefont {Luo}, \citenamefont {Ma}, \citenamefont
  {Nie}, \citenamefont {Pei}, \citenamefont {Piao}, \citenamefont {Qiu},
  \citenamefont {Qu}, \citenamefont {Ren}, \citenamefont {Ren}, \citenamefont
  {Ruan}, \citenamefont {Sha}, \citenamefont {Shao}, \citenamefont {Song},
  \citenamefont {Su}, \citenamefont {Sun}, \citenamefont {Sun}, \citenamefont
  {Tang}, \citenamefont {Wang}, \citenamefont {Wang}, \citenamefont {Wang},
  \citenamefont {Wang}, \citenamefont {Wang}, \citenamefont {Wu}, \citenamefont
  {Wu}, \citenamefont {Xie}, \citenamefont {Xie}, \citenamefont {Xie},
  \citenamefont {Xiong}, \citenamefont {Xu}, \citenamefont {Xu}, \citenamefont
  {Xu}, \citenamefont {Yang}, \citenamefont {You}, \citenamefont {Yu},
  \citenamefont {Yu}, \citenamefont {Zhang}, \citenamefont {Zhang},
  \citenamefont {Zhang}, \citenamefont {Zhang}, \citenamefont {Zhang},
  \citenamefont {Zhang}, \citenamefont {Zhang}, \citenamefont {Zhang},
  \citenamefont {Zhao}, \citenamefont {Zhao}, \citenamefont {Zhou},
  \citenamefont {Zhou}, \citenamefont {Zhu},\ and\ \citenamefont
  {Zou}}]{LLMDeepseek}%
  \BibitemOpen
  \bibfield  {author} {\bibinfo {author} {\bibnamefont {{DeepSeek-AI: X. Bi}}},
  \bibinfo {author} {\bibfnamefont {D.}~\bibnamefont {Chen}}, \bibinfo {author}
  {\bibfnamefont {G.}~\bibnamefont {Chen}}, \bibinfo {author} {\bibfnamefont
  {S.}~\bibnamefont {Chen}}, \bibinfo {author} {\bibfnamefont {D.}~\bibnamefont
  {Dai}}, \bibinfo {author} {\bibfnamefont {C.}~\bibnamefont {Deng}}, \bibinfo
  {author} {\bibfnamefont {H.}~\bibnamefont {Ding}}, \bibinfo {author}
  {\bibfnamefont {K.}~\bibnamefont {Dong}}, \bibinfo {author} {\bibfnamefont
  {Q.}~\bibnamefont {Du}}, \bibinfo {author} {\bibfnamefont {Z.}~\bibnamefont
  {Fu}}, \bibinfo {author} {\bibfnamefont {H.}~\bibnamefont {Gao}}, \bibinfo
  {author} {\bibfnamefont {K.}~\bibnamefont {Gao}}, \bibinfo {author}
  {\bibfnamefont {W.}~\bibnamefont {Gao}}, \bibinfo {author} {\bibfnamefont
  {R.}~\bibnamefont {Ge}}, \bibinfo {author} {\bibfnamefont {K.}~\bibnamefont
  {Guan}}, \bibinfo {author} {\bibfnamefont {D.}~\bibnamefont {Guo}}, \bibinfo
  {author} {\bibfnamefont {J.}~\bibnamefont {Guo}}, \bibinfo {author}
  {\bibfnamefont {G.}~\bibnamefont {Hao}}, \bibinfo {author} {\bibfnamefont
  {Z.}~\bibnamefont {Hao}}, \bibinfo {author} {\bibfnamefont {Y.}~\bibnamefont
  {He}}, \bibinfo {author} {\bibfnamefont {W.}~\bibnamefont {Hu}}, \bibinfo
  {author} {\bibfnamefont {P.}~\bibnamefont {Huang}}, \bibinfo {author}
  {\bibfnamefont {E.}~\bibnamefont {Li}}, \bibinfo {author} {\bibfnamefont
  {G.}~\bibnamefont {Li}}, \bibinfo {author} {\bibfnamefont {J.}~\bibnamefont
  {Li}}, \bibinfo {author} {\bibfnamefont {Y.}~\bibnamefont {Li}}, \bibinfo
  {author} {\bibfnamefont {Y.~K.}\ \bibnamefont {Li}}, \bibinfo {author}
  {\bibfnamefont {W.}~\bibnamefont {Liang}}, \bibinfo {author} {\bibfnamefont
  {F.}~\bibnamefont {Lin}}, \bibinfo {author} {\bibfnamefont {A.~X.}\
  \bibnamefont {Liu}}, \bibinfo {author} {\bibfnamefont {B.}~\bibnamefont
  {Liu}}, \bibinfo {author} {\bibfnamefont {W.}~\bibnamefont {Liu}}, \bibinfo
  {author} {\bibfnamefont {X.}~\bibnamefont {Liu}}, \bibinfo {author}
  {\bibfnamefont {X.}~\bibnamefont {Liu}}, \bibinfo {author} {\bibfnamefont
  {Y.}~\bibnamefont {Liu}}, \bibinfo {author} {\bibfnamefont {H.}~\bibnamefont
  {Lu}}, \bibinfo {author} {\bibfnamefont {S.}~\bibnamefont {Lu}}, \bibinfo
  {author} {\bibfnamefont {F.}~\bibnamefont {Luo}}, \bibinfo {author}
  {\bibfnamefont {S.}~\bibnamefont {Ma}}, \bibinfo {author} {\bibfnamefont
  {X.}~\bibnamefont {Nie}}, \bibinfo {author} {\bibfnamefont {T.}~\bibnamefont
  {Pei}}, \bibinfo {author} {\bibfnamefont {Y.}~\bibnamefont {Piao}}, \bibinfo
  {author} {\bibfnamefont {J.}~\bibnamefont {Qiu}}, \bibinfo {author}
  {\bibfnamefont {H.}~\bibnamefont {Qu}}, \bibinfo {author} {\bibfnamefont
  {T.}~\bibnamefont {Ren}}, \bibinfo {author} {\bibfnamefont {Z.}~\bibnamefont
  {Ren}}, \bibinfo {author} {\bibfnamefont {C.}~\bibnamefont {Ruan}}, \bibinfo
  {author} {\bibfnamefont {Z.}~\bibnamefont {Sha}}, \bibinfo {author}
  {\bibfnamefont {Z.}~\bibnamefont {Shao}}, \bibinfo {author} {\bibfnamefont
  {J.}~\bibnamefont {Song}}, \bibinfo {author} {\bibfnamefont {X.}~\bibnamefont
  {Su}}, \bibinfo {author} {\bibfnamefont {J.}~\bibnamefont {Sun}}, \bibinfo
  {author} {\bibfnamefont {Y.}~\bibnamefont {Sun}}, \bibinfo {author}
  {\bibfnamefont {M.}~\bibnamefont {Tang}}, \bibinfo {author} {\bibfnamefont
  {B.}~\bibnamefont {Wang}}, \bibinfo {author} {\bibfnamefont {P.}~\bibnamefont
  {Wang}}, \bibinfo {author} {\bibfnamefont {S.}~\bibnamefont {Wang}}, \bibinfo
  {author} {\bibfnamefont {Y.}~\bibnamefont {Wang}}, \bibinfo {author}
  {\bibfnamefont {Y.}~\bibnamefont {Wang}}, \bibinfo {author} {\bibfnamefont
  {T.}~\bibnamefont {Wu}}, \bibinfo {author} {\bibfnamefont {Y.}~\bibnamefont
  {Wu}}, \bibinfo {author} {\bibfnamefont {X.}~\bibnamefont {Xie}}, \bibinfo
  {author} {\bibfnamefont {Z.}~\bibnamefont {Xie}}, \bibinfo {author}
  {\bibfnamefont {Z.}~\bibnamefont {Xie}}, \bibinfo {author} {\bibfnamefont
  {Y.}~\bibnamefont {Xiong}}, \bibinfo {author} {\bibfnamefont
  {H.}~\bibnamefont {Xu}}, \bibinfo {author} {\bibfnamefont {R.~X.}\
  \bibnamefont {Xu}}, \bibinfo {author} {\bibfnamefont {Y.}~\bibnamefont {Xu}},
  \bibinfo {author} {\bibfnamefont {D.}~\bibnamefont {Yang}}, \bibinfo {author}
  {\bibfnamefont {Y.}~\bibnamefont {You}}, \bibinfo {author} {\bibfnamefont
  {S.}~\bibnamefont {Yu}}, \bibinfo {author} {\bibfnamefont {X.}~\bibnamefont
  {Yu}}, \bibinfo {author} {\bibfnamefont {B.}~\bibnamefont {Zhang}}, \bibinfo
  {author} {\bibfnamefont {H.}~\bibnamefont {Zhang}}, \bibinfo {author}
  {\bibfnamefont {L.}~\bibnamefont {Zhang}}, \bibinfo {author} {\bibfnamefont
  {L.}~\bibnamefont {Zhang}}, \bibinfo {author} {\bibfnamefont
  {M.}~\bibnamefont {Zhang}}, \bibinfo {author} {\bibfnamefont
  {M.}~\bibnamefont {Zhang}}, \bibinfo {author} {\bibfnamefont
  {W.}~\bibnamefont {Zhang}}, \bibinfo {author} {\bibfnamefont
  {Y.}~\bibnamefont {Zhang}}, \bibinfo {author} {\bibfnamefont
  {C.}~\bibnamefont {Zhao}}, \bibinfo {author} {\bibfnamefont {Y.}~\bibnamefont
  {Zhao}}, \bibinfo {author} {\bibfnamefont {S.}~\bibnamefont {Zhou}}, \bibinfo
  {author} {\bibfnamefont {S.}~\bibnamefont {Zhou}}, \bibinfo {author}
  {\bibfnamefont {Q.}~\bibnamefont {Zhu}},\ and\ \bibinfo {author}
  {\bibfnamefont {Y.}~\bibnamefont {Zou}},\ }\href@noop {} {\bibinfo {title}
  {Deepseek llm: Scaling open-source language models with longtermism}}
  (\bibinfo {year} {2024}),\ \Eprint {https://arxiv.org/abs/2401.02954}
  {arXiv:2401.02954 [cs.CL]} \BibitemShut {NoStop}%
\bibitem [{\citenamefont {Khan}\ and\ \citenamefont
  {Al-Habsi}(2020)}]{MLComputerVision}%
  \BibitemOpen
  \bibfield  {author} {\bibinfo {author} {\bibfnamefont {A.~I.}\ \bibnamefont
  {Khan}}\ and\ \bibinfo {author} {\bibfnamefont {S.}~\bibnamefont
  {Al-Habsi}},\ }\bibfield  {title} {\bibinfo {title} {Machine learning in
  computer vision},\ }\href {https://doi.org/10.1016/j.procs.2020.03.355}
  {\bibfield  {journal} {\bibinfo  {journal} {Procedia Comput. Sci.}\ }\textbf
  {\bibinfo {volume} {167}},\ \bibinfo {pages} {1444} (\bibinfo {year}
  {2020})}\BibitemShut {NoStop}%
\bibitem [{\citenamefont {Stanev}\ \emph {et~al.}(2023)\citenamefont {Stanev},
  \citenamefont {Spagnolo},\ and\ \citenamefont {Sciarrino}}]{DenisCloning}%
  \BibitemOpen
  \bibfield  {author} {\bibinfo {author} {\bibfnamefont {D.}~\bibnamefont
  {Stanev}}, \bibinfo {author} {\bibfnamefont {N.}~\bibnamefont {Spagnolo}},\
  and\ \bibinfo {author} {\bibfnamefont {F.}~\bibnamefont {Sciarrino}},\
  }\bibfield  {title} {\bibinfo {title} {Deterministic optimal quantum cloning
  via a quantum-optical neural network},\ }\href
  {https://doi.org/10.1103/PhysRevResearch.5.013139} {\bibfield  {journal}
  {\bibinfo  {journal} {Phys. Rev. Res.}\ }\textbf {\bibinfo {volume} {5}},\
  \bibinfo {pages} {013139} (\bibinfo {year} {2023})}\BibitemShut {NoStop}%
\bibitem [{\citenamefont {Steinbrecher}\ \emph {et~al.}(2019)\citenamefont
  {Steinbrecher}, \citenamefont {Olson}, \citenamefont {Englund},\ and\
  \citenamefont {Carolan}}]{QONN}%
  \BibitemOpen
  \bibfield  {author} {\bibinfo {author} {\bibfnamefont {G.~R.}\ \bibnamefont
  {Steinbrecher}}, \bibinfo {author} {\bibfnamefont {J.~P.}\ \bibnamefont
  {Olson}}, \bibinfo {author} {\bibfnamefont {D.}~\bibnamefont {Englund}},\
  and\ \bibinfo {author} {\bibfnamefont {J.}~\bibnamefont {Carolan}},\
  }\bibfield  {title} {\bibinfo {title} {Quantum optical neural networks},\
  }\href {https://doi.org/10.1038/s41534-019-0174-7} {\bibfield  {journal}
  {\bibinfo  {journal} {npj Quantum Inf.}\ }\textbf {\bibinfo {volume} {5}},\
  \bibinfo {pages} {60} (\bibinfo {year} {2019})}\BibitemShut {NoStop}%
\bibitem [{\citenamefont {Zhang}\ and\ \citenamefont {Ni}(2020)}]{QMLReview}%
  \BibitemOpen
  \bibfield  {author} {\bibinfo {author} {\bibfnamefont {Y.}~\bibnamefont
  {Zhang}}\ and\ \bibinfo {author} {\bibfnamefont {Q.}~\bibnamefont {Ni}},\
  }\bibfield  {title} {\bibinfo {title} {Recent advances in quantum machine
  learning},\ }\href {https://doi.org/10.1002/que2.34} {\bibfield  {journal}
  {\bibinfo  {journal} {Quantum Eng.}\ }\textbf {\bibinfo {volume} {2}},\
  \bibinfo {pages} {e34} (\bibinfo {year} {2020})}\BibitemShut {NoStop}%
\bibitem [{\citenamefont {Bordoni}\ \emph {et~al.}(2023)\citenamefont
  {Bordoni}, \citenamefont {Stanev}, \citenamefont {Santantonio},\ and\
  \citenamefont {Giagu}}]{DenisAnomaly}%
  \BibitemOpen
  \bibfield  {author} {\bibinfo {author} {\bibfnamefont {S.}~\bibnamefont
  {Bordoni}}, \bibinfo {author} {\bibfnamefont {D.}~\bibnamefont {Stanev}},
  \bibinfo {author} {\bibfnamefont {T.}~\bibnamefont {Santantonio}},\ and\
  \bibinfo {author} {\bibfnamefont {S.}~\bibnamefont {Giagu}},\ }\bibfield
  {title} {\bibinfo {title} {Long-lived particles anomaly detection with
  parametrized quantum circuits},\ }\href
  {https://doi.org/10.3390/particles6010016} {\bibfield  {journal} {\bibinfo
  {journal} {Particles}\ }\textbf {\bibinfo {volume} {6}},\ \bibinfo {pages}
  {297–311} (\bibinfo {year} {2023})}\BibitemShut {NoStop}%
\bibitem [{\citenamefont {Bartlett}\ and\ \citenamefont
  {Fan}(2020)}]{BartlettUniversal}%
  \BibitemOpen
  \bibfield  {author} {\bibinfo {author} {\bibfnamefont {B.}~\bibnamefont
  {Bartlett}}\ and\ \bibinfo {author} {\bibfnamefont {S.}~\bibnamefont {Fan}},\
  }\bibfield  {title} {\bibinfo {title} {Universal programmable photonic
  architecture for quantum information processing},\ }\href
  {https://doi.org/10.1103/PhysRevA.101.042319} {\bibfield  {journal} {\bibinfo
   {journal} {Phys. Rev. A}\ }\textbf {\bibinfo {volume} {101}},\ \bibinfo
  {pages} {042319} (\bibinfo {year} {2020})}\BibitemShut {NoStop}%
\bibitem [{\citenamefont {Suprano}\ \emph {et~al.}(2024)\citenamefont
  {Suprano}, \citenamefont {Zia}, \citenamefont {Innocenti}, \citenamefont
  {Lorenzo}, \citenamefont {Cimini}, \citenamefont {Giordani}, \citenamefont
  {Palmisano}, \citenamefont {Polino}, \citenamefont {Spagnolo}, \citenamefont
  {Sciarrino}, \citenamefont {Palma}, \citenamefont {Ferraro},\ and\
  \citenamefont {Paternostro}}]{QuantumLabExtremeLearning}%
  \BibitemOpen
  \bibfield  {author} {\bibinfo {author} {\bibfnamefont {A.}~\bibnamefont
  {Suprano}}, \bibinfo {author} {\bibfnamefont {D.}~\bibnamefont {Zia}},
  \bibinfo {author} {\bibfnamefont {L.}~\bibnamefont {Innocenti}}, \bibinfo
  {author} {\bibfnamefont {S.}~\bibnamefont {Lorenzo}}, \bibinfo {author}
  {\bibfnamefont {V.}~\bibnamefont {Cimini}}, \bibinfo {author} {\bibfnamefont
  {T.}~\bibnamefont {Giordani}}, \bibinfo {author} {\bibfnamefont
  {I.}~\bibnamefont {Palmisano}}, \bibinfo {author} {\bibfnamefont
  {E.}~\bibnamefont {Polino}}, \bibinfo {author} {\bibfnamefont
  {N.}~\bibnamefont {Spagnolo}}, \bibinfo {author} {\bibfnamefont
  {F.}~\bibnamefont {Sciarrino}}, \bibinfo {author} {\bibfnamefont {G.~M.}\
  \bibnamefont {Palma}}, \bibinfo {author} {\bibfnamefont {A.}~\bibnamefont
  {Ferraro}},\ and\ \bibinfo {author} {\bibfnamefont {M.}~\bibnamefont
  {Paternostro}},\ }\bibfield  {title} {\bibinfo {title} {Photonic quantum
  extreme learning machine},\ }in\ \href
  {https://doi.org/10.1364/QUANTUM.2024.QW4A.2} {\emph {\bibinfo {booktitle}
  {Quantum 2.0 Conference and Exhibition}}}\ (\bibinfo  {publisher} {Optica
  Publishing Group},\ \bibinfo {year} {2024})\ p.\ \bibinfo {pages}
  {QW4A.2}\BibitemShut {NoStop}%
\bibitem [{\citenamefont {Hoch}\ \emph {et~al.}(2025)\citenamefont {Hoch},
  \citenamefont {Caruccio}, \citenamefont {Rodari}, \citenamefont
  {Francalanci}, \citenamefont {Suprano}, \citenamefont {Giordani},
  \citenamefont {Carvacho}, \citenamefont {Spagnolo}, \citenamefont {Koudia},
  \citenamefont {Proietti}, \citenamefont {Liorni}, \citenamefont {Cerocchi},
  \citenamefont {Albiero}, \citenamefont {Di~Giano}, \citenamefont {Gardina},
  \citenamefont {Ceccarelli}, \citenamefont {Corrielli}, \citenamefont
  {Chabaud}, \citenamefont {Osellame}, \citenamefont {Dispenza},\ and\
  \citenamefont {Sciarrino}}]{QuantumLabAdaptiveBS}%
  \BibitemOpen
  \bibfield  {author} {\bibinfo {author} {\bibfnamefont {F.}~\bibnamefont
  {Hoch}}, \bibinfo {author} {\bibfnamefont {E.}~\bibnamefont {Caruccio}},
  \bibinfo {author} {\bibfnamefont {G.}~\bibnamefont {Rodari}}, \bibinfo
  {author} {\bibfnamefont {T.}~\bibnamefont {Francalanci}}, \bibinfo {author}
  {\bibfnamefont {A.}~\bibnamefont {Suprano}}, \bibinfo {author} {\bibfnamefont
  {T.}~\bibnamefont {Giordani}}, \bibinfo {author} {\bibfnamefont
  {G.}~\bibnamefont {Carvacho}}, \bibinfo {author} {\bibfnamefont
  {N.}~\bibnamefont {Spagnolo}}, \bibinfo {author} {\bibfnamefont
  {S.}~\bibnamefont {Koudia}}, \bibinfo {author} {\bibfnamefont
  {M.}~\bibnamefont {Proietti}}, \bibinfo {author} {\bibfnamefont
  {C.}~\bibnamefont {Liorni}}, \bibinfo {author} {\bibfnamefont
  {F.}~\bibnamefont {Cerocchi}}, \bibinfo {author} {\bibfnamefont
  {R.}~\bibnamefont {Albiero}}, \bibinfo {author} {\bibfnamefont
  {N.}~\bibnamefont {Di~Giano}}, \bibinfo {author} {\bibfnamefont
  {M.}~\bibnamefont {Gardina}}, \bibinfo {author} {\bibfnamefont
  {F.}~\bibnamefont {Ceccarelli}}, \bibinfo {author} {\bibfnamefont
  {G.}~\bibnamefont {Corrielli}}, \bibinfo {author} {\bibfnamefont
  {U.}~\bibnamefont {Chabaud}}, \bibinfo {author} {\bibfnamefont
  {R.}~\bibnamefont {Osellame}}, \bibinfo {author} {\bibfnamefont
  {M.}~\bibnamefont {Dispenza}},\ and\ \bibinfo {author} {\bibfnamefont
  {F.}~\bibnamefont {Sciarrino}},\ }\bibfield  {title} {\bibinfo {title}
  {Quantum machine learning with adaptive boson sampling via post-selection},\
  }\href {https://doi.org/10.1038/s41467-025-55877-z} {\bibfield  {journal}
  {\bibinfo  {journal} {Nat. Commun.}\ }\textbf {\bibinfo {volume} {16}},\
  \bibinfo {pages} {902} (\bibinfo {year} {2025})}\BibitemShut {NoStop}%
\bibitem [{\citenamefont {Nielsen}\ \emph {et~al.}(2024)\citenamefont
  {Nielsen}, \citenamefont {Wang}, \citenamefont {Deacon}, \citenamefont
  {Sund}, \citenamefont {Liu}, \citenamefont {Scholz}, \citenamefont {Wieck},
  \citenamefont {Ludwig}, \citenamefont {Midolo}, \citenamefont {Sørensen},
  \citenamefont {Paesani},\ and\ \citenamefont
  {Lodahl}}]{PhotonicNonlinearity}%
  \BibitemOpen
  \bibfield  {author} {\bibinfo {author} {\bibfnamefont {K.~H.}\ \bibnamefont
  {Nielsen}}, \bibinfo {author} {\bibfnamefont {Y.}~\bibnamefont {Wang}},
  \bibinfo {author} {\bibfnamefont {E.}~\bibnamefont {Deacon}}, \bibinfo
  {author} {\bibfnamefont {P.~I.}\ \bibnamefont {Sund}}, \bibinfo {author}
  {\bibfnamefont {Z.}~\bibnamefont {Liu}}, \bibinfo {author} {\bibfnamefont
  {S.}~\bibnamefont {Scholz}}, \bibinfo {author} {\bibfnamefont {A.~D.}\
  \bibnamefont {Wieck}}, \bibinfo {author} {\bibfnamefont {A.}~\bibnamefont
  {Ludwig}}, \bibinfo {author} {\bibfnamefont {L.}~\bibnamefont {Midolo}},
  \bibinfo {author} {\bibfnamefont {A.~S.}\ \bibnamefont {Sørensen}}, \bibinfo
  {author} {\bibfnamefont {S.}~\bibnamefont {Paesani}},\ and\ \bibinfo {author}
  {\bibfnamefont {P.}~\bibnamefont {Lodahl}},\ }\href@noop {} {\bibinfo {title}
  {Programmable nonlinear quantum photonic circuits}} (\bibinfo {year}
  {2024}),\ \Eprint {https://arxiv.org/abs/2405.17941} {arXiv:2405.17941
  [quant-ph]} \BibitemShut {NoStop}%
\bibitem [{\citenamefont {Tang}\ \emph {et~al.}(2022)\citenamefont {Tang},
  \citenamefont {Banchi}, \citenamefont {Wang}, \citenamefont {Shang},
  \citenamefont {Tan}, \citenamefont {Zhou}, \citenamefont {Feng},
  \citenamefont {Pal}, \citenamefont {Li}, \citenamefont {Hu}, \citenamefont
  {Kim},\ and\ \citenamefont {Jin}}]{Tang22}%
  \BibitemOpen
  \bibfield  {author} {\bibinfo {author} {\bibfnamefont {H.}~\bibnamefont
  {Tang}}, \bibinfo {author} {\bibfnamefont {L.}~\bibnamefont {Banchi}},
  \bibinfo {author} {\bibfnamefont {T.-Y.}\ \bibnamefont {Wang}}, \bibinfo
  {author} {\bibfnamefont {X.-W.}\ \bibnamefont {Shang}}, \bibinfo {author}
  {\bibfnamefont {X.}~\bibnamefont {Tan}}, \bibinfo {author} {\bibfnamefont
  {W.-H.}\ \bibnamefont {Zhou}}, \bibinfo {author} {\bibfnamefont
  {Z.}~\bibnamefont {Feng}}, \bibinfo {author} {\bibfnamefont {A.}~\bibnamefont
  {Pal}}, \bibinfo {author} {\bibfnamefont {H.}~\bibnamefont {Li}}, \bibinfo
  {author} {\bibfnamefont {C.-Q.}\ \bibnamefont {Hu}}, \bibinfo {author}
  {\bibfnamefont {M.~S.}\ \bibnamefont {Kim}},\ and\ \bibinfo {author}
  {\bibfnamefont {X.-M.}\ \bibnamefont {Jin}},\ }\bibfield  {title} {\bibinfo
  {title} {Generating haar-uniform randomness using stochastic quantum walks on
  a photonic chip},\ }\href {https://doi.org/10.1103/PhysRevLett.128.050503}
  {\bibfield  {journal} {\bibinfo  {journal} {Phys. Rev. Lett.}\ }\textbf
  {\bibinfo {volume} {128}},\ \bibinfo {pages} {050503} (\bibinfo {year}
  {2022})}\BibitemShut {NoStop}%
\bibitem [{\citenamefont {Whitfield}\ \emph {et~al.}(2010)\citenamefont
  {Whitfield}, \citenamefont {Rodriguez-Rosario},\ and\ \citenamefont
  {Aspuru-Guzi}}]{Whit10}%
  \BibitemOpen
  \bibfield  {author} {\bibinfo {author} {\bibfnamefont {J.}~\bibnamefont
  {Whitfield}}, \bibinfo {author} {\bibfnamefont {C.~A.}\ \bibnamefont
  {Rodriguez-Rosario}},\ and\ \bibinfo {author} {\bibnamefont {Aspuru-Guzi}},\
  }\bibfield  {title} {\bibinfo {title} {Quantum stochastic walks: A
  generalization of classical random walks and quantum walks},\ }\href
  {https://doi.org/10.1103/PhysRevA.81.022323} {\bibfield  {journal} {\bibinfo
  {journal} {Phys. Rev. A}\ }\textbf {\bibinfo {volume} {81}},\ \bibinfo
  {pages} {022323} (\bibinfo {year} {2010})}\BibitemShut {NoStop}%
\bibitem [{\citenamefont {Lambert}\ \emph {et~al.}(2024)\citenamefont
  {Lambert}, \citenamefont {Giguère}, \citenamefont {Menczel}, \citenamefont
  {Li}, \citenamefont {Hopf}, \citenamefont {Suárez}, \citenamefont {Gali},
  \citenamefont {Lishman}, \citenamefont {Gadhvi}, \citenamefont {Agarwal},
  \citenamefont {Galicia}, \citenamefont {Shammah}, \citenamefont {Nation},
  \citenamefont {Johansson}, \citenamefont {Ahmed}, \citenamefont {Cross},
  \citenamefont {Pitchford},\ and\ \citenamefont {Nori}}]{Qutip}%
  \BibitemOpen
  \bibfield  {author} {\bibinfo {author} {\bibfnamefont {N.}~\bibnamefont
  {Lambert}}, \bibinfo {author} {\bibfnamefont {E.}~\bibnamefont {Giguère}},
  \bibinfo {author} {\bibfnamefont {P.}~\bibnamefont {Menczel}}, \bibinfo
  {author} {\bibfnamefont {B.}~\bibnamefont {Li}}, \bibinfo {author}
  {\bibfnamefont {P.}~\bibnamefont {Hopf}}, \bibinfo {author} {\bibfnamefont
  {G.}~\bibnamefont {Suárez}}, \bibinfo {author} {\bibfnamefont
  {M.}~\bibnamefont {Gali}}, \bibinfo {author} {\bibfnamefont {J.}~\bibnamefont
  {Lishman}}, \bibinfo {author} {\bibfnamefont {R.}~\bibnamefont {Gadhvi}},
  \bibinfo {author} {\bibfnamefont {R.}~\bibnamefont {Agarwal}}, \bibinfo
  {author} {\bibfnamefont {A.}~\bibnamefont {Galicia}}, \bibinfo {author}
  {\bibfnamefont {N.}~\bibnamefont {Shammah}}, \bibinfo {author} {\bibfnamefont
  {P.}~\bibnamefont {Nation}}, \bibinfo {author} {\bibfnamefont {J.~R.}\
  \bibnamefont {Johansson}}, \bibinfo {author} {\bibfnamefont {S.}~\bibnamefont
  {Ahmed}}, \bibinfo {author} {\bibfnamefont {S.}~\bibnamefont {Cross}},
  \bibinfo {author} {\bibfnamefont {A.}~\bibnamefont {Pitchford}},\ and\
  \bibinfo {author} {\bibfnamefont {F.}~\bibnamefont {Nori}},\ }\href@noop {}
  {\bibinfo {title} {Qutip 5: The quantum toolbox in python}} (\bibinfo {year}
  {2024}),\ \Eprint {https://arxiv.org/abs/2412.04705} {arXiv:2412.04705
  [quant-ph]} \BibitemShut {NoStop}%
\bibitem [{\citenamefont {Heurtel}\ \emph {et~al.}(2023)\citenamefont
  {Heurtel}, \citenamefont {Fyrillas}, \citenamefont {Gliniasty}, \citenamefont
  {Le~Bihan}, \citenamefont {Malherbe}, \citenamefont {Pailhas}, \citenamefont
  {Bertasi}, \citenamefont {Bourdoncle}, \citenamefont {Emeriau}, \citenamefont
  {Mezher}, \citenamefont {Music}, \citenamefont {Belabas}, \citenamefont
  {Valiron}, \citenamefont {Senellart}, \citenamefont {Mansfield},\ and\
  \citenamefont {Senellart}}]{Perceval}%
  \BibitemOpen
  \bibfield  {author} {\bibinfo {author} {\bibfnamefont {N.}~\bibnamefont
  {Heurtel}}, \bibinfo {author} {\bibfnamefont {A.}~\bibnamefont {Fyrillas}},
  \bibinfo {author} {\bibfnamefont {G.~d.}\ \bibnamefont {Gliniasty}}, \bibinfo
  {author} {\bibfnamefont {R.}~\bibnamefont {Le~Bihan}}, \bibinfo {author}
  {\bibfnamefont {S.}~\bibnamefont {Malherbe}}, \bibinfo {author}
  {\bibfnamefont {M.}~\bibnamefont {Pailhas}}, \bibinfo {author} {\bibfnamefont
  {E.}~\bibnamefont {Bertasi}}, \bibinfo {author} {\bibfnamefont
  {B.}~\bibnamefont {Bourdoncle}}, \bibinfo {author} {\bibfnamefont {P.-E.}\
  \bibnamefont {Emeriau}}, \bibinfo {author} {\bibfnamefont {R.}~\bibnamefont
  {Mezher}}, \bibinfo {author} {\bibfnamefont {L.}~\bibnamefont {Music}},
  \bibinfo {author} {\bibfnamefont {N.}~\bibnamefont {Belabas}}, \bibinfo
  {author} {\bibfnamefont {B.}~\bibnamefont {Valiron}}, \bibinfo {author}
  {\bibfnamefont {P.}~\bibnamefont {Senellart}}, \bibinfo {author}
  {\bibfnamefont {S.}~\bibnamefont {Mansfield}},\ and\ \bibinfo {author}
  {\bibfnamefont {J.}~\bibnamefont {Senellart}},\ }\bibfield  {title} {\bibinfo
  {title} {Perceval: {A} {S}oftware {P}latform for {D}iscrete {V}ariable
  {P}hotonic {Q}uantum {C}omputing},\ }\href
  {https://doi.org/10.22331/q-2023-02-21-931} {\bibfield  {journal} {\bibinfo
  {journal} {{Quantum}}\ }\textbf {\bibinfo {volume} {7}},\ \bibinfo {pages}
  {931} (\bibinfo {year} {2023})}\BibitemShut {NoStop}%
\bibitem [{\citenamefont {Robbins}\ and\ \citenamefont {Monro}(1951)}]{FDSA1}%
  \BibitemOpen
  \bibfield  {author} {\bibinfo {author} {\bibfnamefont {H.}~\bibnamefont
  {Robbins}}\ and\ \bibinfo {author} {\bibfnamefont {S.}~\bibnamefont
  {Monro}},\ }\bibfield  {title} {\bibinfo {title} {A stochastic approximation
  method},\ }\href {https://doi.org/10.1214/aoms/1177729586} {\bibfield
  {journal} {\bibinfo  {journal} {Ann. Math. Stat.}\ }\textbf {\bibinfo
  {volume} {22}},\ \bibinfo {pages} {400–407} (\bibinfo {year}
  {1951})}\BibitemShut {NoStop}%
\bibitem [{\citenamefont {Kiefer}\ and\ \citenamefont
  {Wolfowitz}(1952)}]{FDSA2}%
  \BibitemOpen
  \bibfield  {author} {\bibinfo {author} {\bibfnamefont {J.}~\bibnamefont
  {Kiefer}}\ and\ \bibinfo {author} {\bibfnamefont {J.}~\bibnamefont
  {Wolfowitz}},\ }\bibfield  {title} {\bibinfo {title} {Stochastic estimation
  of the maximum of a regression function},\ }\href
  {https://doi.org/10.1214/aoms/1177729392} {\bibfield  {journal} {\bibinfo
  {journal} {Ann. Math. Stat.}\ }\textbf {\bibinfo {volume} {23}},\ \bibinfo
  {pages} {462–466} (\bibinfo {year} {1952})}\BibitemShut {NoStop}%
\bibitem [{\citenamefont {Amari}(1993)}]{SGD}%
  \BibitemOpen
  \bibfield  {author} {\bibinfo {author} {\bibfnamefont {S.-i.}\ \bibnamefont
  {Amari}},\ }\bibfield  {title} {\bibinfo {title} {Backpropagation and
  stochastic gradient descent method},\ }\href
  {https://doi.org/10.1016/0925-2312(93)90006-o} {\bibfield  {journal}
  {\bibinfo  {journal} {Neurocomputing}\ }\textbf {\bibinfo {volume} {5}},\
  \bibinfo {pages} {185–196} (\bibinfo {year} {1993})}\BibitemShut {NoStop}%
\bibitem [{\citenamefont {Laing}\ and\ \citenamefont
  {O'Brien}(2012)}]{UnitaryTomography1}%
  \BibitemOpen
  \bibfield  {author} {\bibinfo {author} {\bibfnamefont {A.}~\bibnamefont
  {Laing}}\ and\ \bibinfo {author} {\bibfnamefont {J.~L.}\ \bibnamefont
  {O'Brien}},\ }\href@noop {} {\bibinfo {title} {Super-stable tomography of any
  linear optical device}} (\bibinfo {year} {2012}),\ \Eprint
  {https://arxiv.org/abs/1208.2868} {arXiv:1208.2868 [quant-ph]} \BibitemShut
  {NoStop}%
\bibitem [{\citenamefont {Tillmann}\ \emph {et~al.}(2016)\citenamefont
  {Tillmann}, \citenamefont {Schmidt},\ and\ \citenamefont
  {Walther}}]{UnitaryTomography2}%
  \BibitemOpen
  \bibfield  {author} {\bibinfo {author} {\bibfnamefont {M.}~\bibnamefont
  {Tillmann}}, \bibinfo {author} {\bibfnamefont {C.}~\bibnamefont {Schmidt}},\
  and\ \bibinfo {author} {\bibfnamefont {P.}~\bibnamefont {Walther}},\
  }\bibfield  {title} {\bibinfo {title} {On unitary reconstruction of linear
  optical networks},\ }\href {https://doi.org/10.1088/2040-8978/18/11/114002}
  {\bibfield  {journal} {\bibinfo  {journal} {J. Opt.}\ }\textbf {\bibinfo
  {volume} {18}},\ \bibinfo {pages} {114002} (\bibinfo {year}
  {2016})}\BibitemShut {NoStop}%
\bibitem [{\citenamefont {Spagnolo}\ \emph {et~al.}(2017)\citenamefont
  {Spagnolo}, \citenamefont {Maiorino}, \citenamefont {Vitelli}, \citenamefont
  {Bentivegna}, \citenamefont {Crespi}, \citenamefont {Ramponi}, \citenamefont
  {Osellame},\ and\ \citenamefont {Sciarrino}}]{UnitaryTomography3}%
  \BibitemOpen
  \bibfield  {author} {\bibinfo {author} {\bibfnamefont {N.}~\bibnamefont
  {Spagnolo}}, \bibinfo {author} {\bibfnamefont {E.}~\bibnamefont {Maiorino}},
  \bibinfo {author} {\bibfnamefont {C.}~\bibnamefont {Vitelli}}, \bibinfo
  {author} {\bibfnamefont {M.}~\bibnamefont {Bentivegna}}, \bibinfo {author}
  {\bibfnamefont {A.}~\bibnamefont {Crespi}}, \bibinfo {author} {\bibfnamefont
  {R.}~\bibnamefont {Ramponi}}, \bibinfo {author} {\bibfnamefont
  {R.}~\bibnamefont {Osellame}},\ and\ \bibinfo {author} {\bibfnamefont
  {F.}~\bibnamefont {Sciarrino}},\ }\bibfield  {title} {\bibinfo {title}
  {Learning an unknown transformation via a genetic approach},\ }\href
  {https://doi.org/10.1038/s41598-017-14680-7} {\bibfield  {journal} {\bibinfo
  {journal} {Sci. Rep.}\ }\textbf {\bibinfo {volume} {7}},\ \bibinfo {pages}
  {14316} (\bibinfo {year} {2017})}\BibitemShut {NoStop}%
\bibitem [{\citenamefont {Hoch}\ \emph {et~al.}(2023)\citenamefont {Hoch},
  \citenamefont {Giordani}, \citenamefont {Spagnolo}, \citenamefont {Crespi},
  \citenamefont {Osellame},\ and\ \citenamefont
  {Sciarrino}}]{UnitaryTomography4}%
  \BibitemOpen
  \bibfield  {author} {\bibinfo {author} {\bibfnamefont {F.}~\bibnamefont
  {Hoch}}, \bibinfo {author} {\bibfnamefont {T.}~\bibnamefont {Giordani}},
  \bibinfo {author} {\bibfnamefont {N.}~\bibnamefont {Spagnolo}}, \bibinfo
  {author} {\bibfnamefont {A.}~\bibnamefont {Crespi}}, \bibinfo {author}
  {\bibfnamefont {R.}~\bibnamefont {Osellame}},\ and\ \bibinfo {author}
  {\bibfnamefont {F.}~\bibnamefont {Sciarrino}},\ }\bibfield  {title} {\bibinfo
  {title} {Characterization of multimode linear optical networks},\ }\href
  {https://doi.org/10.1117/1.APN.2.1.016007} {\bibfield  {journal} {\bibinfo
  {journal} {Adv. Photonics Nexus}\ }\textbf {\bibinfo {volume} {2}},\ \bibinfo
  {pages} {016007} (\bibinfo {year} {2023})}\BibitemShut {NoStop}%
\bibitem [{\citenamefont {Abbas}\ \emph {et~al.}(2023)\citenamefont {Abbas},
  \citenamefont {King}, \citenamefont {Huang}, \citenamefont {Huggins},
  \citenamefont {Movassagh}, \citenamefont {Gilboa},\ and\ \citenamefont
  {McClean}}]{QuantumBackprop}%
  \BibitemOpen
  \bibfield  {author} {\bibinfo {author} {\bibfnamefont {A.}~\bibnamefont
  {Abbas}}, \bibinfo {author} {\bibfnamefont {R.}~\bibnamefont {King}},
  \bibinfo {author} {\bibfnamefont {H.-Y.}\ \bibnamefont {Huang}}, \bibinfo
  {author} {\bibfnamefont {W.~J.}\ \bibnamefont {Huggins}}, \bibinfo {author}
  {\bibfnamefont {R.}~\bibnamefont {Movassagh}}, \bibinfo {author}
  {\bibfnamefont {D.}~\bibnamefont {Gilboa}},\ and\ \bibinfo {author}
  {\bibfnamefont {J.~R.}\ \bibnamefont {McClean}},\ }\href@noop {} {\bibinfo
  {title} {On quantum backpropagation, information reuse, and cheating
  measurement collapse}} (\bibinfo {year} {2023}),\ \Eprint
  {https://arxiv.org/abs/2305.13362} {arXiv:2305.13362 [quant-ph]} \BibitemShut
  {NoStop}%
\bibitem [{\citenamefont {Chau}\ and\ \citenamefont {Fu}(2015)}]{SAReview}%
  \BibitemOpen
  \bibfield  {author} {\bibinfo {author} {\bibfnamefont {M.}~\bibnamefont
  {Chau}}\ and\ \bibinfo {author} {\bibfnamefont {M.~C.}\ \bibnamefont {Fu}},\
  }\bibinfo {title} {An overview of stochastic approximation},\ in\ \href
  {https://doi.org/10.1007/978-1-4939-1384-8_6} {\emph {\bibinfo {booktitle}
  {Handbook of Simulation Optimization}}},\ \bibinfo {editor} {edited by\
  \bibinfo {editor} {\bibfnamefont {M.~C.}\ \bibnamefont {Fu}}}\ (\bibinfo
  {publisher} {Springer New York},\ \bibinfo {address} {New York, NY},\
  \bibinfo {year} {2015})\ pp.\ \bibinfo {pages} {149--178}\BibitemShut
  {NoStop}%
\bibitem [{\citenamefont {Lewis}\ \emph {et~al.}(2025)\citenamefont {Lewis},
  \citenamefont {Wiersema}, \citenamefont {Carrasquilla},\ and\ \citenamefont
  {Bose}}]{GeodesicUnitary}%
  \BibitemOpen
  \bibfield  {author} {\bibinfo {author} {\bibfnamefont {D.}~\bibnamefont
  {Lewis}}, \bibinfo {author} {\bibfnamefont {R.}~\bibnamefont {Wiersema}},
  \bibinfo {author} {\bibfnamefont {J.}~\bibnamefont {Carrasquilla}},\ and\
  \bibinfo {author} {\bibfnamefont {S.}~\bibnamefont {Bose}},\ }\href@noop {}
  {\bibinfo {title} {Geodesic algorithm for unitary gate design with
  time-independent hamiltonians}} (\bibinfo {year} {2025}),\ \Eprint
  {https://arxiv.org/abs/2401.05973} {arXiv:2401.05973 [quant-ph]} \BibitemShut
  {NoStop}%
\bibitem [{\citenamefont {Lee}\ \emph {et~al.}(2022)\citenamefont {Lee},
  \citenamefont {Hasegawa},\ and\ \citenamefont {Gao}}]{ComplexNN}%
  \BibitemOpen
  \bibfield  {author} {\bibinfo {author} {\bibfnamefont {C.}~\bibnamefont
  {Lee}}, \bibinfo {author} {\bibfnamefont {H.}~\bibnamefont {Hasegawa}},\ and\
  \bibinfo {author} {\bibfnamefont {S.}~\bibnamefont {Gao}},\ }\bibfield
  {title} {\bibinfo {title} {Complex-valued neural networks: A comprehensive
  survey},\ }\href {https://doi.org/10.1109/JAS.2022.105743} {\bibfield
  {journal} {\bibinfo  {journal} {IEEE/CAA J. Autom. Sin.}\ }\textbf {\bibinfo
  {volume} {9}},\ \bibinfo {pages} {1406} (\bibinfo {year} {2022})}\BibitemShut
  {NoStop}%
\bibitem [{\citenamefont {Clevert}\ \emph {et~al.}(2016)\citenamefont
  {Clevert}, \citenamefont {Unterthiner},\ and\ \citenamefont
  {Hochreiter}}]{ELUActivation}%
  \BibitemOpen
  \bibfield  {author} {\bibinfo {author} {\bibfnamefont {D.-A.}\ \bibnamefont
  {Clevert}}, \bibinfo {author} {\bibfnamefont {T.}~\bibnamefont
  {Unterthiner}},\ and\ \bibinfo {author} {\bibfnamefont {S.}~\bibnamefont
  {Hochreiter}},\ }\href@noop {} {\bibinfo {title} {Fast and accurate deep
  network learning by exponential linear units (elus)}} (\bibinfo {year}
  {2016}),\ \Eprint {https://arxiv.org/abs/1511.07289} {arXiv:1511.07289
  [cs.LG]} \BibitemShut {NoStop}%
\bibitem [{\citenamefont {Kingma}\ and\ \citenamefont
  {Ba}(2017)}]{AdamOptimizer}%
  \BibitemOpen
  \bibfield  {author} {\bibinfo {author} {\bibfnamefont {D.~P.}\ \bibnamefont
  {Kingma}}\ and\ \bibinfo {author} {\bibfnamefont {J.}~\bibnamefont {Ba}},\
  }\href@noop {} {\bibinfo {title} {Adam: A method for stochastic
  optimization}} (\bibinfo {year} {2017}),\ \Eprint
  {https://arxiv.org/abs/1412.6980} {arXiv:1412.6980 [cs.LG]} \BibitemShut
  {NoStop}%
\bibitem [{\citenamefont {Srivastava}\ \emph {et~al.}(2014)\citenamefont
  {Srivastava}, \citenamefont {Hinton}, \citenamefont {Krizhevsky},
  \citenamefont {Sutskever},\ and\ \citenamefont
  {Salakhutdinov}}]{Dropout2014}%
  \BibitemOpen
  \bibfield  {author} {\bibinfo {author} {\bibfnamefont {N.}~\bibnamefont
  {Srivastava}}, \bibinfo {author} {\bibfnamefont {G.}~\bibnamefont {Hinton}},
  \bibinfo {author} {\bibfnamefont {A.}~\bibnamefont {Krizhevsky}}, \bibinfo
  {author} {\bibfnamefont {I.}~\bibnamefont {Sutskever}},\ and\ \bibinfo
  {author} {\bibfnamefont {R.}~\bibnamefont {Salakhutdinov}},\ }\bibfield
  {title} {\bibinfo {title} {Dropout: a simple way to prevent neural networks
  from overfitting},\ }\href {https://doi.org/10.5555/2627435.2670313}
  {\bibfield  {journal} {\bibinfo  {journal} {J. Mach. Learn. Res.}\ }\textbf
  {\bibinfo {volume} {15}},\ \bibinfo {pages} {1929–1958} (\bibinfo {year}
  {2014})}\BibitemShut {NoStop}%
\bibitem [{\citenamefont {Smith}\ \emph {et~al.}(2017)\citenamefont {Smith},
  \citenamefont {Kindermans}, \citenamefont {Ying},\ and\ \citenamefont
  {Le}}]{BatchSizeIncrease}%
  \BibitemOpen
  \bibfield  {author} {\bibinfo {author} {\bibfnamefont {S.~L.}\ \bibnamefont
  {Smith}}, \bibinfo {author} {\bibfnamefont {P.-J.}\ \bibnamefont
  {Kindermans}}, \bibinfo {author} {\bibfnamefont {C.}~\bibnamefont {Ying}},\
  and\ \bibinfo {author} {\bibfnamefont {Q.~V.}\ \bibnamefont {Le}},\
  }\href@noop {} {\bibinfo {title} {Don't decay the learning rate, increase the
  batch size}} (\bibinfo {year} {2017}),\ \Eprint
  {https://arxiv.org/abs/1711.00489} {arXiv:1711.00489 [cs.LG]} \BibitemShut
  {NoStop}%
\end{thebibliography}

%apsrev4-2.bst 2019-01-14 (MD) hand-edited version of apsrev4-1.bst
%Control: key (0)
%Control: author (8) initials jnrlst
%Control: editor formatted (1) identically to author
%Control: production of article title (0) allowed
%Control: page (0) single
%Control: year (1) truncated
%Control: production of eprint (0) enabled
%

\end{document}